\documentclass[12pt]{article}%
\usepackage{heck}
\usepackage{cite}
\usepackage{graphicx}
\usepackage{makeidx}
\usepackage{multicol}
\usepackage{geometry}
\usepackage{amsfonts}
\usepackage{mathrsfs}
\usepackage{amssymb}
\usepackage{amsmath}%
\providecommand{\U}[1]{\protect\rule{.1in}{.1in}}
\numberwithin{equation}{section}

\def\^{{\wedge}}
\def\*{{\star}}

\begin{document}

\date{September, 2008}

\preprint{arXiv:0809.1098}

\institution{HarvardU}{Jefferson Physical Laboratory, Harvard University, Cambridge,
MA 02138, USA}%

\title{F-theory, GUTs, and the Weak Scale}
%

\authors{Jonathan J. Heckman\footnote{e-mail: {\tt
jheckman@fas.harvard.edu}} and
Cumrun Vafa\footnote{e-mail: {\tt vafa@physics.harvard.edu}%
}}%

\abstract{In this paper we study a deformation of gauge mediated supersymmetry breaking
in a class of local F-theory GUT models where the scale of supersymmetry breaking
determines the value of the $\mu$ term.  Geometrically correlating these two scales constrains
the soft SUSY breaking parameters of the MSSM.  In this scenario, the hidden SUSY
breaking sector involves an anomalous $U(1)$ Peccei-Quinn symmetry which
forbids bare $\mu$ and $B\mu$ terms.  This sector typically breaks supersymmetry
at the desired range of energy scales through a simple stringy hybrid of a Fayet and
Polonyi model.   A variant of the Giudice-Masiero mechanism generates the value
$\mu \sim 10^{2} - 10^{3}$ GeV when the hidden sector scale of supersymmetry breaking
is ${\sqrt F} \sim 10^{8.5}$ GeV.  Further, the $B\mu$ problem is solved
due to the mild hierarchy between the GUT scale and Planck scale.  These models relate
SUSY breaking with the QCD axion, and solve the strong CP problem through an axion
with decay constant $f_{a} \sim M_{GUT}\cdot \mu / \Lambda$, where $\Lambda \sim 10^{5}$ GeV
is the characteristic scale of gaugino mass unification in gauge mediated models, and
the ratio $\mu / \Lambda \sim M_{GUT}/M_{pl} \sim 10^{-3}$. We find $f_a\sim 10^{12}$ GeV, which is near the high end of the phenomenologically
viable window.  Here, the axino is the goldstino mode which is eaten by the gravitino.  The
gravitino is the LSP with a mass of about $10^{1}-10^{2}$ MeV, and a bino-like neutralino is (typically) the NLSP
with mass of about $10^{2}-10^{3}$ GeV.  Compatibility with electroweak symmetry breaking also
determines the value of $\text{tan}\beta \sim 30 \pm 7$.}%

\maketitle

\tableofcontents

\pagebreak

\section{Introduction}

\ The existence of the string theory landscape with its vast number of
different low energy signatures potentially dilutes the predictive power of
string theory. \ The sheer range of possibilities presents a challenge to
determine which corners of the landscape (if any!) are consistent with experiment.

Decoupling the dynamics of gauge theory from gravity provides an attractive
way to constrain this problem. \ Indeed, at energy scales even a few orders of
magnitude below the Planck scale, gravity will most likely play a subdominant
role compared to other degrees of freedom. \ For considerations at lower
energies it is therefore quite natural to restrict attention to vacua in the
landscape with the correct gauge and matter content, deferring all issues
pertaining to gravity to a later stage of analysis. \ Given the vast size of
the landscape, it is also reasonable to incorporate some additional principles
in our search for vacua which are consistent with observation.

One such constraint is the apparent unification of the gauge coupling
constants of the MSSM at an energy below the Planck scale. \ Here, it is
important to note that the existence of a Grand Unified Theory (GUT) is in
principle compatible with the existence of a decoupling limit. \ For example,
in minimal realizations of GUTs, the gauge theory is typically asymptotically
free. \ Indeed, without asymptotic freedom, the UV completion of the gauge
theory would require incorporating gravity into the theory. \ Moreover, while
string theory can accommodate the generic representations of GUTs, the total
matter content in such models often does not produce an asymptotically free
theory. \ In this regard, the existence of a decoupling limit is especially
helpful in limiting any search for realistic models.

In string compactifications, the requirement that gravity can in principle
decouple translates into the geometric condition that some of the dimensions
of the compactification can in principle expand to a large size while the GUT
model degrees of freedom remain localized on a compact cycle. \ Roughly
speaking, the decoupling principle separates the \textquotedblleft open
string\textquotedblright\ and \textquotedblleft closed
string\textquotedblright\ sectors of the landscape.

In \cite{BHVI,BHVII}, we initiated a study of F-theory GUTs which admit such a
decoupling limit. \ These local F-theory models provide a surprisingly rigid
framework for model building \cite{BHVII}. See see also
\cite{DonagiWijnholt,WatariTATARHETF,IbanezSUSYFTHEORY,MarsanoGMSB,DonagiWijnholtBreak}
for related work on model building in F-theory. \ In F-theory, local
GUT\ models originate from a stack of seven-branes wrapping a del Pezzo
surface equipped with gauge group $SU(5)$, or some larger rank GUT group which
contains $SU(5)$. \ Although the matter content of these models can in
principle originate from either bulk eight-dimensional fields propagating on
the seven-brane, or six-dimensional fields localized at the intersection of
distinct stacks of seven-branes, in minimal $SU(5)$ GUT models, all of the
chiral matter of the MSSM\ localizes at the intersection of seven-branes along
Riemann surfaces in the del Pezzo surface. \ The GUT\ group breaks to the
Standard Model gauge group in the presence of an internal $U(1)$ hypercharge
flux through the del Pezzo surface.

As shown in \cite{BHVII}, simply achieving the correct matter content of the
MSSM\ through an appropriate choice of internal fluxes turns out to
simultaneously address several puzzles in four-dimensional GUT\ models. \ For
example, solving the doublet-triplet splitting problem via an internal $U(1)$
hyperflux required for GUT\ group breaking automatically forbids quartic
superpotential terms which can cause rapid proton decay. \ The presence of
this same hyperflux can also qualitatively explain why the lighter two
generations violate the analogue of the well-known $b$-$\tau$ mass relation.
\ Precisely because this framework is so rigid, it is also possible to
reliably extract predictions for the neutrino masses which are fortuitiously
in accord with current experimental bounds.

To make further contact with observation, any viable model must incorporate a
sector which breaks supersymmetry, and a mediation mechanism which
communicates this to the visible sector. \ From the perspective of the
four-dimensional effective field theory, this mediation mechanism is likely to
originate from gravity/moduli mediation, gauge mediation, or some variant of
these two basic possibilities.\footnote{It is also possible to consider other
scenarios such as anomaly mediation \cite{RSAMSB,GiudiceAMSB} where the scale
of supersymmetry breaking in the hidden sector is typically higher than in
gravity/moduli mediation. \ As another somewhat related possibility, it is
also possible to consider $U(1)$ mediation models as in
\cite{Dermisek:2007qi,Verlinde:2007qk} and its recent implementation in a
compact string based model with F- and D- term mixing \cite{Grimm:2008ed}.
\ To mention just one further possibility, D-brane instantons could perhaps
provide a more stringy mediation mechanism \cite{BuicanINST}.} \ Both
mediation mechanisms contain problems, either with correlating the scale of
supersymmetry breaking with the superpotential contribution to the Higgs mass
$\mu H_{u}H_{d}$ known as the $\mu$ term, or from large flavor changing
neutral currents (FCNCs). \ In gravity/moduli mediation scenarios, the $\mu$
problem can be solved via the Giudice-Masiero mechanism \cite{GiudiceMasiero},
but the generic pattern of soft masses will often generate large FCNCs, which
is typically not a problem in models of gauge mediated supersymmetry breaking.
\ Nevertheless, gravity/moduli mediation has recently been investigated in
F-theory models as in the recent attractive proposal of
\cite{IbanezSUSYFTHEORY}. \ See
\cite{Conlon:2005ki,Conlon:2006tj,Conlon:2007dw,Conlon:2007gk} and especially
\cite{Conlon:2006wz} for further details on moduli mediation scenarios in type
IIB compactifications.

By contrast, the primary stumbling block to realistic phenomenology in most
models of gauge mediation is the $\mu/B\mu$ problem. \ Indeed, perturbative
gauge interactions do not generate superpotential terms involving the Higgs
fields. \ Precisely because gauge fields mediate the effects of supersymmetry
breaking to the visible sector, correlating the size of the $\mu$ term with
supersymmetry breaking remains somewhat obscure in such a
scenario.\ \ Logically speaking, however, the values of $\mu$ and $B\mu$ may
have nothing to do with supersymmetry breaking. \ For example, in
\cite{BHVII}, exponential wave function suppression near the GUT model
seven-brane could also generate small values for the $\mu$ term. \ Although
this relieves much of the tension present in gauge mediated scenarios, it is
also not very predictive! \ In this paper, we shall therefore not consider
this possibility further.

A central goal of the present paper is to address the $\mu/B\mu$ problem in
variants of gauge mediated supersymmetry breaking which can arise in F-theory.
\ Focussing on gauge mediation is particularly reasonable in the context of
decoupling the \textquotedblleft open\textquotedblright\ and \textquotedblleft
closed\textquotedblright\ string sectors of the landscape. However, as
explained above, correlating the origin of $\mu/B\mu$ with supersymmetry
breaking requires some deformation away from a minimal realization of the
gauge mediation scenario. \ In fact, we find that solving the $\mu$ problem in
F-theory sometimes \textit{requires} that the mediation mechanism can decouple
from gravity.

Regardless of the mediation mechanism, one natural way to correlate the value
of $\mu$ with the scale of supersymmetry breaking is through variants of the
Giudice-Masiero mechanism. \ In this solution to the $\mu$ problem, the Higgs
chiral superfields $H_{u}$ and $H_{d}$ couple to a GUT\ group singlet $X$
through an interaction term of the form:%
\begin{equation}
O_{X^{\dagger}H_{u}H_{d}}=\gamma\int d^{4}\theta\frac{X^{\dagger}H_{u}H_{d}%
}{M_{X}}\text{,}%
\end{equation}
where $\gamma$ is an order one constant which depends on the details of the
model, and $M_{X}$ is roughly the scale at which this operator is generated.
\ When $X$ develops a supersymmetry breaking vev $\left\langle X\right\rangle
=x+\theta^{2}F$, the resulting contribution to the $\mu$ term is $\gamma
\cdot\overline{F}/M_{X}$. \ In gravity-mediated scenarios, $F\sim
10^{21}-10^{22}$ GeV$^{2}$, and $M_{X}$ is identified with the Planck scale.
\ For lower values of $F\leq10^{19}$ GeV$^{2}$ consistent with gauge
mediation, the induced value of $\mu$ is near the weak scale provided $M_{X}$
is near or below the GUT scale. \ This curious numerology has been observed in
\cite{HallLykkenWeinberg} and has recently formed the basis for the
\textquotedblleft sweet spot\textquotedblright\ model of supersymmetry
breaking \cite{KitanoIbeSweetSpot} (also see \cite{Lalak:2008bc}). \ See
\cite{MarsanoGMSB} for a recent effort in replicating the effective field
theory of the sweet spot scenario in the context of a local F-theory model.

One of the primary results of this paper is that integrating out the
Kaluza-Klein modes associated with $X$ generates the operator $O_{X^{\dagger
}H_{u}H_{d}}$ with $M_{X}$ near the GUT scale. \ To a certain extent, this is
to be expected because if this operator is present in a local model where
gravity can in principle decouple, the scale of suppression will be set by the
GUT, rather than Planck scale. \ As a consequence, the value of $F$ required
in gravity mediated scenarios will generate a value for the $\mu$ term which
is far too large. \ Hence, crude considerations already reveal that in a broad
class of models, some variant of gauge mediation must transmit supersymmetry
breaking to the MSSM. \ As in most perturbatively realized gauge mediated
models, this implies that $X$ must also couple to at least one vector-like
pair of messenger fields $Y$ and $Y^{\prime}$ through the F-term:
\begin{equation}
O_{XYY^{\prime}}=\lambda\int d^{2}\theta XYY^{\prime}\text{.}%
\end{equation}
\ Although different from the models we present, for recent related work on
gauge mediated models of supersymmetry breaking in string theory, see
\cite{FloratosKokorelis,CveticGMSB,KumarGMSB,MarsanoGMSB}.

Of course, in order for this variant on the Giudice-Masiero mechanism to solve
the $\mu$ problem, additional unwanted contributions to this term must also be
absent. \ For example, to prevent the presence of a large bare $\mu$ term, it
is natural for all fields to transform under a $U(1)$ Peccei-Quinn (PQ)
symmetry with charges:%
\begin{equation}%
\begin{tabular}
[c]{|r|r|r|r|r|}\hline
& $\Phi$ & $H_{u},H_{d}$ & $Y,Y^{\prime}$ & $X$\\\hline
$U(1)_{PQ}$ & $+1$ & $-2$ & $+2$ & $-4$\\\hline
\end{tabular}
\label{PQTABLE}%
\end{equation}
where in the above table, $\Phi$ denotes any chiral superfield of the MSSM
other than the Higgs up/down. \ This symmetry allows all requisite interaction
terms of the MSSM as well as the operators $O_{X^{\dagger}H_{u}H_{d}}$ and
$O_{XYY^{\prime}}$, but forbids the typically problematic superpotential terms
such as a bare $\mu$ term $\mu H_{u}H_{d}$, as well as $XH_{u}H_{d}$. \ As
explained in \cite{BHVII}, the existence of such $U(1)$ symmetries is quite
common in F-theory compactifications because matter fields originate from the
intersection of distinct stacks of seven-branes.

We also find that the presence of the $U(1)_{PQ}$ symmetry generates
\textit{additional} soft term contributions beyond those present in a gauge
mediated model. \ This $U(1)_{PQ}$ symmetry is typically anomalous and is
therefore Higgsed at high energies. \ In this case, heavy $U(1)_{PQ}$ gauge
boson exchange between $X$ and a generic chiral superfield $\Psi$ charged
under $U(1)_{PQ}$ generates the term:%
\begin{equation}
O_{X^{\dag}X\Psi^{\dag}\Psi}=-4\pi\alpha_{PQ}\frac{e_{X}e_{\Psi}}%
{M_{U(1)_{PQ}}^{2}}\int d^{4}\theta X^{\dag}X\Psi^{\dag}\Psi
\end{equation}
where in the above, $M_{U(1)_{PQ}}$ is the mass of the heavy gauge boson which
is typically on the order of the GUT scale, and the $e$'s denote the
$U(1)_{PQ}$ charges of the chiral superfields $X$ and $\Psi$. \ When $X$
develops a supersymmetry breaking vev, this term can also contribute to the
soft mass terms of the $\Psi$ fields. \ Depending on its precise value, this
contribution can lead to an interesting \textit{predictive deformation of the
gauge mediation scenario} which we also study.

Although seemingly unrelated, the physics of the $X$ field and the anomalous
$U(1)_{PQ}$ can also solve the strong CP problem. \ This comes about because
in gauge mediated models, the phase for the scalar component of $X$ couples to
the QCD instanton density through the axion-like coupling:%
\begin{equation}
L_{ax}\supset\frac{\arg x}{32\pi^{2}}\varepsilon^{\mu\nu\rho\sigma}TrF_{\mu
\nu}F_{\rho\sigma}\text{.} \label{argxQCD}%
\end{equation}
If the $U(1)_{PQ}$ symmetry had been non-anomalous, the vev of $X$ would have
broken this symmetry and the corresponding Goldstone mode $\arg x$ would
correspond to the longitudinal component of the massive $U(1)_{PQ}$ gauge
boson, thus eliminating $\arg x$ as a candidate axion field. \ Precisely
because this $U(1)$ symmetry is already Higgsed at high scales, a linear
combination of $\arg x$ with another bulk mode axion can play the role of the
QCD axion, solving the strong CP\ problem. \ A variant of this same idea for
four-dimensional effective field theories has also appeared in
\cite{KaplanChengCP}, where an additional $X$ field also participates. \ In
that model, one linear combination of the two phases corresponds to the
Goldstone mode of the broken $U(1)_{PQ}$ symmetry, while another linear
combination plays the role of the QCD axion.

As should now be clear, the natural appearance of an anomalous $U(1)_{PQ}$
symmetry in this class of models has many benefits. Such symmetries also fit
quite naturally with the programme of unifying the gauge interactions of the
MSSM. \ This fact is not new, and has been appreciated for some time in the
context of four-dimensional $E_{6}$ GUT models where the chiral matter of the
MSSM\ and Higgs fields all descend from the $27$. \ The $U(1)_{PQ}$ embeds as
the abelian factor of the subgroup $SO(10)\times U(1)\subset E_{6}$ because
the $27$ decomposes as:%
\begin{align}
E_{6}  &  \supset SO(10)\times U(1)\\
27  &  \rightarrow1_{+4}+10_{-2}+16_{+1}%
\end{align}
which contains both the Higgs fields $(10_{-2})$ and the chiral matter
$(16_{+1})$ of the MSSM. \ Note, however, that four-dimensional models will
also contain many additional states beyond those required to accommodate the
MSSM. \ These additional states significantly reduce the appeal of starting
with such large four-dimensional gauge groups.

The situation is more flexible in local F-theory models precisely because
matter fields can organize into representations of a group which contains the
four-dimensional gauge group as a proper subset. \ Geometrically, this is a
consequence of the fact that the rank of the singularity type can jump by more
than one rank along matter curves. \ When this occurs, we find that the matter
content of the local F-theory GUT\ models typically come charged under a
$U(1)$ Peccei-Quinn symmetry, rendering the matter content of our proposal
technically natural. \ While most four-dimensional GUTs focus on the role of
the $27$ of $E_{6}$, we note that the messenger fields can come from the
$10_{+2}\subset\overline{27}$ and the $X$ field can come from the
$1_{-4}\subset\overline{27}$. \ One bonus feature of this scenario is that
matter parity -- a symmetry which is often invoked to prevent rapid nucleon
decay in the MSSM -- can also be identified as a $%
\mathbb{Z}
_{2}$ subgroup of $U(1)_{PQ}$.

In any complete model of supersymmetry breaking, it is also important to
specify the mechanism by which supersymmetry is broken. \ In keeping with the
philosophy of decoupling, we present one stringy/higher dimensional gauge
theoretic realization of supersymmetry breaking which only utilizes degrees of
freedom associated with the $U(1)_{PQ}$ seven-brane. This $U(1)$ symmetry is
anomalous and instanton effects will generate a term linear in the $X$ field
which violates this symmetry. \ Nevertheless, this gauge symmetry leaves
behind an important remnant in the form of a D-term potential with a
non-trivial Fayet-Iliopoulos term determined by the background field strength
on the Peccei-Quinn seven-brane. \ We find that this hybrid Fayet-Polonyi
model typically breaks supersymmetry in the range required for our local
F-theory models. \ Further details on such instanton generated effects in
local F-theory models have recently been studied in \cite{HMSSNV} (see also
\cite{MarsanoToolbox}).

As noted above, the numerology of these local F-theory GUTs is remarkably
constrained and we have used this fact to extract definite predictions for the
low energy spectrum. \ We find that our model satisfies all of the required
crude phenomenological constraints when:%
\begin{align}
F  &  \sim10^{17}\text{ GeV}^{2}\\
M_{mess}  &  \sim x\sim10^{12}\text{ GeV}\\
M_{X}  &  \sim10^{15.5}\text{ GeV}%
\end{align}
where $M_{mess}$ is the mass of the messenger fields. \ Moreover we will show
that obtaining these specific values does not involve any unnatural
fine-tunings. The resulting values for the universal gaugino contribution
$F/x$, $\mu$ term and axion decay constant are:%
\begin{align}
\Lambda &  =\frac{F}{x}\sim10^{5}-10^{6}\text{ GeV}\\
\mu &  =\gamma\frac{\overline{F}}{M_{X}}\sim10^{2}-10^{3}\text{ GeV}\\
f_{a}  &  =\sqrt{2}\left\vert x\right\vert =\sqrt{2}\frac{M_{X}}{\gamma}%
\frac{\mu}{\Lambda}\sim10^{12}\text{ GeV,}%
\end{align}
where $\gamma\sim10$. \ We also find that $B\mu\sim0$ and the A-terms vanish
near the messenger scale. \ The small hierarchy in scales determined by the
ratio $\mu/ \Lambda\sim10^{-3}$ is in fact correlated with the small hierarchy
between the GUT scale and Planck scale $M_{GUT} / M_{pl} \sim10^{-3}$. \ As we
show in the explicit realization of the Fayet-Polonyi model, this same ratio
also determines the value of the intermediate scale $10^{12}$ GeV.

Armed with the specific UV\ boundary conditions of our model, it is in fact
possible to extract detailed properties of the low energy theory. \ The reason
this is possible is primarily due to the predictive features of gauge
mediation scenarios as well as the low energy requirement that electroweak
symmetry breaking takes place. \ Even the PQ\ deformation away from gauge
mediation is sufficiently simple to retain much of the predictive power of
this scenario. \ Iterating all parameters under renormalization group flow
between the weak scale and messenger scale allows us to constrain both the
UV\ and IR\ behavior of the theory. \ To perform this analysis, we have used
the program \texttt{SOFTSUSY} \cite{SOFTSUSYAllanach}.

As in most gauge mediated scenarios, the gravitino is the LSP, and in our
models it has a mass of order $10^{1}-10^{2}$ MeV. \ We find that consistent
electroweak symmetry breaking requires that $\tan\beta\sim30\pm7$ for
$\Lambda\sim10^{5.5\mp0.5}$ GeV. \ In the case of a single vector-like pair of
messenger fields, the NLSP\ is typically a bino-like neutralino, while for a
larger number of vector-like pairs $(>3$ vector-like pairs of $5
\oplus\overline{5})$, the stau will instead become the NLSP. \ We also find
that the NLSP\ can correspond to the stau in a single messenger model when a
large PQ\ deformation is present.

The organization of this paper is as follows. \ In section \ref{FTHREVIEW} we
briefly review some features of the local F-theory GUT\ models found in
\cite{BHVI,BHVII}. \ Next, in section \ref{ReviewMU} we review the $\mu/B\mu$
problem and in particular its relation to gauge mediated supersymmetry
breaking. \ In section \ref{HIGHERDIMOPS} we show that a broad class of local
F-theory models contain Giudice-Masiero operators which require the scale of
supersymmetry breaking to be lower than in gravity/moduli mediation scenarios.
\ Motivated by this result, in section \ref{DIAMONDRING}, we present a minimal
realization of gauge mediation in this context. \ In section \ref{AXIONGMSB}
we show that this class of models typically contain a candidate QCD axion with
decay constant in the phenomenologically viable range. \ Section
\ref{E6MOTIVATION} presents a sketch of a less minimal GUT model based on
embedding $U(1)_{PQ}$ in $E_{6}$, and in section \ref{FPMODEL} we study the
dynamics of the anomalous $U(1)_{PQ}$ theory and show that it can break
supersymmetry in the range required from bottom up considerations. \ We next
determine in section \ref{MSSMREGION} the region of MSSM parameter space for
this class of models, and section \ref{CONCLUDE} presents our conclusions and
potential directions for future investigation.

\section{Review of Minimal F-theory GUTs\label{FTHREVIEW}}

We now briefly review the basic features of local F-theory GUT\ models studied
in \cite{BHVII}. \ F-theory can be viewed as a strong coupling limit of
type\ IIB\ string theory where the axio-dilaton is interpreted as the complex
structure modulus of an elliptic curve. \ F-theory compactified on an
elliptically fibered Calabi-Yau fourfold leads to a four-dimensional
$\mathcal{N}=1$ theory below the scale of compactification. \ This elliptic
fibration can degenerate to a singularity of $ADE$ type over complex
codimension one subspaces of the threefold base $B_{3}$. These loci are
interpreted as the location of seven-branes with corresponding gauge group of
$ADE$ type. \ The singularity type can enhance further over complex
codimension two subspaces. \ This is interpreted as the intersection of
distinct seven-branes, which we shall refer to as matter curves. \ In this
way, it is possible to achieve matter content such as the spinor
representation of $SO(10)$, something which cannot be realized in perturbative
type IIB\ constructions. \ Finally, the singularity type can enhance even
further at points of $B_{3}$, where three matter curves meet. \ These terms
lead to additional contributions to the superpotential through wave function
triple overlaps. \ Although forbidden in perturbative type II\ constructions,
GUT\ model interaction terms such as $5_{H}\times10_{M}\times10_{M}$ naturally
occur at $E$-type enhancements of the compactification. \ See \cite{BHVI} for
a detailed discussion of the relevant interaction terms.

As found in \cite{BHVII}, the existence of a limit where the gauge dynamics of
the GUT model can decouple from gravity turns out to impose surprisingly
powerful restrictions on the ultraviolet behavior of the gauge theory. \ To
decouple the GUT, the seven-brane must wrap a del Pezzo surface. \ There is
essentially one such surface, called del Pezzo eight which is defined by
$\mathbb{P}^{2}$ blown up at eight points. \ All other del Pezzo surfaces can
be obtained from this one by blowing down appropriate two-cycles. \ As shown
in \cite{BHVI}, the zero mode content for del Pezzo models never contains an
adjoint valued chiral superfield, and the GUT group instead breaks to the
Standard Model group through an internal flux in the hypercharge direction of
the Standard Model gauge group. \ In these GUT models, the Higgs fields
localize on curves where the net hyperflux is non-trivial, whereas the chiral
matter localizes on curves with vanishing net hyperflux. \ In this way,
general arguments based on index theory show that the chiral matter will
organize into complete GUT multiplets while for an appropriate choice of flux,
doublet triplet splitting will automatically occur. \ Index theory also
requires the Higgs fields to localize on distinct curves, and this turns out
to automatically forbid quartic superpotential terms responsible for proton
decay. \ This framework is rigid enough that reliable estimates of neutrino
masses can be achieved. \ Remarkably, the resulting values for the light
neutrinos are in accord with current experimental bounds. \ To extract further
low energy predictions, it is necessary to specify the mechanism by which
supersymmetry is broken as well as the way in which this breaking communicates
to the MSSM. \ The aim of this paper is to study this question in the context
of gauge mediated supersymmetry breaking.

\subsection{Supersymmetry Breaking and Local Models\label{susyLOCAL}}

Before proceeding to more specific aspects of supersymmetry breaking, we first
briefly discuss the primary assumptions under which we shall operate
throughout this paper. As indicated in the Introduction, the primary objective
of this paper is to develop a self-consistent scenario of supersymmetry
breaking in the context of a local model. Indeed, one of the advantages of
working within local models is that many aspects of Planck scale physics can
be deferred to a later stage of analysis. Within this framework, our goal will
be to determine how various aspects of supersymmetry breaking are correlated.

On the other hand, it is well known that in some cases, issues of moduli
stabilization can directly feed into aspects of supersymmetry breaking. In
order for such effects not to overwhelm the contributions from purely local
considerations, it is therefore necessary to assume that some supersymmetric
mechanism stabilizes most moduli. It is beyond the scope of the present paper
to address all aspects of moduli stabilization, and we shall therefore simply
assume that an appropriate mechanism is available.

In this regard, it is worth pointing out that moduli stabilization is also
intimately connected with the value of the cosmological constant. In keeping
with the spirit of the present class of models where gravity can in principle
decouple, we shall view our local considerations as imposing constraints on
the form of candidate global models, and in particular moduli stabilization scenarios.

\section{Review of Gauge Mediation and the $\mu/B\mu$ Problem\label{ReviewMU}}

In this section we briefly review gauge mediation and the $\mu/B\mu$ problem.
\ This material is primarily review and can safely be skipped by the reader
who is familiar with the relevant issues. \ Even so, our emphasis will be
slightly different than what is sometimes stressed in the literature. \ To
frame the discussion, we first recall the $\mu/B\mu$ problem of the MSSM, and
then proceed to describe how the Giudice-Masiero mechanism addresses this
issue in gravity/moduli mediated supersymmetry breaking. \ Motivated by the
potential presence of large FCNCs, we next review the salient features of
gauge mediated supersymmetry breaking models, and explain why the $\mu/B\mu$
problem is potentially more severe in that case.

The soft supersymmetry breaking terms of the MSSM\ Lagrangian determine the
low energy particle content. \ In order to solve the hierarchy problem without
significant fine-tuning, the soft masses must be within an order of magnitude
of the weak scale. \ An extensive phenomenological review of this sector of
the MSSM\ may be found in \cite{KaneLykken}. \ There are typically at least
three sectors in any viable model of supersymmetry breaking. \ These consist
of the visible sector, defined by the fields and interaction terms of the
MSSM, a hidden sector where supersymmetry breaking occurs, and a messenger
sector which communicates the breaking of supersymmetry in the hidden sector
to the visible sector. \ In direct mediation models, the messenger and
\textquotedblleft hidden\textquotedblright\ sectors\ are combined.

Supersymmetry breaking can originate from a violation of either the F-term
equations of motion, the D-term equations of motion, or some combination of
the two. \ In this paper we shall always assume that the effects of
supersymmetry breaking can be parameterized by the expectation value of an
MSSM gauge singlet chiral superfield $X$ with vev:%
\begin{equation}
\left\langle X\right\rangle =x+\theta^{2}F\text{, } \label{susbreakvev}%
\end{equation}
so that the scale of supersymmetry breaking associated with the chiral
superfield $X$ is $\sqrt{F}$. \ After integrating out the messenger sector,
$X$ will couple to the fields of the MSSM. \ When $X$ attains the vev of
equation (\ref{susbreakvev}), it will subsequently induce various soft
supersymmetry breaking terms in the MSSM\ Lagrangian.

Although technically speaking the $\mu$ term of the MSSM is defined as a
contribution to the superpotential of the MSSM, as we now review, the
$\mu/B\mu$ problem suggests that this term and the soft supersymmetry breaking
terms of the MSSM\ possess a common origin. \ Because the $H_{u}$ and $H_{d}$
fields form a vector-like pair with respect to the quantum numbers of the
Standard Model gauge group $G_{std}$, there is a priori no reason to exclude
terms in the superpotential of the MSSM\ of the form:%
\begin{equation}
W\supset\mu H_{u}H_{d}\text{.}%
\end{equation}
There are various refinements of the $\mu$ problem, but at the crudest level,
it is the puzzling fact that the mass of a vector-like pair is generically
closer to the GUT scale rather than the weak scale. \ Similarly, there is no
reason to exclude large terms in the effective potential involving the scalar
components $h_{u}$ and $h_{d}$ of the respective chiral superfields $H_{u}$
and $H_{d}$:%
\begin{equation}
V_{eff}\supset B\mu h_{u}h_{d}\text{.}%
\end{equation}
Combined, these two issues define the weakest version of the $\mu/B\mu$ problem.

At the level of effective field theory, the bare $\mu$ term can be forbidden
by assuming that the Higgs superfields are both charged under a $U(1)$
Peccei-Quinn symmetry. \ In order to allow all of the necessary interaction
terms of the MSSM, all fields which couple to the Higgs fields must also be
appropriately charged under $U(1)_{PQ}$. \ As explained in \cite{BHVII}, such
symmetries are quite common in GUT models based on F-theory.

In F-theory, vector-like pairs interact by coupling with gauge singlet fields
localized on matter curves normal to the surface wrapped by the GUT\ model
seven-brane. \ For example, letting $X_{\bot}$ denote the singlet which
interacts with the Higgs fields, the corresponding contribution to the
superpotential is:%
\begin{equation}
W_{X_{\bot}H_{u}H_{d}}=\kappa X_{\bot}H_{u}H_{d}\text{.} \label{attractrepel}%
\end{equation}
Depending on the sign of fluxes through the curve supporting the $X_{\bot}$
field, the singlet wave function will either be attracted or repelled away
from the GUT\ model seven-brane. \ When it is repelled, the value of the
Yukawa coupling $\kappa$ can naturally be much smaller than an order one
number, so that even if the vev of $X_{\bot}$ is near the GUT\ scale, the
resulting $\mu$ term could still be quite small. \ Note in particular that
$B=0$ at high scales, and is only generated at lower energies by radiative
corrections to the effective potential for the Higgs scalars.

Even so, this solution is somewhat unsatisfactory because it does not explain
why $\mu$ is so close to the scale of supersymmetry breaking. \ This is a
stronger version of the $\mu/B\mu$ problem. \ On the other hand, it could well
be that these two scales are simply uncorrelated, in which case the above
mechanism provides a simple mechanism by which to solve the weak version of
the $\mu$ problem. \ Solving the strong version of the $\mu/B\mu$ problem is
one of the primary aims of this paper.

Beyond aesthetic concerns, there is a potentially more serious problem when
the coupling of equation (\ref{attractrepel}) contributes to the
superpotential. \ Indeed, when $X_{\bot}=X$, note that the vev of equation
(\ref{susbreakvev}) will generate contributions to the $\mu$ and $B\mu$ terms:%
\begin{align}
\mu &  =\kappa x\label{muind}\\
B\mu &  =\kappa F\text{.}%
\end{align}
In other words, when $\kappa$ is an order one coefficient, the vevs of $x$ and
$\sqrt{F}$ must already be quite close to the weak scale to avoid removing the
Higgs fields from the low energy spectrum. \ This is problematic because as we
will review shortly, in both gravity/moduli mediation and gauge mediation
scenarios, $\sqrt{F}$ is typically greater than the weak scale.

\subsection{Interpolating from Gravity/Moduli Mediation to Gauge Mediation}

Starting from higher values of the supersymmetry breaking scale set by
$\sqrt{F}$, we now review how a phenomenologically viable mediation mechanism
will interpolate from gravity/moduli mediation to gauge mediation at lower
values of $\sqrt{F}$. \ One conveniant way to parameterize the dominant
mediation mechanism is in terms of the gravitino mass:%
\begin{equation}
m_{3/2}=\sqrt{\frac{4\pi}{3}}\frac{\left\vert F \right\vert }{M_{pl}}%
\end{equation}
where $M_{pl}\sim1.2\times10^{19}$ GeV. \ In gravity mediation models, the
gravitino mass is around $10^{2}-10^{3}$ GeV, while in gauge mediation models,
this value is at the very most $1$ GeV. \ In certain cases, it is possible to
further increase the scale of supersymmetry breaking in anomaly mediated
scenarios. \ For further review, see for example, \cite{KitanoIbeSweetSpot}.

Gravity mediated supersymmetry breaking refers to any class of models where
the soft-breaking terms originate from Planck suppressed higher-dimension
operators. \ As such, in string theory it is more appropriate to refer to this
class of possibilities as gravity/moduli mediated supersymmetry breaking.
\ The $X$ field couples to the chiral superfields $\Psi$ of the MSSM\ through
terms such as:%
\begin{equation}
L_{soft}\supset\int d^{4}\theta\left(  \gamma_{pl}\frac{X^{\dag}H_{u}H_{d}%
}{M_{pl}}+\gamma_{\Psi}\frac{X^{\dag}X\Psi^{\dag}\Psi}{M_{pl}^{2}}\right)
\label{GiudiceMasTerm}%
\end{equation}
where the $\gamma$'s denote order one coefficients. \ When $X$ develops a vev
as in equation (\ref{susbreakvev}), the resulting theory will automatically
contain a contribution to the $\mu$-term and the soft scalar mass terms:%
\begin{align}
\mu_{eff}  &  \sim\frac{\overline{F}}{M_{pl}}\label{gravinducedmassterm}\\
m_{\Psi}^{2}  &  \sim\frac{\left\vert F \right\vert ^{2}}{M_{pl}^{2}}\text{.}%
\end{align}
In other words, the scale $\mu_{eff}$ correlates with the energy scale of the
soft mass terms, solving the $\mu$-problem. \ This means of generating the
$\mu$ term is known as the Giudice-Masiero mechanism \cite{GiudiceMasiero} and
we shall\ sometimes refer to $X^{\dag}H_{u}H_{d}$ as the Giudice-Masiero
operator. \ Note that by construction, the value of $\mu$ correlates with the
scale of supersymmetry breaking.

In order for the Higgs up/down fields to retain masses near the weak scale,
the contribution to the $\mu$ term from the Giudice-Masiero mechanism must be
at most $10^{3}$ GeV. \ Hence, $\left\vert F \right\vert $ is bounded above
by:\footnote{We note that this value can be increased even further in certain
scenarios such as anomaly mediation.}
\begin{equation}
\left\vert F \right\vert \lesssim10^{21}-10^{22}\text{ GeV}^{2}\text{.}%
\end{equation}

It is certainly appealing that in gravity/moduli mediated models, the $\mu$
term automatically correlates with the scale of the soft breaking terms. \ But
in gravity/moduli mediated scenarios, the coefficients $\gamma_{\Psi}$ in
equation (\ref{GiudiceMasTerm}) are also typically generic order one
coefficients, so that large flavor changing neutral currents (FCNCs) will be
present. \ For this reason, it is quite common to specify additional flavor
symmetries in order for such models to remain viable.

Nevertheless, the Planck suppressed contributions proportional to
$\gamma_{\Psi}$ in equation (\ref{GiudiceMasTerm}) will always be present, and
in the absence of an appropriate theory of flavor, will always generate FCNCs
when $X$ develops a supersymmetry breaking vev. \ Perhaps the simplest way to
avoid any problem with FCNCs is to lower the scale of supersymmetry breaking
so that $F/M_{pl}$ is at most $1$ GeV. \ As reviewed in
\cite{KitanoIbeSweetSpot}, for example, this bound is low enough to remain in
accord with observation.

If the contribution of the Planck suppressed operators to the soft mass terms
falls below the weak scale, other mediation mechanisms must account for the
soft breaking terms of the MSSM. \ Gauge mediation
\cite{DineFischlerSusyTech,DimopoulosRabySupercolor,DineFischlerII,NappiOvrut,DineFischlerIII,AlvarezGaumeClaudsonWiseLow,DimopoulosRabyGeometric,DineNelsonGMSB,DineNelsonShirman,DineNelsonNirShirman}
is a mechanism where the gauge fields of the Standard Model communicate
supersymmetry breaking to the\ MSSM. \ See \cite{GiudiceSUSYReview} for a
review of gauge mediation. \ In gauge mediation, the $X$ field couples in the
superpotential to \textquotedblleft messenger fields\textquotedblright\ which
either transform in a real representation of the Standard Model gauge group
$G_{std}=SU(3)_{C}\times SU(2)_{L}\times U(1)_{Y}$, or in vector-like pairs of
complex representations. \ In this paper we shall always assume that the
messenger fields transform in vector-like pairs of complex representations,
and we shall denote these pairs as $Y$ and $Y^{\prime}$.

As a brief aside, we note that this simplifying assumption is well-motivated
in F-theory models. \ For example in the minimal $SU(5)$ GUT\ models studied
in \cite{BHVII}, the only available representations are the $24$ for bulk
modes, and the $5$, $10$ or complex conjugates for modes which localize on
matter curves. \ Furthermore, when the GUT\ model seven-brane wraps a del
Pezzo surface, the vanishing theorem of \cite{BHVI} establishes that no zero
modes transform in the $24$ of $SU(5)$. \ Hence, all chiral superfields of the
four-dimensional effective theory charged under the GUT group must descend
from complex representations of $SU(5)$.

In this paper we shall further restrict attention to single messenger models
where the gauge singlet $X$ interacts with the messenger fields through the
superpotential term:%
\begin{equation}
W_{XYY}=\lambda XYY^{\prime} \label{MessYUK}%
\end{equation}
where $\lambda$ is generically an order one coefficient.\footnote{While it is
certainly possible to also consider models where the messenger sector is
strongly coupled, in the context of F-theory models, our expectation is that a
perturbative treatment will suffice for most purposes. \ The study of strongly
coupled messenger sectors is indeed a topic of current interest, see for
example \cite{MeadeSeibergShih,DistlerRobbins}. \ In the context of local
F-theory models with perturbative gauge dynamics, we shall make the reasonable
assumption that the messenger sector is sufficiently weakly coupled that a
perturbative treatment is available.} \ Even in this simple case, there is
additional structure present in most F-theory models because the wave
functions of different components of a complete GUT\ representation will in
general be different \cite{BHVII}.

More generally, it is in principle possible to consider models with a larger
number of messenger fields. \ On the other hand, in order to maintain the
existence of a decoupling limit, we must also require that the running of the
couplings in the zero mode sector should preserve asymptotic freedom. \ In an
$SU(5)$ model with three generations in the $\overline{5}_{M}\oplus10_{M}$, a
single vector-like pair of Higgs fields and $N_{mess}$ vector-like pairs of
messengers in the $5\oplus\overline{5}$ of $SU(5)$, the beta function is:%
\begin{align}
b_{GUT}  &  =3C_{2}(24)-3\left(  C_{2}(10_{M})+C_{2}(\overline{5}_{M})\right)
-2C_{2}(5_{H})-2N_{mess}\left(  C_{2}(5_{Y})\right) \\
&  =8-N_{mess}%
\end{align}
where in the above, $C_{2}$ denotes the quadratic Casimir of various
representations. \ In other words, asymptotic freedom limits the number of
messengers to $N_{mess}<8$. \ This is somewhat stronger than the usual
condition typically considered in the GUT\ literature where the coupling
constant must simply remain perturbative up to the GUT\ scale.

Once $X$ develops a vev as in equation (\ref{susbreakvev}), the coupling
between the gauge fields and the messenger fields will induce soft breaking
terms in the MSSM. \ Integrating out the $Y$ fields, the soft breaking terms
will contain the contributions:%
\begin{align}
L  &  \supset\int d^{4}\theta\left(  \underset{i=1}{\overset{3}{%
{\displaystyle\sum}
}}-\alpha_{i}^{2}C_{2}(R_{\Psi}^{i})\left(  \log\left\vert X\right\vert
^{2}\right)  ^{2}\Psi^{\dag}\Psi\right) \label{phisoftmass}\\
&  +\int d^{2}\theta\operatorname{Re}\left(  \underset{i=1}{\overset{3}{%
{\displaystyle\sum}
}}\frac{1}{8\pi i}\left(  \tau_{YM}^{(i)}+\frac{1}{2\pi i}\log X\right)
Tr_{G_{(i)}}W_{(i)}^{\alpha}W_{(i)\alpha}\right)  \label{gauginoterm}%
\end{align}
where $\Psi$ is shorthand for any chiral superfield of the MSSM, and
$\tau_{YM}^{(i)}$ denotes the holomorphic Yang-Mills coupling of the
$i^{\text{th}}$ gauge group factor $G_{(i)}$ so that:
\begin{equation}
\tau_{YM}=\frac{4\pi i}{g_{YM}^{2}}+\frac{\theta_{YM}}{2\pi}=\frac{i}%
{\alpha_{YM}}+\frac{\theta_{YM}}{2\pi}\text{.}%
\end{equation}

In gauge mediation, the soft masses of the gauginos follow from equation
(\ref{gauginoterm}) so that:%
\begin{equation}
m_{i}=\frac{\alpha_{i}}{4\pi}\frac{F}{x}\equiv\frac{\alpha_{i}}{4\pi}\Lambda
\end{equation}
where in the above, we have introduced the ratio $\Lambda=F/x$. \ Assuming
that the mass of the bino is on the order of $10^{2}$ GeV and originates
predominantly from gauge mediation requires:%
\begin{equation}
\Lambda=\frac{F}{x}\sim10^{5}\text{ GeV.}%
\end{equation}

One of the most important features of gauge mediation is that in the absence
of other sources of flavor violation, the soft masses depend only on the gauge
quantum numbers of a given field. \ In particular, this means that the
potentially dangerous FCNCs of gravity mediated models are quite suppressed.

Turning the discussion around, the presence of messenger fields in the low
energy spectrum could in principle be incompatible with a higher scale of
supersymmetry breaking such as that associated with gravity/moduli mediation.
\ For example, if the singlet field $X$ interacts with messenger fields
transforming in GUT\ multiplets, increasing the ratio $\Lambda=F/x$ will
simply increase the contribution to the soft masses due to gauge mediation.

Unfortunately, generating an appropriate value of the $\mu$ term in gauge
mediation is somewhat problematic. \ Indeed, if the only contribution to the
effective $\mu$ term is given by equation (\ref{gravinducedmassterm}), the
resulting value of $\mu$ would be far too low. \ The simplest attempts to
solve the $\mu/B\mu$ problem also fail. \ For example, returning to the
discussion near equation (\ref{muind}), directly coupling the $X$ field to the
Higgs fields through a term such as:%
\begin{equation}
W_{XH_{u}H_{d}}=\kappa XH_{u}H_{d} \label{XHUHD}%
\end{equation}
can indeed induce an appropriate value for the $\mu$ term. \ Note, however,
that $B$ is insensitive to the value of $\kappa$ because:%
\begin{equation}
B=\frac{B\mu}{\mu}=\frac{\kappa F}{\kappa x}=10^{5}\text{ GeV.}%
\end{equation}
Thus, unless $\kappa$ is quite small, generic values of $x$ and $F$ will
typically generate mass terms for the Higgs fields far above the weak scale.
\ While it is indeed possible in F-theory constructions to exponentially
suppress $\kappa$ so that both these terms are sufficiently small, the
alternative and conceptually simpler point of view which we shall adopt in
this paper is that the coupling of line (\ref{XHUHD}) should be absent from
the low energy superpotential.

\section{Variants of the Giudice-Masiero Mechanism and Higher Dimension
Operators\label{HIGHERDIMOPS}}

Although the Giudice-Masiero mechanism solves the $\mu$ problem in
gravity/moduli mediated scenarios, in gauge mediated models, the same
contribution to the $\mu$ term would be far below the weak scale. \ But while
this contribution to the low energy theory is somewhat innocuous, we have also
seen that the superpotential term $XH_{u}H_{d}$ should not be present. \ In
models with a $U(1)_{PQ}$ symmetry, the Giudice-Masiero mechanism and the
presence of the term $XH_{u}H_{d}$ are in fact mutually exclusive. \ Indeed,
when $H_{u}$, $H_{d}$ and $X$ are all charged under $U(1)_{PQ}$, the F-term
$XH_{u}H_{d}$ is forbidden whenever the D-term $X^{\dag}H_{u}H_{d}$ is
allowed, and the converse statement holds as well. \ Thus, the existence of
the Giudice-Masiero operator effectively frustrates the $B\mu$ term.

The Giudice-Masiero operator:%
\begin{equation}
O_{X^{\dag}H_{u}H_{d}}=\gamma_{pl}\int d^{4}\theta\frac{X^{\dag}H_{u}H_{d}%
}{M_{pl}}%
\end{equation}
can also generate a value of $\mu$ near the weak scale in gauge mediated
scenarios when the coefficient $\gamma_{pl}$ is sufficiently large. \ Said
differently, this is equivalent to replacing $\gamma_{pl}$ by an order one
coefficient and the suppression scale $M_{pl}$ by some lower energy scale
$M_{X}$, such as the GUT scale so that:%
\begin{equation}
O_{X^{\dag}H_{u}H_{d}}=\gamma\int d^{4}\theta\frac{X^{\dag}H_{u}H_{d}}{M_{X}%
}\text{.}%
\end{equation}
This observation is not new and has, for example, formed the basis of the
\textquotedblleft sweet spot\textquotedblright\ model of supersymmetry
breaking \cite{KitanoIbeSweetSpot}.

In a generic effective field theory, the typical situation is not as simple.
\ If we perform the natural identification $M_{X}=x$, this theory will also
generate a large $B\mu$ term via the operator:%
\begin{equation}
\int d^{4}\theta\frac{X^{\dag}XX^{\dag}H_{u}H_{d}}{M_{X}^{3}}\text{.}%
\end{equation}
Indeed, saturating the $d^{4}\theta$ integrals through the $X$ superfields
yields:%
\begin{equation}
B\mu=-\frac{|F|^{2}\overline{x}}{M_{X}^{3}}\text{.}%
\end{equation}
When $M_{X}=|x|$, this yields a problematically large value of the $B\mu$
term:%
\begin{equation}
\left\vert B\mu\right\vert =\left\vert \frac{F}{x}\right\vert ^{2}\text{.}%
\end{equation}
Thus, in order to solve the $\mu/B\mu$ problem, we must also explain why the
suppression scale in the above operators can be greater in magnitude than $x$.

Moreover, while intriguing, replacing $M_{pl}$ by $M_{X}$ in a
four-dimensional effective field theory is quite ad hoc. \ Nevertheless, in
this paper we will show that this replacement is quite natural in local
F-theory models and in certain cases unavoidable. \ For example, when
$F\sim10^{17}$ GeV$^{2}$ and $M_{X}\sim10^{15.5}$ GeV, this induces an
effective $\mu$ term:%
\begin{equation}
\mu_{eff}=\gamma\frac{\overline{F}}{M_{X}}\sim\gamma\cdot10^{1.5}\text{ GeV}%
\end{equation}
which without any fine-tuning is already quite close to the weak scale.
\ Further, as we show later in this paper, fluxes can naturally allow the
scale of $\left\vert x \right\vert $ to be much smaller than $M_{X}$.
\ Indeed, the mild hierarchy of scales $\left\vert x\right\vert /M_{X}%
\sim10^{-3}$ provides a solution to the $B\mu$ problem in the class of models
we consider.

Summarizing the bottom up considerations described above, we are interested in
four-dimensional effective field theories where the F-term coupling
$XH_{u}H_{d}$ is forbidden, but the D-term coupling $X^{\dag}H_{u}H_{d}$ is
allowed. \ In F-theory constructions, these two features are not only
compatible, but are in fact quite tightly correlated! \ Geometrically, the
matter curves supporting the $X$, $H_{u}$ and $H_{d}$ fields must form a
triple intersection in order for the D-term coupling to be gauge-invariant.
\ Remarkably, integrating out the Kaluza-Klein modes associated with $X$
generates a Giudice-Masiero operator suppressed by an energy scale close to
$M_{GUT}$.

This has important consequences for F-theory models where the $X$ field is
charged under a Peccei-Quinn symmetry and interacts with the Higgs fields
through a Giudice-Masiero term. \ Because the suppression scale of this
operator will at most be a few orders of magnitude below the Planck scale, it
follows that the resulting $\mu$ term would be far too large in gravity/moduli
mediated scenarios. \ In other words, we deduce that in a large class of
F-theory compactifications, the scale of supersymmetry breaking must be
sufficiently \textit{low} to remain in accord with electroweak symmetry breaking!

The presence of the $U(1)_{PQ}$ symmetry has another consequence in the low
energy theory. \ Whereas in a four-dimensional effective field theory
$U(1)_{PQ}$ can be treated as a global symmetry, in a quantum theory of
gravity, this symmetry must be gauged. \ When this $U(1)$ is non-anomalous, it
will also contribute to the soft scalar mass terms via the usual gauge
mediation mechanism. \ However, in F-theory constructions, this $U(1)$ is
typically anomalous, and the corresponding gauge boson will develop a large
mass via the Green-Schwarz mechanism. Precisely because all of the fields of
the MSSM must be charged under $U(1)_{PQ}$, heavy $U(1)_{PQ}$ gauge boson
exchange between MSSM fields and $X$ will generate additional contributions to
the soft scalar mass terms of the MSSM\ fields.

The rest of this section is organized as follows. \ First, we show that in
local F-theory models where the $U(1)_{PQ}$ symmetry allows the $X$ field to
couple to the Higgs fields through a Giudice-Masiero operator, integrating out
the Kaluza-Klein modes of the higher dimensional theory automatically
generates this operator. \ This establishes that in a large class of F-theory
models, gravity/moduli mediation would yield a value of the $\mu$ term which
is phenomenologically unviable. \ Next, we show that when the $U(1)_{PQ}$
symmetry is anomalous, heavy $U(1)_{PQ}$ gauge boson exchange can in some
cases generate important corrections to the soft scalar masses.

\subsection{Giudice-Masiero Operators\label{GMOP}}

In this subsection we show that in a broad class of F-theory models,
integrating out the Kaluza-Klein modes associated with $X$ generates a
Giudice-Masiero term which is suppressed by the Kaluza-Klein scale. \ We
emphasize that this is essentially a tree level computation, and that once
additional details of the compactification have been specified, the
coefficient of this higher dimension operator is completely calculable. \ This
is in contrast to standard arguments from effective field theory, which must
typically appeal to estimates based on genericity considerations.

To generate a Giudice-Masiero operator, we shall assume that the matter curves
$\Sigma_{X}$, $\Sigma_{H_{u}}$ and $\Sigma_{H_{d}}$ which support the fields
$X$, $H_{u}$ and $H_{d}$ form a triple intersection such that $XH_{u}H_{d}$ is
\textit{not} a gauge invariant operator. \ In this case, the term:%
\begin{equation}
O_{X^{\dag}H_{u}H_{d}}=\gamma\int d^{4}\theta\frac{X^{\dag}H_{u}H_{d}}{M_{X}}%
\end{equation}
\textit{does} correspond to a gauge invariant operator. This is because, as
explained in \cite{BHVI}, the six-dimensional fields from which $X$, $H_{u}$
and $H_{d}$ descend organize into vector-like pairs of four-dimensional
$\mathcal{N}=1$ chiral superfields $\mathbb{X\oplus X}^{c}$, $\mathbb{H}%
_{u}\mathbb{\oplus H}_{u}^{c}$ and $\mathbb{H}_{d}\mathbb{\oplus H}_{d}^{c}$
labelled by points on the matter curves. Thus, only one of the two interaction
terms $\mathbb{X}\mathbb{H}_{u}\mathbb{H}_{d}$ or $\mathbb{X}^{c}%
\mathbb{H}_{u}\mathbb{H}_{d}$ can descend to a superpotential term for the
zero modes. \ Assuming the $\mathbb{X}^{c}\mathbb{H}_{u}\mathbb{H}_{d}$
interaction is present, the fact that $\mathbb{X}^{\dag}$ and $\mathbb{X}^{c}$
have identical gauge quantum numbers implies that $O_{X^{\dag}H_{u}H_{d}}$ is
gauge invariant. \ To establish that $O_{X^{\dag}H_{u}H_{d}}$ is generated, we
first present the Lagrangian density for this system in four-dimensional
$\mathcal{N} = 1$ superspace:
\begin{align}
L  &  =\underset{\Sigma_{X}}{\int}d^{4}\theta\left[  \mathbb{X}^{\dag
}e^{V^{^{\prime}}}\mathbb{X}e^{V^{\prime\prime}}+\left(  \mathbb{X}%
^{c}\right)  ^{\dag}e^{-V^{^{\prime}}}\mathbb{X}^{c}e^{-V^{\prime\prime}%
}\right] \label{LEFFGiudiceMasiero}\\
&  +\underset{\Sigma_{u}}{\int}d^{4}\theta\left[  \left(  \mathbb{H}%
_{u}\right)  ^{\dag}e^{V^{\prime}}\mathbb{H}_{u}e^{+V}+\left(  \mathbb{H}%
_{u}^{c}\right)  ^{\dag}e^{-V^{\prime}}\mathbb{H}_{u}^{c}e^{-V}\right] \\
&  +\underset{\Sigma_{d}}{\int}d^{4}\theta\left[  \left(  \mathbb{H}%
_{d}\right)  ^{\dag}e^{+V^{\prime\prime}}\mathbb{H}_{d}e^{-V}+\left(
\mathbb{H}_{d}^{c}\right)  ^{\dag}e^{-V^{\prime\prime}}\mathbb{H}_{d}%
^{c}e^{+V}\right] \\
&  +\underset{\Sigma_{X}}{\int}d^{2}\theta\mathbb{X}^{c}\left(  \overline
{\partial}+\mathbb{A}^{\prime}+\mathbb{A}^{\prime\prime}\right)  \mathbb{X}\\
&  +\underset{\Sigma_{u}}{\int}d^{2}\theta\mathbb{H}_{u}^{c}\left(
\overline{\partial}+\mathbb{A}+\mathbb{A}^{\prime}\right)  \mathbb{H}_{u}\\
&  +\underset{\Sigma_{d}}{\int}d^{2}\theta\mathbb{H}_{d}^{c}\left(
\overline{\partial}+\mathbb{A}-\mathbb{A}^{\prime\prime}\right)
\mathbb{H}_{d}\\
&  +\underset{\Sigma}{\int}d^{2}\theta\left[  \delta_{p}\mathbb{X}%
^{c}\mathbb{H}_{u}\mathbb{H}_{d}+\delta_{p}\mathbb{XH}_{u}^{c}\mathbb{H}%
_{d}^{c}\right]  +h.c.\text{.}%
\end{align}
In the above, we have organized the gauge field contribution from the various
seven-branes into four-dimensional $\mathcal{N}=1$ supermultiplets so that the
$V$'s denote the contribution from vector multiplets which transform as bulk
mode scalars on the associated K\"{a}hler surface and the $\mathbb{A}$'s
denote the contribution from chiral superfields which transform as bulk mode
one-forms on the appropriate K\"{a}hler surface. \ Finally, $\delta_{p}$
denotes a delta function with support at a point of triple intersection. \ The
F-term equations of motion for the six-dimensional fields are:%
\begin{align}
\frac{\partial W}{\partial\mathbb{X}}  &  =-\overline{\partial}_{\mathbb{A}%
^{\prime}+\mathbb{A}^{\prime\prime}}\mathbb{X}^{c}+\delta_{p}\mathbb{H}%
_{u}^{c}\mathbb{H}_{d}^{c}=0\\
\frac{\partial W}{\partial\mathbb{X}^{c}}  &  =\overline{\partial}%
_{\mathbb{A}^{\prime}+\mathbb{A}^{\prime\prime}}\mathbb{X}+\delta
_{p}\mathbb{H}_{u}\mathbb{H}_{d}=0
\end{align}
with similar expressions for the $\mathbb{H}$ equations of motion. \ Expanding
about a fixed supersymmetric background gauge field configuration for the
fields $\mathbb{A}^{\prime}$ and $\mathbb{A}^{\prime\prime}$ which we denote
by $A^{\prime}$ and $A^{\prime\prime}$, solving for $\mathbb{X}^{c}$ and
$\mathbb{X}$ yields:%
\begin{align}
\mathbb{X}^{c}  &  =\frac{1}{\overline{\partial}_{A^{\prime}+A^{\prime\prime}%
}}\left(  \delta_{p}\mathbb{H}_{u}^{c}\mathbb{H}_{d}^{c}\right)
\label{XCEOM}\\
\mathbb{X}  &  =X-\frac{1}{\overline{\partial}_{A^{\prime}+A^{\prime\prime}}%
}\left(  \delta_{p}\mathbb{H}_{u}\mathbb{H}_{d}\right)  \label{XEOM}%
\end{align}
where in the above, we have included the zero mode $X$ which by definition, is
annihilated by $\overline{\partial}_{A^{\prime}+A^{\prime\prime}}$.

Substituting these expressions into equation (\ref{LEFFGiudiceMasiero}) is
equivalent to integrating out the Kaluza-Klein modes of the $\mathbb{X}$ and
$\mathbb{X}^{c}$ fields. \ The resulting effective action for the zero modes
therefore contains the term:%
\begin{align}
L_{eff}  &  \supset\underset{\Sigma_{X}}{\int}d^{4}\theta\left[  \left(
X-G_{A^{\prime}+A^{\prime\prime}}\left(  z,p\right)  H_{u}H_{d}\right)
^{\dag}e^{V^{^{\prime}}}\left(  X-G_{A^{\prime}+A^{\prime\prime}}\left(
z,p\right)  H_{u}H_{d}\right)  e^{V^{\prime\prime}}\right] \label{Leff}\\
&  +\underset{\Sigma_{u}}{\int}d^{4}\theta\left[  H_{u}^{\dag}e^{V^{\prime}%
}H_{u}e^{+V}\right]  +\underset{\Sigma_{d}}{\int}d^{4}\theta\left[
H_{d}^{\dag}e^{+V^{\prime\prime}}H_{d}e^{-V}\right]
\end{align}
where in the above, $G_{A^{\prime}+A^{\prime\prime}}$ denotes the Green's
function defined by the relation:%
\begin{equation}
\overline{\partial}_{A^{\prime}+A^{\prime\prime}}G_{A^{\prime}+A^{\prime
\prime}}\left(  z,p\right)  =\delta_{p}=\delta^{(2)}(z-p)
\end{equation}
where $z$ is a local coordinate on the Riemann surface $\Sigma_{X}$. \ Using
the approximation:%
\begin{equation}
\underset{\Sigma_{X}}{\int}G_{A^{\prime}+A^{\prime\prime}}\left(  z,p\right)
\sim\frac{M_{\ast}^{2}Vol(\Sigma_{X})}{M_{X}}%
\end{equation}
where $M_{X}$ denotes the mass scale of the Kaluza-Klein modes on the curve
$\Sigma_{X}$, canonically normalizing all Kinetic terms yields the Lagrangian
density of the $X$/Higgs system:%
\begin{align}
L_{eff}  &  \supset\int d^{4}\theta\left[  H_{u}^{\dag}e^{V^{\prime}}%
H_{u}e^{+V}+H_{d}^{\dag}e^{+V^{\prime\prime}}H_{d}e^{-V}+X^{\dag}e^{V^{\prime
}}Xe^{-V^{\prime\prime}}\right] \\
&  +\int d^{4}\theta\left[  \frac{\sqrt{M_{\ast}^{2}Vol(\Sigma_{X})}}%
{\sqrt{M_{\ast}^{2}Vol(\Sigma_{u})}\sqrt{M_{\ast}^{2}Vol(\Sigma_{d})}}%
\frac{X^{\dag}e^{V^{^{\prime}}}H_{u}H_{d}e^{V^{\prime\prime}}}{M_{X}}\right]
+h.c.\text{.}%
\end{align}
When $X$ develops a non-supersymmetric vev, this induces an effective $\mu
$-term:%
\begin{equation}
\mu_{eff}=\frac{\sqrt{M_{\ast}^{2}Vol(\Sigma_{X})}}{\sqrt{M_{\ast}%
^{2}Vol(\Sigma_{u})}\sqrt{M_{\ast}^{2}Vol(\Sigma_{d})}}\times\frac
{\overline{F}}{M_{X}}\equiv\gamma\frac{\overline{F}}{M_{X}}\text{.}
\label{MUEFFHERE}%
\end{equation}
A similar expression holds when additional $X$ field zero modes contribute to
the low energy theory. \ Assuming that $Vol(\Sigma_{u})=Vol(\Sigma
_{d})=M_{GUT}^{-2}$, the relations $M_{\ast}^{4}/M_{GUT}^{4}=\alpha_{GUT}$ and
$Vol(\Sigma_{X})=M_{X}^{-2}$ imply:%
\begin{equation}
\gamma=\alpha_{GUT}^{1/4}\frac{M_{GUT}}{M_{X}}\text{.} \label{gammaest}%
\end{equation}
As estimated in \cite{BHVII}, it is most natural for the Kaluza-Klein scale
$M_{X}$ be close to the GUT scale. \ When convenient, throughout this paper we
will use the representative values $M_{GUT}\sim3\times10^{16}$ GeV and
$M_{X}\sim10^{15.5}$ GeV. \ This implies that the parameter $\gamma\sim10$,
and that $\mu_{eff}\sim300$ GeV. \ Thus, this class of higher-dimensional
models realizes a variant of the Giudice-Masiero mechanism, but for gauge
mediated models!

To close this section, let us also address the value of the $B\mu$ term in
this case. Due to the presence of the $U(1)_{PQ}$ symmetry, the leading order
contribution originates from the higher dimension operator:%
\begin{equation}
\int d^{4}\theta\frac{X^{\dag}XX^{\dag}H_{u}H_{d}}{M_{X}^{3}}\rightarrow
\mu\frac{\overline{x}}{M_{X}}h_{u}h_{d},
\end{equation}
where $h_{u}$ and $h_{d}$ denote the scalar components of the Higgs chiral
superfields. Since $\overline{x}/M_{X}$ is by construction small, the
resulting value of $B\mu$ at the messenger scale can effectively be set to
zero. In addition, let us also note that additional corrections to $B\mu$ will
be suppressed because the matter fields localize on curves in the higher
dimensional geometry. Similar considerations apply for other contributions,
such as the A-terms of the MSSM.

\subsection{$U(1)_{PQ}$ Induced Soft Mass Terms\label{PQDEF}}

In a quantum theory of gravity, the $U(1)$ Peccei-Quinn symmetry must be
gauged. \ Thus, because the $X$ field is charged under $U(1)_{PQ}$, it will
interact via gauge boson exchange with all other fields charged under
$U(1)_{PQ}$. \ In the present case of interest, the $U(1)_{PQ}$ symmetry is
anomalous and the corresponding gauge boson will develop a large mass via the
Green-Schwarz mechanism. \ Heavy $U(1)_{PQ}$ gauge boson exchange generates
the operators \cite{ArkaniHamedANOM}:
\begin{equation}
\label{KAHLERCORR}O_{X^{\dag}X\Psi^{\dag}\Psi}=-4\pi\alpha_{PQ}\frac
{e_{X}e_{\Psi}}{M_{U(1)_{PQ}}^{2}}\cdot\int d^{4}\theta X^{\dag}X\Psi^{\dag
}\Psi
\end{equation}
where in the above, $\Psi$ denotes a generic chiral superfield charged under
$U(1)_{PQ}$, and $X$ and $\Psi$ are canonically normalized chiral superfields.
\ In addition, $e_{X}=+4$, $e_{\Psi}=+2$ for the Higgs fields, and $e_{\Psi
}=-1$ for all other chiral superfields of the MSSM. \ An important feature of
this contribution is that the overall sign of the corresponding operator is
completely fixed by the charges of the various fields. \ In addition, this
term is diagonal in a gauge quantum number basis of eigenstates. \ This
implies that this class of operators will not introduce additional FCNCs.
\ Finally, we note that we have implicitly assumed that the dominant
contribution to this operator comes from this heavy $U(1)_{PQ}$ gauge boson.
\ In higher dimensional theories such as this one, additional terms could
potentially yield corrections to this result when the mass of this gauge boson
is sufficiently large.

Once $X$ develops a supersymmetry breaking vev, the operator of equation
\eqref{KAHLERCORR} will induce a contribution to the soft masses squared:
\begin{equation}
\delta_{PQ}m_{\Psi}^{2}(M_{mess})=4\pi\alpha_{PQ}\frac{e_{\Psi}e_{X}%
}{M_{U(1)_{PQ}}^{2}}\left\vert F\right\vert ^{2}\text{.}
\label{scalarcorrection}%
\end{equation}
Note that the relative strengths of this contribution to the matter fields
versus Higgs fields is set by their relative PQ charges:
\begin{equation}
\frac{\delta_{PQ}m_{\Phi}^{2}(M_{mess})}{\delta_{PQ}m_{H}^{2}(M_{mess}
)}=-\frac{1}{2}%
\end{equation}
where here, $H$ refers to the Higgs superfields and $\Phi$ refers to all other
chiral superfields of the MSSM. \ This provides a universal prediction for a
specific deformation away from the usual gauge mediated supersymmetry breaking
scenario.\footnote{That this is the leading contribution to this higher
dimension operator follows from the fact that the $X$ and $\Psi$ fields
localize on different curves.} \ In particular, we also note that the PQ
deformation will lower the mass of the sleptons and squarks, and will increase
the soft mass squared of the Higgs fields at the messenger scale. \ Note that
to leading order, the masses of the gauginos are not altered by this
contribution.\footnote{As emphasized in \cite{ArkaniHamedANOM}, there will in
general be a contribution to the masses of the gauginos in such models via
D-term breaking effects through the coupling of the gauge fields to the
dilaton. \ Typically, however, such contributions can only make a substantial
contribution to the gaugino masses when the value of $F$ is larger than we
consider here, and we shall therefore neglect this contribution throughout
this paper.}

As a crude order of magnitude estimate, we use the values for $\alpha_{PQ}$
and $M_{U(1)_{PQ}}$ for external $U(1)$ factors obtained in \cite{BHVII}:%
\begin{align}
\alpha_{PQ}  &  \sim10^{-2}\\
M_{U(1)_{PQ}}  &  \sim10^{15}\text{ GeV.}%
\end{align}
Assuming that $F\leq10^{17}$ GeV$^{2}$, the soft mass terms are roughly given
by:%
\begin{equation}
\Delta_{PQ}\sim\sqrt{4\pi\alpha_{PQ}}\frac{F}{M_{U(1)_{PQ}}}\sim30\text{ GeV.}%
\end{equation}
These estimates depend on various assumptions about the sizes of $\alpha_{PQ}$
and $M_{U(1)_{PQ}}$, and a slight change in their values can potentially
induce important contributions to the soft mass terms of the MSSM. \ Indeed,
we will study some of the consequences of this deformation in section
\ref{MSSMREGION}.

\section{Diamond Ring Model\label{DIAMONDRING}}

In this section we present a minimal F-theory realization of gauge mediated
supersymmetry breaking where $X$ couples to the Higgs fields through a
Giudice-Masiero operator. \ As we have seen in previous sections, the presence
of an additional $U(1)_{PQ}$ symmetry plays an especially prominent role in
the low energy effective field theory. \ In F-theory, such symmetries can
originate from either Kaluza-Klein reduction of bulk gravity modes, or from
the worldvolume gauge group of seven-branes in the compactification. \ In
keeping with the general principle of decoupling, we shall only consider the
latter possibility. \

We consider a model where $X$ localizes on a matter curve defined by the
intersection of two distinct seven-branes wrapping K\"{a}hler surfaces which
we denote by $S^{\prime}$ and $S^{\prime\prime}$. \ To denote the charge of
$X$ under these two seven-branes which generically have $U(1)$ gauge groups,
we shall sometimes write $X^{+,-}$. \ In order for $X^{+,-}$ to interact with
both the messenger fields and the Higgs fields, the curve $\Sigma_{X}$ must
form one triple intersection with the messenger curves and another triple
intersection with the Higgs curves.

\begin{figure}
[ptb]
\begin{center}
\includegraphics[
height=4.0369in,
width=3.8311in
]%
{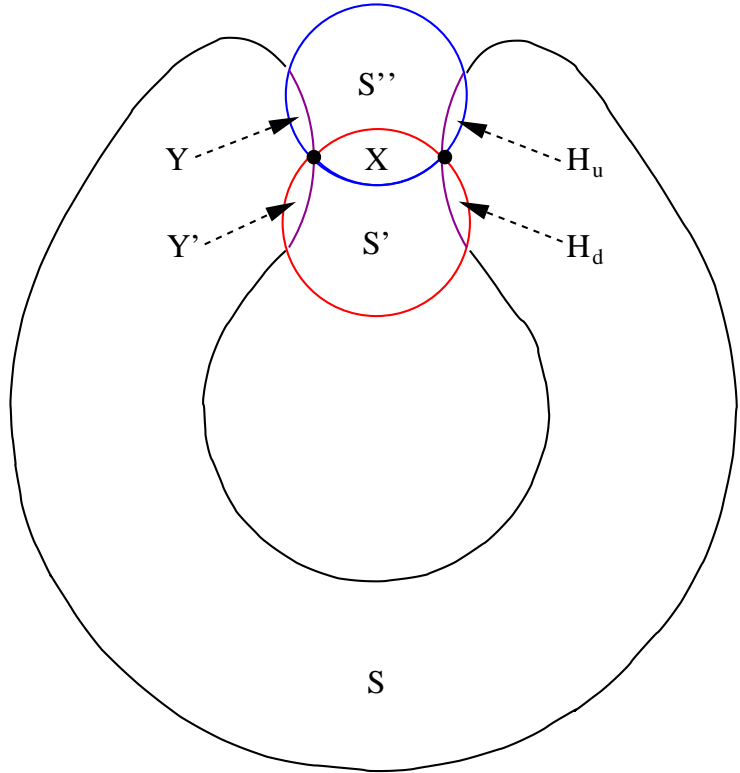}%
\caption{Depiction of the diamond ring model. \ The GUT\ model seven-brane
wraps the K\"{a}hler surface $S$, while the $X$ field localizes at the
intersection of two additional seven-branes wrapping the surfaces $S^{\prime}$
and $S^{\prime\prime}$.}%
\label{diamondring}%
\end{center}
\end{figure}

In the four-dimensional effective field theory, the superpotential coupling
$XYY^{\prime}$ is allowed and the coupling $XH_{u}H_{d}$ is forbidden when the
$Y$, $H$ and GUT\ model chiral matter fields have charges:
\begin{equation}%
\begin{tabular}
[c]{|r|r|r|r|}\hline
& $U(1)^{\prime}$ & $U(1)^{\prime\prime}$ & $U(1)_{PQ}$\\\hline
$X$ & $+2$ & $-2$ & $+4$\\\hline
$Y$ & $-2$ & $0$ & $-2$\\\hline
$Y^{\prime}$ & $0$ & $+2$ & $-2$\\\hline
$H_{u}$ & $+2$ & $0$ & $+2$\\\hline
$H_{d}$ & $0$ & $-2$ & $+2$\\\hline
$10_{M}$ & $-1$ & $0$ & $-1$\\\hline
$\overline{5}_{M}$ & $+1$ & $+2$ & $-1$\\\hline
\end{tabular}
\end{equation}
where in the above, we have also defined the $U(1)_{PQ}$ charge as the linear
combination:%
\begin{equation}
U(1)_{PQ}=U(1)^{\prime}-U(1)^{\prime\prime}\text{.}%
\end{equation}
Geometrically, the curve $\Sigma_{X}$ must intersect $S$ at two distinct
points. \ For this reason, we shall refer to this construction as the
\textquotedblleft diamond ring model\textquotedblright. \ See figure
\ref{diamondring} for a depiction of this intersecting brane configuration.%

By construction, the low energy effective theory contains the terms:%
\begin{equation}
L\supset\gamma\int d^{4}\theta\frac{X^{\dag}H_{u}H_{d}}{M_{X}}+\int
d^{2}\theta\lambda XYY^{\prime}\text{,}%
\end{equation}
where as explained in section \ref{HIGHERDIMOPS}, the first term originates
from integrating out Kaluza-Klein modes associated with $X$.

Given the prominent role that the $X$ field Kaluza-Klein modes play in this
class of models, it is important to analyze whether integrating out the other
Kaluza-Klein modes of the other fields play a similar role in the low energy
effective action. To this end, let $\widetilde{X}$ denote the Kaluza-Klein
modes associated with the $X$ field and similarly let $\widetilde{\Xi}$ denote
the Kaluza-Klein modes for the Higgs fields. \ In this somewhat condensed
notation, the $\widetilde{\Xi}$ denote fields are charged under all three
gauge group factors of the MSSM. \ The messenger sector superpotential is:%
\begin{align}
W  &  \supset\lambda_{XYY^{\prime}}\left(  X+\widetilde{X}\right)
(Y+\widetilde{Y})\left(  Y^{\prime}+\widetilde{Y^{\prime}}\right)
+\lambda_{X^{c}Y^{c}Y^{\prime c}}\widetilde{X^{c}}\widetilde{Y^{c}}%
\widetilde{Y^{\prime c}}\label{XHHONE}\\
&  +\lambda_{X^{c}H_{u}H_{d}}\widetilde{X^{c}}\left(  H_{u}+\widetilde{\Xi
_{u}}\right)  \left(  H_{d}+\widetilde{\Xi_{d}}\right)  +\lambda_{XH_{u}%
^{c}H_{d}^{c}}\left(  X+\widetilde{X}\right)  \widetilde{\Xi_{u}^{c}%
}\widetilde{\Xi_{d}^{c}}\\
&  +M_{X}\widetilde{X}\widetilde{X^{c}}+M_{u}\widetilde{\Xi_{u}}\widetilde
{\Xi_{u}^{c}}+M_{d}\widetilde{\Xi_{d}}\widetilde{\Xi_{d}^{c}}+M_{Y}%
\widetilde{Y}\widetilde{Y^{c}}+M_{Y^{\prime}}\widetilde{Y^{\prime}}%
\widetilde{Y^{\prime c}}\text{.} \label{XHHTHREE}%
\end{align}
Typically the wave functions for the Kaluza-Klein modes and the zero modes
will differ, so that the evaluation of these wave functions at a point of
triple overlap can induce further structure in the superpotential.

The Kaluza-Klein modes $\widetilde{\Xi_{u}^{c}}$ and $\widetilde{\Xi_{d}^{c}}$
have the same gauge quantum numbers as the messenger fields, and it is
therefore tempting to economize the field content of the model. \ However, in
gauge mediated scenarios, these fields do not communicate supersymmetry
breaking to the MSSM. \ To see this, note that to leading order, the
additional cubic interaction terms of the superpotential can be neglected and
we can fully determine the contribution to the soft masses by computing the
log of the effective masses of the messenger fields:%
\begin{equation}
\log(M_{tot})=\underset{i}{\sum}\log\left(  M_{mess}^{i}\right)
\label{MTOTcontr}%
\end{equation}
where the sum $i$ runs over the contributions from the various candidate
messenger fields. \ Returning to equations (\ref{XHHONE})-(\ref{XHHTHREE}),
the mass terms are:%
\begin{align}
W_{Y}^{(2)}  &  \supset\frac{1}{2}\left[
\begin{array}
[c]{cc}%
Y & Y^{\prime}%
\end{array}
\right]  \left[
\begin{array}
[c]{cc}%
0 & \lambda_{XYY^{\prime}}\left\langle X\right\rangle \\
\lambda_{XYY^{\prime}}\left\langle X\right\rangle  & 0
\end{array}
\right]  \left[
\begin{array}
[c]{c}%
Y\\
Y^{\prime}%
\end{array}
\right] \\
&  +\frac{1}{2}\left[
\begin{array}
[c]{cccc}%
\widetilde{Y} & \widetilde{Y^{\prime}} & \widetilde{Y^{c}} & \widetilde
{Y^{\prime c}}%
\end{array}
\right]  \left[
\begin{array}
[c]{cccc}%
0 & \lambda_{XYY^{\prime}}\left\langle X\right\rangle  & M_{Y} & 0\\
\lambda_{XYY^{\prime}}\left\langle X\right\rangle  & 0 & 0 & M_{Y^{\prime}}\\
M_{Y} & 0 & 0 & 0\\
0 & M_{Y^{\prime}} & 0 & 0
\end{array}
\right]  \left[
\begin{array}
[c]{c}%
\widetilde{Y}\\
\widetilde{Y^{\prime}}\\
\widetilde{Y^{c}}\\
\widetilde{Y^{\prime c}}%
\end{array}
\right]  \text{,}\\
W_{\Xi}^{(2)}  &  =\frac{1}{2}\left[
\begin{array}
[c]{cccc}%
\widetilde{\Xi_{u}^{c}} & \widetilde{\Xi_{d}^{c}} & \widetilde{\Xi_{u}} &
\widetilde{\Xi_{d}}%
\end{array}
\right]  \left[
\begin{array}
[c]{cccc}%
0 & \lambda_{XH_{u}^{c}H_{d}^{c}}\left\langle X\right\rangle  & M_{u} & 0\\
\lambda_{XH_{u}^{c}H_{d}^{c}}\left\langle X\right\rangle  & 0 & 0 & M_{d}\\
M_{u} & 0 & 0 & 0\\
0 & M_{d} & 0 & 0
\end{array}
\right]  \left[
\begin{array}
[c]{c}%
\widetilde{\Xi_{u}^{c}}\\
\widetilde{\Xi_{d}^{c}}\\
\widetilde{\Xi_{u}}\\
\widetilde{\Xi_{d}}%
\end{array}
\right]  \text{.}%
\end{align}
\newline By inspection, the mass matrix which determines $W_{Y}^{(2)}$
decomposes as the block diagonal sum of a $2\times2$ matrix, and a $4\times4$
matrix. \ Further, in the choice of basis made above, the location of non-zero
entries in this $4\times4$ matrix is identical to the $4\times4$ mass matrix
which determines $W_{\Xi}^{(2)}$. \ Because the determinant of each $4\times4$
matrix is independent of $X$, we conclude that the Higgs field Kaluza-Klein
modes cannot play the role of the messenger fields. \ Hence, this class of
models in fact does require a separate messenger sector.

\section{Strong CP\ and Gauge Mediation\label{AXIONGMSB}}

Up to this point, we have primarily focussed on the magnitude of the vevs for
the chiral superfield $X$. \ The phase of $X$ will also couple to some of the
fields of the MSSM. \ In this section we show that in the generic case where
the $U(1)_{PQ}$ gauge theory is anomalous, a linear combination of this phase
and a bulk gravity mode can potentially play the role of the QCD\ axion. \ We
find that when the gauge mediation variant of the Giudice-Masiero mechanism
solves the $\mu$ problem, the axion decay constant automatically falls within
the experimentally allowed range of values.

To set notation, we now review some of the features of axion models which are
germane. \ See \cite{Khlopov:1999rs,Kim:2008hd} for general reviews on axion
physics, and \cite{SvrcekWittenAxions} and references therein for an extensive
discussion of axions in string theory. \ The strong CP\ problem is the fact
that although the CP\ violating operator:
\begin{equation}
\frac{\overline{\theta}}{32\pi^{2}}\varepsilon^{\mu\nu\rho\sigma}%
Tr_{SU(3)}F_{\mu\nu}F_{\rho\sigma}%
\end{equation}
can in principle contribute to the Standard Model Lagrangian, the effective
$\overline{\theta}$ angle must satisfy the constraint $\left\vert
\overline{\theta}\right\vert \lesssim10^{-10}$ in order to remain in accord
with observation \cite{Harrisetal,RomalisLimit}. \ Here, $\overline{\theta}$
is defined as the net contribution from the bare theta angle and the phase
from the determinant of the mass matrix for all fermions charged under
$SU(3)_{C}$.

The axion solution to the strong CP\ problem promotes $\overline{\theta}$ to a
dynamical field $a$ with Lagrangian density:%
\begin{equation}
L_{axion}=\frac{f_{a}^{2}}{2}\partial_{\mu}a\partial^{\mu}a+\frac{a}{32\pi
^{2}}\varepsilon^{\mu\nu\rho\sigma}Tr\left(  F_{\mu\nu}F_{\rho\sigma}\right)
\end{equation}
where in the above $f_{a}$ is defined as the axion decay constant. \ The
crucial point is that QCD instantons will generate an effective potential for
$a$ which has a minimum at zero, thus solving the strong CP\ problem. \ More
explicitly, the effective potential can be approximated using the pion
Lagrangian, see for example section 23.6 of \cite{WeinbergII}:%
\begin{equation}
V_{ax}(a)=m_{\pi}^{2}f_{\pi}^{2}\left(  1-\cos a\right)
\end{equation}
where $f_{\pi}\sim90$ MeV is the pion decay constant and $m_{\pi}\sim130$ MeV
is the mass of the pion. \ The effective mass of the canonically normalized
field $f_{a}\cdot a$ is therefore:%
\begin{equation}
m_{a}=\frac{m_{\pi}f_{\pi}}{f_{a}}\sim\frac{10^{16}\text{ eV}^{2}}{f_{a}%
}\text{.}%
\end{equation}

Current bounds on the value of $f_{a}$ only allow a narrow window of available
energy scales:%
\begin{equation}
10^{9}\text{ GeV}<f_{a}<10^{12}\text{ GeV.} \label{axwindow}%
\end{equation}
The lower bound is from estimates on supernova cooling and is difficult to
evade. \ The upper bound is somewhat flexible because it assumes a standard cosmology.

In supersymmetric models, the axion is only one real degree of freedom of a
complex scalar. \ The other real degree of freedom in the complex scalar of
the corresponding supermultiplet is the saxion which may be viewed as simply
another modulus which must develop a suitable mass to avoid cosmological
problems. \ As emphasized in \cite{BanksDineGraesser}, stabilizing the saxion
is a potentially more severe issue than attaining the correct axion decay
constant. \ Moreover, it is possible that in some cases $f_{a}$ could be as
high as $10^{15}$ GeV \cite{BanksDineGraesser}. \ In keeping with the
principle of decoupling, however, we shall defer all issues concerning moduli
stabilization to a later stage of analysis. \ For this reason, we shall assume
the conventional upper bound on the axion decay constant.

In string theory models, there are potentially many candidate axion fields
because various moduli fields will generically couple to the QCD\ instanton
density. \ Note, however, that if the compactification scale sets the dynamics
of the axion, the resulting axion decay constant will be above the available
window defined by line (\ref{axwindow}). \ In fact, the ubiquity of such
axion-like couplings is also potentially problematic for the strong
CP\ problem. \ As an example, consider a model where the axion field receives
contributions to its effective potential from sources other than
QCD\ instanton effects so that:%
\begin{equation}
V_{ax}(a)=V_{QCD}(a)+V(a)\text{.}%
\end{equation}
A priori, the minima of $V_{QCD}$ and $V$ are uncorrelated so that the overall
minimum of $V_{ax}$ may shift away from zero. \ To estimate the size of this
shift, let $\theta_{0}$ denote the minimum of $V$. \ Expanding $V_{ax}$ to
quadratic order in $a$ therefore yields:%
\begin{equation}
V_{ax}(a)-\left(  V_{QCD}(0)+V(\theta_{0})\right)  =\frac{1}{2}V_{QCD}%
^{\prime\prime}(0)a^{2}+\frac{1}{2}V^{\prime\prime}(\theta_{0})\left(
a-\theta_{0}\right)  ^{2}%
\end{equation}
it follows that the minimum of $V_{ax}$ shifts to:%
\begin{equation}
\theta=\frac{V^{\prime\prime}(\theta_{0})}{V_{QCD}^{\prime\prime}%
(0)+V^{\prime\prime}(\theta_{0})}\cdot\theta_{0}\text{.}%
\end{equation}
Assuming $\theta_{0}$ is an order one constant, this implies:%
\begin{equation}
\frac{V^{\prime\prime}(\theta_{0})}{V_{QCD}^{\prime\prime}(0)}<10^{-10}
\label{VTHETABOUND}%
\end{equation}
in order for $a$ to solve the strong CP problem.

We now identify possible axion fields in the present class of
compactifications. \ Independent of the details of the particular GUT model
seven-brane theory, the QCD instanton density will always couple to the
background four-form potential through the seven-brane worldvolume coupling
$C_{(4)}\wedge Tr_{SU(3)}(F\wedge F)$. \ Letting $c_{S}$ denote the integral
of $C_{(4)}$ over the cycle wrapped by $S$, the effective action for $c_{S}$
is:%
\begin{equation}
L_{c}=\frac{M_{\ast}^{8}Vol(B_{3})}{2}\partial_{\mu}c_{S}\partial^{\mu}%
c_{S}+\frac{c_{S}}{32\pi^{2}}\varepsilon^{\mu\nu\rho\sigma}Tr\left(  F_{\mu
\nu}F_{\rho\sigma}\right)  \text{.} \label{LC}%
\end{equation}
As shown in \cite{BHVII}, $M_{\ast}^{8}Vol(B_{3})\sim M_{pl}^{2}$, which
establishes that this field cannot play the role of the QCD axion. This is in
accord with the general observations in \cite{SvrcekWittenAxions} where the
most natural axion fields one would think of in the context of string theory
end up having too large a decay constant to be phenomenologically viable to be
identified with the QCD axion. \ In fact, as we now argue, a specific linear
combination of $c_{S}$ and the phase of the scalar component for the $X$ field
can play the role of the QCD axion, thus providing a solution to this problem
in the context of string theory. \ Note that here, the axino is the goldstino
mode which is eaten by the gravitino.\footnote{For earlier discussion on
potential connections between the axion and the gravitino, see for example
\cite{Fayet:1979qi,Fayet:1979yb}, and for other discussions on the connection
between supersymmetry breaking and the axion, see for example
\cite{Fayet:1980ad}.}

Returning to equations (\ref{phisoftmass})-(\ref{gauginoterm}), note that
whereas the contribution to the soft scalar masses only involves $\left\vert
X\right\vert ^{2}$, the coupling between $X$ and the gauge field strength
kinetic terms $TrW^{\alpha}W_{\alpha}$ also includes a coupling to the phase
of $X$. \ With notation as before so that $\langle X\rangle=x+\theta^{2}F$, we
shall denote the phase of the vev $x$ by the $2\pi$ periodic variable $a_{x}$
so that:
\begin{equation}
x =\left\vert x\right\vert \exp\left(  ia_{x}\right)  \text{.}%
\end{equation}
Expanding equations (\ref{phisoftmass})-(\ref{gauginoterm}) yields:
\begin{align}
L  &  \supset\int d^{4}\theta\left(  \underset{i=1}{\overset{3}{%
{\displaystyle\sum}
}}-\alpha_{i}^{2}C_{2}(R_{\Psi}^{i})\left(  \log\left\vert X\right\vert
^{2}\right)  ^{2}\Psi^{\dag}\Psi\right) \\
&  +\int d^{2}\theta\operatorname{Re}\left(  \underset{i=1}{\overset{3}{%
{\displaystyle\sum}
}}\frac{1}{8\pi i}\left(  \tau_{YM}^{(i)}+\frac{1}{2\pi i}\log X\right)
Tr_{G_{i}}W_{(i)}^{\alpha}W_{(i) \alpha}\right) \\
&  =\int d^{4}\theta\left(  \underset{i=1}{\overset{3}{%
{\displaystyle\sum}
}}-\alpha_{i}^{2}C_{2}(R_{\Psi}^{i})\left(  \log\left\vert x\right\vert
^{2}+\theta^{2}\frac{F}{x} + \overline{\theta}^{2}\frac{\overline{F}%
}{\overline{x}} \right)  ^{2} \Psi^{\dag}\Psi\right) \\
&  +\int d^{2}\theta\operatorname{Re}\left(  \underset{i=1}{\overset{3}{%
{\displaystyle\sum}
}}\frac{1}{8\pi i}\left(  \tau_{YM}^{(i)}+\frac{1}{2\pi i}\left[
\log\left\vert x\right\vert +ia_{x}+\theta^{2}\frac{F}{x} \right]  \right)
Tr_{G_{i}}W_{(i)}^{\alpha}W_{(i)\alpha}\right)  \text{.} \label{axionax}%
\end{align}
By inspection, the angle $a_{x}$ couples to the QCD instanton density, as
required for a candidate axion.

As we have seen in previous sections, $X$ will generically be charged under at
least one $U(1)$ group which must be gauged in a quantum theory of gravity.
\ For example, in the diamond ring model, $X$ transforms in the bifundamental
of a $U(1)\times U(1)$ gauge group defined by two intersecting seven-branes.
\ When these $U(1)$ factors are non-anomalous, the vev of $X$ will
spontaneously break the gauge symmetry, and the phase of $X$ will be eaten by
the gauge boson, eliminating $a_{x}$ as a candidate axion. \ Alternatively,
when these $U(1)$ factors are anomalous, the corresponding gauge bosons will
develop a mass via the Green-Schwarz mechanism, leaving behind global
symmetries which may potentially be violated by instanton effects. \ The vev
of $X$ spontaneously breaks this symmetry and the corresponding Goldstone mode
will persist as a candidate axion.\footnote{More precisely, the Goldstone mode
is given by a linear combination of the phase of $X$, with a small
contribution from another axion-like field which enters in the Green-Schwarz
mechanism.} \ We note that independent of any discussion of supersymmetry
breaking, the idea that such low energy global symmetries could lead to lower
values for the axion decay constant in string based models was already noted
in \cite{BarrHarmless}. \ See section 5 of \cite{SvrcekWittenAxions} for a
review of the potential role of anomalous $U(1)$ factors in string-motivated
axion physics.

Having identified two fields which possess axion-like couplings to the
QCD\ instanton density, the axion Lagrangian is given by:%
\begin{align}
L_{ax}  &  = \left\vert x\right\vert ^{2}\partial_{\mu}a_{x}\partial^{\mu
}a_{x} + \frac{1}{2}M_{\ast}^{8}Vol(B_{3})\partial_{\mu}c\partial^{\mu}%
c+\frac{a_{x}+c}{32\pi^{2}}\varepsilon^{\mu\nu\rho\sigma}Tr\left(  F_{\mu\nu
}F_{\rho\sigma}\right) \\
&  \equiv\frac{f_{a}^{2}}{2}\partial_{\mu}\left(  a_{x}+c\right)
\partial^{\mu}\left(  a_{x}+c\right)  +\frac{f_{\bot}^{2}}{2}\partial_{\mu
}(a_{x}+\beta c)\partial^{\mu}(a_{x}+\beta c)\\
&  +\frac{(a_{x}+c)}{32\pi^{2}}\varepsilon^{\mu\nu\rho\sigma}Tr\left(
F_{\mu\nu}F_{\rho\sigma}\right)
\end{align}
where the linear combination $a_{x}+c$ defines the candidate axion field, and
$a_{x} + \beta c$ is an orthogonal linear combination of fields. \ In the
above, the Planck scale enters as $M_{pl}^{2}=M_{\ast}^{8}Vol(B_{3})$.
\ Solving for the axion decay constant $f_{a}$ yields:
\begin{equation}
f_{a}=\frac{\sqrt{2}M_{pl}\left\vert x\right\vert }{\sqrt{M_{pl}%
^{2}+2\left\vert x\right\vert ^{2}}}= \sqrt{2}\left\vert x\right\vert
+O\left(  \frac{\left\vert x\right\vert ^{2}}{M_{pl}^{2}}\right)  \text{,}
\label{axcon}%
\end{equation}
where in the final line we have expanded to leading order in the parameter $x$
which is generically smaller than the Planck scale.

Equation (\ref{axcon}) has important consequences for the parameters of the
low energy theory. The effective $\mu$ term, messenger mass and axion decay
constant are given by the relations:%
\begin{align}
\mu_{eff}  &  =\gamma\cdot\frac{\overline{F}}{M_{X}}\\
M_{mess}  &  =\lambda_{mess}\left\vert x\right\vert \\
f_{a}  &  =\sqrt{2}\left\vert x\right\vert \\
\Lambda &  =\frac{F}{x}\sim10^{5}\text{ GeV.}%
\end{align}
This yields the intriguing relation:%
\begin{equation}
f_{a}=\sqrt{2}\frac{M_{X}}{\gamma}\cdot\frac{\mu_{eff}}{\Lambda}\text{,}%
\end{equation}
which connects the GUT scale and weak scale to both the axion decay constant
and the scale of supersymmetry breaking. \ We note that this same relation
persists when the number of messenger fields is greater than one.

To estimate the value of the axion decay constant, we use the same
representative values obtained near equation \eqref{gammaest} so that
$\gamma\sim10$, $\mu_{eff}\sim300$ GeV and $M_{X}\sim10^{15.5}$. \ Setting
$\Lambda\sim10^{5}$ GeV, the resulting value of $f_{a}$ is then:
\begin{equation}
f_{a}\sim10^{12}\text{ GeV.}%
\end{equation}
Remarkably, this is within the axion window of line (\ref{axwindow})! \ Given
the fact that this is only an order of magnitude estimate, we find it
encouraging that with at most a mild fine-tuning of parameters, connecting
weak scale phenomenology to supersymmetry breaking in F-theory automatically
produces a viable value for the axion decay constant!

\section{$E_{6}$, $U(1)_{PQ}$ and F-theory\label{E6MOTIVATION}}

In previous sections we have shown that when the curve $\Sigma_{X}$ supporting
$X$ forms one triple intersection with the Higgs curves and another triple
intersection with the messenger curves, interactions with the Kaluza-Klein
modes of the $X$ field generate a value for the $\mu$ term in gauge mediated
scenarios which is strikingly close to the weak scale. \ From the perspective
of F-theory, however, the identification of the $U(1)_{PQ}$ symmetry appears
somewhat accidental. \ In terms of the particular geometric realization of
these matter curves, this translates into the fact that the $\Sigma_{X}$ curve
defined by the intersection of two external surfaces $S^{\prime}$ and
$S^{\prime\prime}$ must intersect $S$ at two distinct points. \ While this can
certainly be arranged for certain geometries, it appears ad hoc.

These problems can be viewed as symptoms of the fact that the GUT\ model can
undergo a further unification to a higher $E$-type GUT group. \ Indeed,
perhaps one of the most compelling features of GUT\ models is the elegant
packaging of the representation content of the MSSM into three generations of
the $\overline{5}\oplus10$ of $SU(5)$, and the even further unification to the
$16$ of $SO(10)$ once right-handed neutrinos are included. \ Aside from any
theoretical bias in favor of this aesthetically appealing structure, the
qualitative expectation that the seesaw mechanism can naturally generate small
neutrino masses in the Standard Model is at the very least intriguing
circumstantial evidence that this type of structure is quite natural for
phenomenology as well. \ In fact, unifying the Higgs and chiral matter into
$E_{6}$ allows a further unification into the $27$ of $E_{6}$.

Each additional stage of unification equips the low energy effective theory
with additional $U(1)$ symmetries which are of interest phenomenologically.
\ For example, $SO(10)$ unification contains an additional $U(1)_{B-L}$ factor
which can increase the lifetime of the proton. Matter parity can be viewed as
a discrete $%
\mathbb{Z}
_{2}$ subgroup of $U(1)_{B-L}$ so that a chiral generation of the MSSM
organizes into a copy of the $16$ of $SO(10)$ and has parity $-1$, while the
Higgs fields of the MSSM which descend from the $10$ of $SO(10)$ have parity
$+1$.

Further unification to $E_{6}$ also endows the low energy theory with a
$U(1)_{PQ}$ symmetry in the low energy effective theory. \ To see how this
comes about, consider the decomposition of the $78$ and $27$ of $E_{6}$ into
representations of $SO(10)\times U(1)$:
\begin{align}
E_{6}  &  \supset SO(10)\times U(1)\\
78  &  \rightarrow45_{0}+1_{0}+16_{-3}+\overline{16}_{+3}\\
27  &  \rightarrow1_{4}+10_{-2}+16_{1}\text{.}%
\end{align}
In traditional four-dimensional GUT\ models, it is common to organize all of
the matter content of the MSSM\ into copies of the $27$. \ It is immediate
that the additional $U(1)$ charge assignment is consistent with that
associated to the $U(1)_{PQ}$ charge for the Higgs fields and chiral matter
content of the MSSM. Moreover note that matter parity can be viewed as a
discrete $%
\mathbb{Z}
_{2}$ subgroup of this additional $U(1)_{PQ}$.

While traditional four-dimensional GUT models focus on the role of the $27$ of
$E_{6}$, the $\overline{27}$ plays an equally important role for the mediation
sector of the model. \ Indeed, the $X$ field and messengers can both descend
from the $\overline{27}$ of $E_{6}$ upon making the identifications:%
\begin{align}
X  &  :1_{-4}\in\overline{27}\\
Y,Y^{\prime}  &  :10_{+2}\in\overline{27}\text{.}%
\end{align}
By inspection, these charge assignments allow an interaction term between $X$
and the messengers, while forbidding a similar interaction term with the Higgs
fields, exactly as in the charge assignments we have been assuming in the
context of the diamond ring model!

As a brief aside, we recall that in \cite{BHVI,BHVII}, our minimal
realizations of GUT models were based on geometries where the matter content
of the MSSM descends from the $78$ of $E_{6}$. \ While it is certainly
possible to assign a $U(1)_{PQ}$ charge consistent with the $78^{3}$
interaction term, note that in this intermediate decomposition, the Higgs
field descends from the $\overline{16}$ of $SO(10)$. \ On the other hand,
continuing with the natural progression of $E$-type groups, it is also quite
natural to package the matter content of the MSSM in terms of the $27$.

Perhaps unfortunately, there are also well-known problems in four-dimensional
GUT models based on the gauge group $G_{S}=E_{6}$. \ For example, although the
Higgs fields and chiral matter naturally package into the $27$ of $E_{6}$,
there are three generations of chiral matter, but only one \textquotedblleft
generation\textquotedblright\ of Higgs fields. \ To remove the extraneous
Higgs fields from the other two generations of $27$'s it is necessary to
include either higher dimensional representations of $E_{6}$, or higher
dimension operators. \ This additional complexity throws into question the
economy of $E_{6}$ as a GUT group.

This issue is in fact more severe in models based on local del Pezzo
compactifications of F-theory. As explained in \cite{BHVII}, GUT\ group
breaking via fluxes will typically generate additional exotic fields in the
zero mode spectrum unless the bulk gauge group is $G_{S}=SU(5)$, or $SO(10)$
when the model descends to a flipped $SU(5)$ model in four dimensions.
\ Although \cite{BHVII} does not contain a complete proof that direct breaking
from $G_{S}=E_{6}$ to $G_{std}$ will always generate exotics, the number of
independent instanton configurations is generically a smaller number than the
number of different exotic representations which must be excluded from the low
energy spectrum. \ In a certain sense, this is in fact a welcome restriction
on the structure of the low energy theory, but appears to clash with the
elegant packaging of the matter fields into representations of $SO(10)$ or
$E_{6}$.

There is an important loophole to the above considerations which demonstrates
the flexibility of local models in F-theory. \ Recall that in F-theory, a
matter curve is defined as a complex codimension one subspace in $S$ where the
singularity type enhances by at least one rank from $G_{S}$ to $G_{\Sigma}$
such that $G_{S}\varsubsetneq G_{\Sigma}$. \ The matter content localized
along the matter curve can be viewed as the intersection of two seven-branes
with bulk gauge groups $G_{S}$ and $G_{S^{\prime}}$ such that $G_{S}\times
G_{S^{\prime}}\subset G_{\Sigma}$. \ Many of the examples in \cite{BHVI,BHVII}
focussed on rank one enhancement configurations where $G_{S^{\prime}}=U(1)$ or
$SU(2)$ because this is the minimal allowed singularity enhancement in
F-theory. More generally, however, the singularity type can enhance by more
than one rank.

In order to retain the appealing rigidity of local GUT\ models found in
\cite{BHVII}, we shall focus exclusively on local del Pezzo models where the
bulk gauge group $G_{S}=SU(5)$ breaks to the Standard Model gauge group via an
internal $U(1)$ hyperflux. \ The matter content of the GUT model can exhibit
further unification along curves when the rank of the singularity type
increases by more than the minimal required amount. \ To see how this
unification works in practice, first recall that six-dimensional
hypermultiplets in the $5$ or $10$ of $SU(5)$ originate from curves where the
bulk singularity $G_{S}=SU(5)$ respectively enhances to $SU(6)$ or $SO(10)$.
\ On the other hand, when $G_{S}=SO(10)$ enhances to $E_{6}$ along a matter
curve, a six-dimensional field in the $16$ of $SO(10)$ will localize along the
same matter curve. \ Because the $16$ of $SO(10)$ unifies the $\overline{5}$
and $10$ of $SU(5)$, our expectation is that a local enhancement from $SU(5)$
directly to $E_{6}$ corresponds to a six-dimensional field in the
$1+\overline{5}+10$ localized on this curve. \ To establish this result, we
can decompose the adjoint representation of $E_{6}$ to $E_{5}\times
U(1)_{2}=SO(10)\times U(1)_{2}$, and then further decompose to irreducible
representations of $E_{4}\times U(1)_{1}\times U(1)_{2}=SU(5)\times
U(1)_{1}\times U(1)_{2}$. \ Following the general philosophy of
\cite{KatzVafa}, the matter fields which localize on the curve of $E_{6}$
enhancement must be simultaneously charged under these two $U(1)$ subgroup
factors. \ In particular, it follows that a six-dimensional hypermultiplet in
the $1_{-5,-3}+\overline{5}_{3,-3}+10_{-1,-3}$ localizes on this matter curve.
\ A similar analysis establishes that a six-dimensional hypermultiplet with
matter content specified by a $27$ of $E_{6}$ localizes on a curve where the
singularity type enhances to $E_{7}$.

\begin{figure}
[ptb]
\begin{center}
\includegraphics[
height=3.0545in,
width=4.5662in
]%
{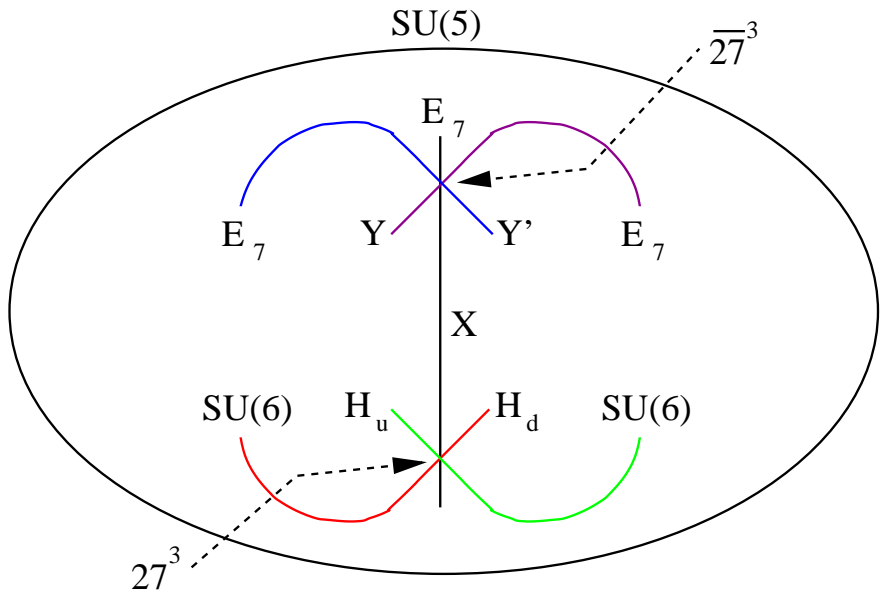}%
\caption{Depiction of the messenger sector of a local $SU(5)$ model where the
messengers and $X$ field originate from local enhancements to $E_{7}$ and
embed in the $\overline{27}$ of $E_{6}$. \ The $XYY^{\prime}$ interaction term
descends from a local enhancement to $E_{8}$ at a point of triple
intersection. \ The Higgs fields originate from local enhancement to $SU(6)$.
\ In this case, the Higgs fields embed in the $27$ of $E_{6}$ and can
therefore only participate in a $27^{3}$ interaction so that a direct coupling
with the $X$ field via the superpotential is forbidden, but an interaction
term via the K\"{a}hler potential is allowed.}%
\label{e6pq}%
\end{center}
\end{figure}

This same logic also holds for higher rank enhancements at points of $S$. \ As
in \cite{BHVI}, we can start from the adjoint representation of this higher
singularity type, and by decomposing the matter fields along each curve which
is neutral under a particular subset of generators, we find the expected
enhancement in singularity type. \ For example, we can consider a geometry
where $SU(5)$ enhances up to $E_{8}$ at a point of $S$ and such that $E_{8}$
only decreases by one rank along the various matter curves of the geometry.
\ In this case, decomposing the adjoint representation of $E_{8}$ to
$E_{6}\times U(1)_{1}\times U(1)_{2}$ yields:%
\begin{align}
E_{8}  &  \supset E_{6}\times U(1)_{1}\times U(1)_{2}\\
248  &  \rightarrow78_{0,0}+1_{0,0}+27_{2,0}+27_{-1,-1}+27_{1,-1}%
+\overline{27}_{-2,0}+\overline{27}_{1,1}+\overline{27}_{-1,1}\text{,}%
\end{align}
which implies that there are three curves where the singularity type enhances
to $E_{7}$. \ Further, the low energy theory contains a $27^{3}$ interaction
term, where each $27$ corresponds to a local enhancement in $SU(5)$ to $E_{7}%
$. \ It is now immediate that there are many further combinations of rank
enhancement which are potentially of interest. \ As usual, our guiding
principal to determine the matter content and allowed interaction terms relies
heavily on the close connection between the various ways of partially Higgsing
the singularity type at various subspaces of additional enhancement, and the
corresponding deformation theory of the geometry. \ See figure \ref{e6pq} for
a depiction of how various $SU(5)$ enhancements to $E_{7}$ can potentially
accommodate the interaction terms of the messenger sector.%

There is a subtlety in reading off the matter content in the context of higher
rank enhancements which we now explain. In the above enhancement from $SU(5)$
to $E_{7}$, we have mainly concentrated on how the $27$ obtained from the last
step of enhancement from $E_{6}$ to $E_{7}$ decomposes into representations of
$SU(5)$. \ In fact, there are additional contributions from the adjoint of
$E_{6}$ which come from Higgsing $E_{6}$ down to $SU(5)$. \ If these
additional matter fields localize on the same curve as the $27$, as would be
true in the simplest such setup, then we would not be able to control the zero
mode content of these contributions independently by adjusting fluxes on the
curves. \ We would simply have too many constraints to satisfy with too few
fluxes to choose from. \ One way to avoid these extra matter fields is to
require that they develop a vev, partially Higgsing the corresponding gauge
group. \ In F-theory, this corresponds to brane recombination \cite{BHVI}.
\ It is also possible to consider geometries where a surface of general type
locally behaves as a rigid divisor, effectively Higgsing the theory to a local
del Pezzo model. \ Some examples of this type of phenomenon for rigid
$\mathbb{P}^{1}$'s in a local Calabi-Yau threefold can be found in
\cite{CIV,CachazoKatzVafa}. In the context of F-theory models, the matter
content localized on a curve in the original theory could therefore simply
descend to the local del Pezzo model. \ Finally, it is also possible that an
appropriately engineered effective field theory could generate a GUT\ scale
mass for these additional states, removing them from the low energy spectrum.
\ At any rate, we will not address this issue further in this paper and leave
a more detailed analysis to future work. \ Instead, we will focus on the
interplay between the matter content of such theories with the PQ symmetry
provided by higher rank enhancements in the local singularity type.

In general, when the rank of the singularity enhances by more than one rank,
the adjoint representation will contain more than one distinct representation
charged under the bulk gauge group $G_{S}$ so that distinct irreducible
representations of $G_{S}$ may localize on the same curve. \ In local models
where $U(1)_{PQ}$ embeds in $E_{6}$, the zero mode content along such curves
will typically automatically include not just a generation of the MSSM, but
also additional $X$ fields, messengers or Higgs fields depending on the
overall bundle assignment.

As an explicit example of these considerations, consider the matter content of
a curve where the singularity type enhances from $SU(5)$ to $E_{7}$ so that a
six-dimensional hypermultiplet in the $27_{1}$ in the $E_{6}\times U(1)$
localizes along this curve. \ Decomposing $E_{6}$ to the subgroup
$SO(10)\times U(1)_{PQ}$, the $27_{1}$ decomposes as:%
\begin{align}
E_{6}\times U(1)  &  \supset SO(10)\times U(1)_{PQ}\times U(1)\\
27_{1}  &  \rightarrow1_{4,1}+10_{-2,1}+16_{1,1}\text{.}%
\end{align}
It thus follows that the bundle assignments for each representation are:%
\begin{align}
1_{4,1}  &  \in K_{\Sigma}^{1/2}\otimes L_{PQ}^{4}\otimes L^{1}\\
10_{-2,1}  &  \in K_{\Sigma}^{1/2}\otimes L_{PQ}^{-2}\otimes L^{1}\\
16_{1,1}  &  \in K_{\Sigma}^{1/2}\otimes L_{PQ}^{1}\otimes L^{1}%
\end{align}
with similar bundle assignments for the complex conjuagate representations.
\ To avoid any subtleties having to do with non-trivial holonomies, we now
restrict our analysis to the case where $\Sigma$ is a genus zero curve.
\ Assuming genericity of all bundles, a similar argument will hold in the more
general case by analyzing the degrees of the various bundles. \ The number of
fields in the $1_{4,1}$ is completely determined by the number of zero modes
in the $10_{-2,1}$ and $16_{1,1}$. \ Letting $N$ with an appropriate subscript
denote the number of fields of a given representation, the bundle assignments
for the $10_{-2,1}$ and $16_{1,1}$ are:
\begin{align}
10_{-2,1}  &  \in K_{\Sigma}^{1/2}\otimes\mathcal{O}_{\Sigma}(N_{10_{-2,1}})\\
16_{1,1}  &  \in K_{\Sigma}^{1/2}\otimes\mathcal{O}_{\Sigma}(N_{16_{1,1}%
})\text{.}%
\end{align}
Solving for the bundles $L_{PQ}$ and $L$ yields:%
\begin{align}
L  &  =\mathcal{O}_{\Sigma}(2N_{16_{1,1}}+N_{10_{-2,1}})^{1/3}\\
L_{PQ}  &  =\mathcal{O}_{\Sigma}(N_{16_{1,1}}-N_{10_{-2,1}})^{1/3}%
\end{align}
so that the bundle assignment for fields in the $1_{4,1}$ is:%
\begin{equation}
1_{4,1}\in K_{\Sigma}^{1/2}\otimes\mathcal{O}_{\Sigma}\left(  2N_{16_{1,1}%
}-N_{10_{-2,1}}\right)  \text{.}%
\end{equation}
In other words, the net matter content on this curve is:%
\begin{equation}
R_{net}=\left(  2N_{16_{1,1}}-N_{10_{-2,1}}\right)  \times1_{4,1}%
+N_{10_{-2,1}}\times10_{-2,1}+N_{16_{1,1}}\times16_{1,1} \label{NETMULT}%
\end{equation}
where a negative multiplicity factor indicates that the resulting fields
transform in the complex conjugate representation.

The above result demonstrates that when a full generation of the
MSSM\ transforming in the $16$ of $SO(10)$ localizes on a matter curve where
$SU(5)$ enhances to $E_{7}$, the zero mode content of the theory will
generically contain additional matter fields. \ Of particular relevance are
matter curves where $N_{16_{1,1}}>0$ and $N_{16_{1,1}}=0$. \ First consider
the case where a matter curve contains at least one chiral generation of the
MSSM. \ In order to solve the doublet triplet splitting problem using an
internal $U(1)$ hyperflux, it follows that none of the Higgs fields can also
localize on the same curve so that $N_{10_{-2,1}}<0$. \ Returning to equation
(\ref{NETMULT}), this implies that such matter curves also support
$2\left\vert N_{16_{1,1}}\right\vert +\left\vert N_{10_{-2,1}}\right\vert $
zero mode GUT\ group singlets with $U(1)_{PQ}$ charge opposite to that of the
$X$ fields. \ These singlets are essentially harmless in the low energy theory
so long as they do not develop a vev, which can typically be arranged. \ Next
consider matter curves where $N_{16_{1,1}}=0$. \ In this case, equation
(\ref{NETMULT}) implies that candidate Higgs fields and $X$ fields generically
localize on the same matter curve. \ In particular, the messenger fields do
not appear to localize on the same curve as the $X$ field zero modes.

On the other hand, in order to retain the doublet triplet splitting mechanism
via hyperflux proposed in \cite{BHVII}, the Higgs fields must localize on a
curve of $SU(6)$ enhancement. \ Indeed, when additional matter localizes on a
curve, the $U(1)$ hyperflux will generate additional zero modes. \ Assuming
that the Higgs field localizes on a curve where the singularity type enhances
to $SU(6)$, the requisite interaction term between the Higgs field and chiral
matter of the MSSM\ now requires that a local $SU(6)$ enhancement must form a
triple intersection with two curves which enhance to $E_{7}$. \ Note that if
such an intersection can be realized, the Higgs fields will automatically come
equipped with the correct $U(1)$ PQ\ charge.

We now present a local construction of such an enhancement. \ At the point of
triple intersection, the singularity type must necessarily enhance to $E_{8}$.
\ The bulk $SU(5)$ gauge group can therefore be viewed as one of the factors
in the decomposition of $E_{8}$ to $SU(5)\times SU(5)$. \ With respect to this
subgroup, the adjoint representation of $E_{8}$ decomposes as:%
\begin{align}
E_{8}  &  \supset SU(5)_{1}\times SU(5)_{2}\\
248  &  \rightarrow(24,1)+(1,24)+(5,10)+(\overline{5},\overline{10}%
)+(10,\overline{5})+(\overline{10},5)
\end{align}
where we shall take the bulk gauge group to correspond to the $SU(5)_{1}$
factor. \ A local enhancement in singularity type of the $SU(5)_{1}$
corresponds to the locus where some directions in the Cartan of $SU(5)_{2}$
combine with $SU(5)_{1}$ to form a higher rank singularity. \ To set notation,
let $t_{1},...,t_{5}$ denote the local generators of the Cartan of $SU(5)_{2}$
subject to the tracelessness condition $t_{1}+...+t_{5}=0$. \ Additional
massless states of $SU(5)_{2}$ will contribute along various directions in the
Cartan subalgebra. \ Some examples of rank one enhancements can be achieved
when the $5$ or $10$ of $SU(5)_{2}$ contribute to the massless states of
$SU(5)_{1}$. \ Indeed, the $5$ contributes when $t_{i}=0$, while the $10$
contributes when $t_{i}+t_{j}=0$ for $i\neq j$, corresponding to the local
enhancements:%
\begin{align}
SO(10)  &  :t_{i}=0\\
SU(6)  &  :t_{i}+t_{j}=0\text{.}%
\end{align}
Along $t_{i}=0$, we find $10\times2+1=21$ additional states contribute so that
the $24$ of $SU(5)$ enhances to the $45$ of $SO(10)$. \ Moreover, along
$t_{i}+t_{j}=0$, we find $5\times2+1=11$ additional states so that the adjoint
of $SU(5)$ instead enhances to the adjoint of $SU(6)$. \ A local enhancement
to $E_{7}$ corresponds to the direction in the Cartan of $SU(5)_{2}$ where an
entire $SU(3)\times U(1)$ subgroup of $SU(5)_{2}$ combines with $SU(5)_{1}$.
\ In terms of the local generators of the Cartan of $SU(5)_{2}$ this reads:%
\begin{align}
E_{7}  &  :t_{1}+t_{2}=t_{3}=t_{4}=t_{5}=0\label{ESO}\\
E_{7}^{\prime}  &  :t_{4}+t_{5}=t_{1}=t_{2}=t_{3}=0 \text{.} \label{EST}%
\end{align}
Decomposing various representations of $SU(5)_{2}$ into irreducible
representations of the maximal subgroup $SU(3)\times SU(2)\times U(1)$
yields:
\begin{align}
SU(5)_{2}  &  \supset SU(3)\times SU(2)\times U(1)\\
24  &  \rightarrow(8,1)_{0}+(1,3)_{0}+(3,\overline{2})_{-5}+(\overline
{3},2)_{+5}+(1,1)_{0}\\
10  &  \rightarrow(1,1)_{6}+(3,2)_{1}+(\overline{3},1)_{-4}\\
5  &  \rightarrow(3,1)_{-2}+(1,2)_{+3}\text{.}%
\end{align}
In particular, we conclude that along the loci defined by lines (\ref{ESO})
and (\ref{EST}), the $5$ contributes three states ($t_{i}=0$ for $i\geq3$),
the $10$ contributes four states ($t_{i}+t_{j}=0$ for $i,j\geq3$ and $i\neq
j$), and the $24$ contributes nine states for a total of $2\times
10\times3+2\times5\times4+9=109$ additional states. \ Combined with the
adjoint of $SU(5)_{1}$, this indeed yields the $133$ expected states of the
adjoint of $E_{7}$.

To be completely explicit, we now present a local model which exhibits the
desired enhancement type. \ To this end, let $z_{1}$ and $z_{2}$ denote two
local coordinates in the vicinity of the local $E_{8}$ enhancement at the
point $z_{1}=z_{2}=0$. \ In terms of these coordinates, one realization of the
desired triple intersection is:%
\begin{align}
t_{1}  &  =z_{2}(z_{1}+z_{2})\\
t_{2}  &  =z_{2}(z_{1}-z_{2})\\
t_{3}  &  =-4z_{1}z_{2}\\
t_{4}  &  =z_{1}(z_{1}+z_{2})\\
t_{5}  &  =z_{1}(-z_{1}+z_{2})\text{.}%
\end{align}
Along the locus $z_{1}=0$, we note that $t_{3}=t_{4}=t_{5}=0$, and
$t_{1}+t_{2}=0$, which corresponds to a local enhancement of $SU(5)$ to
$E_{7}$. \ Similarly, along $z_{2}=0$, we obtain another local enhancement
from $SU(5)$ to $E_{7}$. \ There are also several curves along which $SU(5)$
enhances to $SU(6)$. \ For example, $t_{1}+t_{3}=z_{2}(z_{2}-3z_{1})$ vanishes
along $z_{2}=0$, and $z_{2}=3z_{1}$. \ Along this second locus, the local
behavior of the $t_{i}$'s is:%
\begin{align}
t_{1}  &  =12z_{1}^{2}\\
t_{2}  &  =-6z_{1}^{2}\\
t_{3}  &  =-12z_{1}^{2}\\
t_{4}  &  =4z_{1}^{2}\\
t_{5}  &  =2z_{1}^{2}\text{.}%
\end{align}
which establishes that no additional states from a $10$ or $5$ of $SU(5)_{2}$
beyond those expected become massless. \ Thus, the required triple
intersection of matter curves is realized by the local coordinates:%
\begin{align}
E_{7}  &  :z_{1}=0\\
E_{7}^{\prime}  &  :z_{2}=0\\
SU(6)  &  :z_{2}=3z_{1}\\
E_{8}  &  :z_{1}=z_{2}=0\text{,}%
\end{align}
as desired. \ In this particular case, we note that the local enhancement to
$E_{7}$ may contain additional matter fields beyond the $27$ of $E_{6}$. \ In
this regard, this example should be viewed as a starting point for a more
complete analysis.

\section{Fayet-Polonyi Model of Supersymmetry Breaking\label{FPMODEL}}

On general grounds, there are likely to be several dynamical mechanisms
available which generate a vev for $X$ consistent with gauge mediated
supersymmetry breaking. \ Rather than posit the existence of an entirely new
sector which would inevitably dilute the predictive power of the theory, in
this section we show that the anomalous $U(1)_{PQ}$ gauge theory which has
already figured prominently in this paper will in many cases break
supersymmetry. \ Along these lines, we also show that this sector of the
theory can naturally accommodate a \textit{low scale of supersymmetry breaking}
consistent with the other results of this paper.

The anomalous $U(1)_{PQ}$ gauge theory is effectively a combination of a Fayet
model D-term potential which sets the value of $x$, and a Polonyi term which
sets the value of $F$ in the vev $\left\langle X\right\rangle =x+\theta^{2}F$.
\ In this regard, it is important that the $U(1)_{PQ}$ symmetry is both gauged
\textit{and} anomalous. \ The Fayet-Iliopoulos parameter $\xi_{PQ}$ is
determined by the background flux through the Peccei-Quinn seven-brane. \ By
appealing to a variant of the Bousso-Polchinski flux-scanning argument of
\cite{BoussoPolchinski}, we show that $x\sim\Delta_{\min}\cdot M_{X}$ where
$\Delta_{\min}\sim M_{X}/M_{pl}$ is the minimal flux spacing. \ The fact that
the $U(1)_{PQ}$ symmetry is anomalous is more crucial for the Polonyi term.
\ Indeed, precisely because the $U(1)_{PQ}$ symmetry is anomalous, instanton
effects will generate superpotential terms which violate the global
$U(1)_{PQ}$ symmetry of the low energy theory.

The rest of this section is organized as follows. \ First, we spell out in
field theory terms what we mean by the Fayet-Polonyi model. \ Next, we focus
on the specific realization of this model in our context. \ To this end, we
determine the D-term potential of the anomalous $U(1)_{PQ}$ theory and further
show that instanton effects generate a linear superpotential term which breaks
supersymmetry. \ To complete our analysis, we
next show that the Fayet-Polonyi model breaks supersymmetry at a scale
consistent with solving both the $\mu$ problem and strong CP problem. \ Finally, we show how the bosonic partner of the axion, the saxion is stabilized in the present class of models. Some
additional more technical discussion of higher order instanton effects and
their potential relevance for the axion potential is deferred to Appendix
\ref{AXCONT}.

\subsection{Generalities of Fayet-Polonyi Model}

Here we briefly specify in field theoretic terms what we mean by a
Fayet-Polonyi model of SUSY breaking. First we recall each of the two models
and then present the hybrid model.

We start with the Fayet model \cite{FayetModel}.\footnote{See
\cite{Fayet:1976et,Fayet:1977yc,Fayet:1979sa} for early work on supersymmetry
breaking which exploits the presence of related $U(1)$ symmetries.} \ This is a
model of an $\mathcal{N}=1$ supersymmetric $U(1)$ gauge theory coupled to some
charged fields $X,Y$ with respective charges $+1$ and $-1$. \ In this model,
the FI parameter $\xi$ is non-zero, and the superpotential is given by:
\begin{equation}
W=mXY\text{.}%
\end{equation}
The physical potential is a combination of the F-term contribution and the
D-term contribution:%
\begin{equation}
V=\left\vert \frac{\partial W}{\partial X}\right\vert ^{2}+\left\vert
\frac{\partial W}{\partial Y}\right\vert ^{2}+{\frac{1}{2}}g^{2}D^{2}%
\end{equation}
where $g$ is the gauge coupling and:%
\begin{equation}
D=|X|^{2}-|Y|^{2}-\xi\text{.}%
\end{equation}
Supersymmetry is preserved only when $\partial W/\partial X=\partial
W/\partial Y=D=0$. The F-flatness conditions require the vev of the fields
$X,Y$ to be zero. On the other hand, this is incompatible with the vanishing
$D$-term which requires one of the two fields $X$ or $Y$ to cancel the
contribution from the FI term $\xi$. \ For large enough $\xi$ (i.e. if $\xi\gg
m^{2}/g^{2}$) the minimum of the potential will be dictated by screening the
D-term. Without loss of generality, consider the case where $\xi$ is large a
positive so that $X$ attains a non-zero vev. \ In this case, the top component
of the $Y$ field $F_{Y}=mX\not =0$ so that supersymmetry is broken. \ The
explicit value of $F_{Y}$ is:%
\begin{equation}
F_{Y}=m\sqrt{\xi-{\frac{m^{2}}{g^{2}}}}\text{.}%
\end{equation}
Note this mechanism works even if there are more charged fields and that these
additional fields typically pick up contributions to their masses on the order
of $m$, due to the imperfect screening of the D-term.

We now turn to the Polonyi model \cite{PolonyiModel}. This is a model of a
chiral superfield with a superpotential of the form
\begin{equation}
W=\kappa X\text{.}%
\end{equation}
Since $\kappa=\partial W/\partial X\neq0$, this model breaks supersymmetry
because for a generic interacting field theory where the K\"{a}hler potential
$K(X,{\overline{X}})$ is non-trivial, this SUSY breaking cannot be absorbed
away by an overall shift in the vacuum energy density.

As its name suggests, the hybrid Fayet-Polonyi model combines elements from
both models. \ Consider again a theory of an $\mathcal{N}=1$ supersymmetric
$U(1)$ gauge theory coupled to various charged matter fields. Let $X$ denote
one such field. Moreover let us assume that the $U(1)$ is anomalous and is
Higged through the Green-Schwarz mechanism, which in particular requires the
existence of an FI-term $\xi$. \ Because the low energy global $U(1)$ symmetry
is anomalous, instanton effects will likely generate $U(1)$ violating terms in
the superpotential:%
\begin{equation}
W(X)=\kappa X+...
\end{equation}
where in particular the linear term $\kappa$ is not zero. Note that $\kappa$
is expected to be very small, if it is an instanton generated effect, of the
order of $\Lambda_{0}^{2}\mathrm{\exp}(-1/g^{2})$ for some cutoff mass scale
$\Lambda_{0}$. In the application we will have for the PQ brane, the higher
powers of $X$ could also potentially appear in the superpotential but they are
much smaller than the leading term, and so we have effectively a Polonyi-like
superpotential. Note that the fact that the superpotential seems to violate
$U(1)$ charge is not inconsistent with gauge symmetry, because the instanton
corrections, captured by $\kappa$ pick up anomalous $U(1)$ charge exactly to
neutralize the term. The same comment applies to the higher monomials as well.
The mechanism of supersymmetry breaking is now very similar to that of the
Fayet model: The non-vanishing $\xi$ term drives $X$ to have a non-trivial VEV
of the order of $\sqrt{\xi-a}$ where $|a|\sim(|W^{\prime}W^{\prime\prime
}/g^{2}X|)_{|X|=\sqrt{\xi}}$. The term $a$ also sets the scale for the mass
contributions from the VEV of $X$ to the other fields charged under this
$U(1)$ (due to the imperfect screening of the D-term). In the application
below such mass corrections are rather small and do not make significant
contributions. Indeed the first contribution to these mass terms will come
from the coefficient of $X^{2}$ and if generated by instanton effects will
generically be far smaller in scale than the linear term. Although ultimately
different, see \cite{DudasPokorski} for a related model of supersymmetry
breaking which combines elements from the Fayet and Polonyi models.

\subsection{Fayet-Polonyi Model from a $U(1)_{PQ}$ Seven-Brane}

In this subsection we describe the explicit realization of the Fayet-Polonyi
model based on a PQ seven-brane. \ Although the $U(1)_{PQ}$ symmetry can
sometimes correspond to a linear combination of distinct $U(1)$ factors on
different seven-branes, this is irrelevant for the purposes of the present
analysis, and we shall therefore always consider the case where $U(1)_{PQ}$ is
realized on a single seven-brane.

Now, as we have seen in previous sections, all of the fields of the MSSM\ are
necessarily charged under $U(1)_{PQ}$. \ Hence, the $U(1)_{PQ}$ seven-brane
theory will contain several different matter curves which will all contribute
to the four-dimensional effective theory. \ The zero mode content of the
four-dimensional theory is then determined by an appropriate choice of
background flux on the GUT model seven-brane, the $U(1)_{PQ}$ seven-brane, and
possibly other seven-branes of the compactification. \ To set notation, let
$\Psi$ denote a generic chiral superfield charged under $U(1)_{PQ}$ which
localizes on a curve $\Sigma_{\Psi}$, and $X$ a chiral superfield localized on
a curve $\Sigma_{X}$ which can potentially develop a supersymmetry breaking vev.

Before proceeding to the explicit realization of the\ Fayet-Polonyi model, we
now describe some general features of the $U(1)_{PQ}$ gauge theory realized on
a seven-brane. \ For a generic choice of background fluxes on the $U(1)_{PQ}$
seven-brane, the zero mode content of the theory will be such that the
four-dimensional $U(1)_{PQ}$ symmetry is anomalous. \ As a consequence, the
corresponding gauge boson in the four-dimensional effective theory will
develop a mass due to the Green-Schwarz mechanism. \ More explicitly, in the
eight-dimensional worldvolume theory of the PQ\ seven-brane, there is an
axion-like coupling between the RR\ four-form potential and the $U(1)_{PQ}$
gauge field strength of the form:%
\begin{equation}
L_{PQ}\supset M_{\ast}^{8}\underset{\mathbb{R}^{3,1}\times B_{3}}{\int
}dC_{(4)}\wedge\ast_{10}dC_{(4)}+M_{\ast}^{4}\underset{\mathbb{R}^{3,1}\times
S_{PQ}}{\int}C_{(4)}\wedge F_{%
\mathbb{R}
^{3,1}}\wedge\left\langle F_{PQ}\right\rangle \text{.}%
\end{equation}
Here, $F_{%
\mathbb{R}
^{3,1}}$ and $F_{PQ}$ respectively denote the $U(1)_{PQ}$ field strengths in
the four-dimensional spacetime directions and the K\"{a}hler surface $S_{PQ}$
wrapped by the PQ\ seven-brane. \ Reducing to the four-dimensional effective
theory, we note that under a $U(1)_{PQ}$ gauge transformation, the four-form
will in general shift by an amount which exactly cancels the contribution from
the anomaly. \ Letting $\mathcal{C}$ denote the four-dimensional superfield
associated with the reduction of the four-form $C_{(4)}$ and $V_{PQ}$ the
vector multiplet for the $U(1)_{PQ}$ gauge boson so that under the gauge
transformation $V_{PQ}\mapsto V_{PQ}+i\Lambda_{PQ}-i\Lambda_{PQ}^{\dag}$,
$\mathcal{C}\mapsto\mathcal{C}+i\Lambda_{PQ}$, the corresponding action in
superspace is:%
\begin{equation}
L_{PQ}\supset\int d^{4}\theta K\left(  \mathcal{C+C}^{\dag}-V_{PQ}\right)
\text{.}%
\end{equation}
where $K$ denotes an appropriate K\"{a}hler potential. Expanding in powers of
$V_{PQ}$, this is given as:%
\begin{equation}
L_{PQ}\supset\int d^{4}\theta K\left(  \mathcal{C+C}^{\dag}\right)
-K^{\prime}\left(  \mathcal{C+C}^{\dag}\right)  V_{PQ}+\frac{1}{2}%
K^{\prime\prime}\left(  \mathcal{C+C}^{\dag}\right)  V_{PQ}^{2}+...\text{.}%
\end{equation}

Fixing a background value of $\mathcal{C}$ as $\mathcal{C}_{0}$ so that
$\mathcal{C=C}_{0}+c$, in terms of component fields, we conclude that the
four-dimensional effective action contains the terms:%
\begin{equation}
L_{PQ}\supset\frac{1}{2}\xi_{\ast}\left(  \partial_{\mu}c+A_{\mu}\right)
^{2}+\xi_{\ast}D_{PQ} \label{PQCOMP}%
\end{equation}
where $\sqrt{\xi_{\ast}}$ is a mass scale associated with the PQ\ seven-brane,
and $D_{PQ}$ denotes the usual auxiliary field of the vector multiplet. \ In
the above, we have dropped irrelevant constant multiplicative factors of order
one which will not play any role in the discussion to follow. \ By inspection,
the first term in equation (\ref{PQCOMP}) will contribute to the mass of the
$U(1)_{PQ}$ gauge boson, while the second term contributes to the
FI\ parameter of the four-dimensional theory.

Depending on the particular flux data through the seven-brane theory, there
could potentially be additional contributions to the effective action. \ In
particular, in subsection \ref{DTERMPOT} we show that there is an additional
contribution to the FI\ parameter from fluxes when the K\"{a}hler form is not
orthogonal to $\left\langle F_{PQ}\right\rangle $. \ Moreover, precisely
because the $X$ field is charged under the anomalous $U(1)_{PQ}$ symmetry, it
follows that the phase of the $X$ field $a_{x}$ studied in section
\ref{AXIONGMSB} as a candidate axion field will also shift under a $U(1)_{PQ}$
gauge transformation. \ Including the contribution from these terms, we have:%
\begin{equation}
L_{PQ}\supset\left\vert x\right\vert ^{2}\left(  \partial_{\mu}a_{x}+A_{\mu
}\right)  ^{2}+\frac{1}{2}\xi_{\ast}\left(  \partial_{\mu}c+A_{\mu}\right)
^{2}+\left(  \xi_{flux}+\xi_{\ast}\right)  D_{PQ}\text{.}%
\end{equation}
In the following, we shall refer to the net FI\ parameter as:%
\begin{equation}
\xi_{PQ}\equiv\xi_{flux}+\xi_{\ast}\text{.} \label{FINET}%
\end{equation}
In general, $\left\vert \xi_{\ast}\right\vert \gg\left\vert x\right\vert $, so
that the mass of the $U(1)_{PQ}$ gauge boson will typically be close to the
GUT\ scale. \ On the other hand, in subsection \ref{DTERMPOT} we will show
that by scanning over all fluxes, the minimal non-zero value of the mass scale
$\sqrt{\xi_{PQ}}$ is typically in the range required for $a_{x}$ to play the
role of the QCD axion.

From the perspective of the four-dimensional effective theory, it may at first
appear puzzling that $x$ can develop any vev at all due to the D-term
potential. Indeed, since the gauge boson is very heavy, it can effectively be
integrated out. In this context, there is no D-term potential to speak of. The
essential point is that at high energy scales, the D-term potential is more
appropriately written as:%
\begin{equation}
V_{D}=2\pi\alpha_{PQ}\left(  \left\vert x\right\vert ^{2}-K^{\prime
}(\mathcal{C}+\mathcal{C}^{\dag})+\xi_{flux}\right)  ^{2}.
\end{equation}
The condition $V_{D}=0$ corresponds to a background field configuration where
in general both $x$ and $\mathcal{C}$ develop non-zero vevs. The essential
point we will exploit later is that the vev of $x$ can be tuned to be lower
than the mass scale of the PQ\ gauge boson. The actual axion will then be
given as a linear combination of the phase of $x$, with a small contribution
from $c$ as well. Indeed, this corresponds to a flat direction of the
potential $V_{D}$. Besides the axion, the other bosonic component of the
corresponding supermultiplet is the saxion. We now show how the D-term and
F-term potentials of the Fayet-Polonyi model can be generated in the PQ
seven-brane theory.

\subsubsection{D-Term Potential\label{DTERMPOT}}

As shown above, the four-dimensional anomalous $U(1)_{PQ}$ gauge boson will
develop a mass through a Green-Schwarz mechanism via an axion-like coupling
between the RR\ four-form and the eight-dimensional $U(1)_{PQ}$ field
strength. \ In the four-dimensional effective theory, the strength of this
coupling is proportional to the net flux $F_{PQ}$ through $S_{PQ}$. \ The
presence of this flux can induce additional contributions to the
FI\ parameters when $F_{PQ}$ is not orthogonal to the K\"{a}hler form on
$S_{PQ}$. \ For ease of discussion, we will present this analysis in the
simplified case where there is only a single matter curve $\Sigma_{X}$
corresponding to the $X$ field of the $U(1)_{PQ}$ theory.

First recall that the D-term equation of motion for the $U(1)_{PQ}$
seven-brane theory is \cite{BHVI}:%
\begin{equation}
\omega_{PQ}\wedge F_{PQ}=\frac{1}{2}\omega_{PQ}\wedge\delta_{\Sigma_{X}%
}\left[  \mu\left(  \sigma,\overline{\sigma}\right)  -\mu\left(
\overline{\sigma^{c}},\sigma^{c}\right)  \right]  \label{DEOM}%
\end{equation}
which holds pointwise on the surface $S_{PQ}$ wrapped by the Peccei-Quinn
seven-brane. \ In the above, $\omega_{PQ}$ denotes the K\"{a}hler form on
$S_{PQ}$ and $F_{PQ}$ is the internal gauge field strength. \ In addition, at
each point of $\Sigma_{X}$, $\sigma$ and $\sigma^{c}$ respectively label the
scalar component of the four-dimensional $\mathcal{N}=1$ chiral superfields
$\mathbb{X}$ and $\mathbb{X}^{c}$ and $\mu$ denotes the natural moment map
pairing defined on $\Sigma_{X}$ \cite{BHVI}. \ Expanding about a background
gauge field configuration on $\Sigma_{X}$, let $X_{i}$ denote the zero modes
of $\mathbb{X}$ with similar conventions for $X_{i_{c}}^{c}$. \ Integrating
equation (\ref{DEOM}) over the curve $\Sigma_{X}$ implies:%
\begin{align}
\underset{\Sigma_{X}}{\int}\left(  \omega_{PQ}\wedge F_{PQ}\right)
|_{\Sigma_{X}}  &  =\frac{1}{2}\left[  \underset{i}{\sum}\mu\left(
x_{i},\overline{x}_{i}\right)  -\underset{i_{c}}{\sum}\mu\left(  \overline
{x}_{i_{c}}^{c},x_{i_{c}}^{c}\right)  \right] \\
&  \sim e_{X}\frac{M_{\ast}^{2}Vol(\Sigma_{X})}{2}\left[  \underset{i}{\sum
}\left\vert x_{i}\right\vert ^{2}-\underset{i_{c}}{\sum}\left\vert x_{i_{c}%
}^{c}\right\vert ^{2}\right]
\end{align}
where $e_{X}$ denotes the integral charge of $X$ under $U(1)_{PQ}$. \ In the
above, we have used the fact that the zero modes are orthogonal in the sense
that $\mu\left(  x_{i},\overline{x}_{j}\right)  =0$ for $i\neq j$.
\ Canonically normalizing the kinetic terms for $X$ and $X^{c}$, the D-term
equation of motion in the four-dimensional effective theory is:%
\begin{equation}
\xi_{flux}\equiv\underset{\Sigma_{X}}{\int}\left(  \omega_{PQ}\wedge
F_{PQ}\right)  |_{\Sigma_{X}}=\frac{e_{X}}{2}\left[  \underset{i}{\sum
}\left\vert x_{i}\right\vert ^{2}-\underset{i_{c}}{\sum}\left\vert x_{i_{c}%
}^{c}\right\vert ^{2}\right]  \label{FIPQ}%
\end{equation}
where by abuse of notation, we have labelled the rescaled $x$'s by the same
variable. \ Equation (\ref{FIPQ}) demonstrates that the FI\ parameter of the
$U(1)_{PQ}$ gauge theory is given by integrating the background flux over
$\Sigma_{X}$. \ Letting $V_{PQ}$ denote the vector multiplet of the
$U(1)_{PQ}$ gauge theory, in four-dimensional $\mathcal{N}=1$ superspace, the
effective action for the $X$ zero modes contains the terms:%
\begin{equation}
L_{PQ}^{(0)}\supset\int d^{4}\theta\left[  \left(  X_{i}\right)  ^{\dag
}e^{tV_{PQ}}X_{i}+\left(  X_{i_{c}}^{c}\right)  ^{\dag}e^{-tV_{PQ}}X_{i_{c}%
}^{c}+\frac{2t}{e_{X}}\xi_{flux}V_{PQ}\right]  \label{LPQZERO}%
\end{equation}
where $t=e_{X}\cdot g_{PQ}/2$. \ Returning to equation (\ref{FINET}), we
conclude that the net FI\ term is given by the sum of the bulk contribution
$\xi_{\ast}$, and contributions from matter curves, $\xi_{flux}$.

Having identified the origin of the Fayet-Iliopoulos term in the anomalous
$U(1)_{PQ}$ theory, the D-term potential is therefore given by:%
\begin{equation}
V_{\text{Fayet}}=2\pi\alpha_{PQ}\cdot D_{PQ}^{2}=2\pi\alpha_{PQ}\cdot\left(
\left\vert X\right\vert ^{2}+\underset{\Psi}{\sum}e_{\Psi}\left\vert
\Psi\right\vert ^{2}-K^{\prime}(\mathcal{C}+\mathcal{C}^{\dag})+\xi
_{flux}\right)  ^{2}%
\end{equation}
where $\alpha_{PQ}$ is the fine structure constant of the $U(1)_{PQ}$ gauge
theory, $\xi_{PQ}$ is given by equation (\ref{FINET}), the $\Psi$'s denote all
fields besides $X$ charged under $U(1)_{PQ}$, and $e_{\Psi}$ denotes the
integral charge of $\Psi$ in a normalization where $e_{X}=2$. Here, we have
also included the presence of the field $\mathcal{C}$.

\subsubsection{F-Term Potential\label{FTERMPOT}}

In this subsection we sketch the form of the superpotential generated by
Euclidean three-branes wrapping a del Pezzo surface $S_{PQ}$. \ Instanton
effects in type II\ string theory have recently been investigated, for
example, in \cite{BlumenhagenWeigandINST,IbanezUrangaMajorana,SaulinaKachru}.
\ Particular applications to local F-theory models have been studied in
\cite{HMSSNV,MarsanoToolbox}. \ For simplicity, we shall assume that instanton
effects from Euclidean three-branes wrapping other K\"{a}hler surfaces in the
geometry are sufficiently small that they can safely be neglected.
\ Instantons will generically contribute to any $U(1)$ seven-brane theory with
matter localized on curves. \ To see why this is so, consider the reduction of
the eight-dimensional seven-brane theory to the six-dimensional theory defined
by one such matter curve. \ The net contribution to the six-dimensional gauge
anomaly is proportional to $TrF^{4}$ so that all contributions to the anomaly
will contribute with the same sign. \ Because this symmetry is anomalous, we
expect instantons of the six-dimensional theory to contribute to the effective superpotential.

Of course, the particular form of the instanton generated superpotential will
depend on the zero mode content of the four-dimensional effective theory.
\ The $k$-instanton sector contribution to the superpotential is given by
summing over all internal fluxes of $k$ Euclidean three-branes wrapping the
surface $S_{PQ}$. \ Letting $X_{1},...,X_{p}$ denote the various zero modes of
the four-dimensional effective theory, the net contribution from instantons
is:%
\begin{equation}
W_{inst}^{tot}=\underset{I,k,f_{inst,k,m}}{\sum}c_{I,k,f_{inst,k,m}}q^{k}\cdot
w^{\nu(f_{inst,k,I})}X^{I}\text{,}%
\end{equation}
where $I=(i_{1},...,i_{p})$ is a multi-index, and $X^{I}\equiv X_{1}^{i_{1}%
}\cdot\cdot\cdot X_{p}^{i_{p}}$. \ Letting $\tau_{PQ}$ denote the complexified
gauge coupling constant of the PQ gauge theory and $\tau_{IIB}$ the
axio-dilaton, $q=\exp(2\pi i\tau_{PQ})$ is the generic instanton contribution
from a three-brane wrapping $S_{PQ}$, and $w=\exp(2\pi i\tau_{IIB})$ is a
contribution which depends on the internal instanton number $\nu
(f_{inst,k,I})$ through a given three-brane configuration defined by the
supersymmetric internal flux $f_{inst,k,I}$. Here, we have indicated the
schematic form of the contribution, because in general F-theory backgrounds,
$\tau_{IIB}$ will have non-trivial position dependence. In such cases, the
contribution from the analogue of $w$ will correspond to the integration of
$\tau_{IIB}$ against the instanton density defined by the internal flux.

The background choice of fluxes can in many cases lead to a Polonyi-like
superpotential, and here we shall assume that this is realized in the present class of models.
In the related explicit example of \cite{HMSSNV}, it was shown that
this is indeed the case for a configuration of seven-branes wrapping
appropriate del Pezzo surfaces which intersect along a rigid genus zero matter
curve. When a single zero mode $X$ localizes on this curve, instantons of the
higher dimensional gauge theory will generate the leading order superpotential
terms:
\begin{equation}
W_{inst}^{tot}=M_{PQ}^{2}\kappa_{1}qX+O(q^{2})\text{.} \label{POLLIKE}%
\end{equation}
where the $\kappa_{i}$ correspond to moduli dependent worldvolume determinant
factors, and we have also absorbed the contribution from $w$ into $\kappa_{i}$
as well. \ More generally, for an appropriate choice of background flux
through the PQ\ seven-brane, a similar analysis establishes that this same
contribution will also be present even when additional zero modes localize on
other curves. \ Although the other fields of the MSSM\ are charged under
$U(1)_{PQ}$, our expectation is that these instanton effects will be dominated
by other contributions to the tree level superpotential. \ Some additional
technical details on this point as well as some discussion on higher order
instanton corrections are deferred to Appendix \ref{AXCONT}.

Including all terms which involve a non-trivial dependence on $X$, the full
superpotential is therefore of the form:
\begin{equation}
W(X,Y)=\lambda XYY^{\prime}+M_{PQ}^{2}\kappa_{1}qX+O(M_{PQ}q^{2})+O(q)
\label{FIRSTSEC}%
\end{equation}
In the above, the $O(M_{PQ}q^{2})$ term corresponds to possible
multi-instanton contributions involving only the $X$ field, and the $O(q)$
term refers to possible contributions involving MSSM fields. \ At leading
order, the F-term potential is therefore:
\begin{align}
V_{\text{Polonyi}}  &  =\left\vert \frac{\partial W(X,Y)}{\partial
X}\right\vert ^{2}+\left\vert \frac{\partial W(X,Y)}{\partial Y}\right\vert
^{2}+\left\vert \frac{\partial W(X,Y)}{\partial Y^{\prime}}\right\vert ^{2}\\
&  =\left\vert \lambda YY^{\prime}+M_{PQ}^{2}\kappa_{1}\cdot q\right\vert
^{2}+\left\vert \lambda XY^{\prime}\right\vert ^{2}+\left\vert \lambda
XY\right\vert ^{2}\text{.}%
\end{align}

\subsection{Fayet-Polonyi Model\label{FayetPolonyi}}

We now determine the scale of supersymmetry breaking in our realization of the
Fayet-Polonyi model on a PQ seven-brane. \ In particular, we show that this
theory can accommodate the vev of $X$ required for the gauge mediation
scenario explored in this paper.

The effective potential for the $X$ field is given by the sum of
$V_{\text{Fayet}}$ and $V_{\text{Polonyi}}$:%
\begin{align}
V_{PQ}  &  =V_{\text{Fayet}}+V_{\text{Polonyi}}\\
&  =2\pi\alpha_{PQ}\cdot\left(  \left\vert X\right\vert ^{2}+\underset{\Psi
}{\sum}e_{\Psi}\left\vert \Psi\right\vert ^{2}-K^{\prime}(\mathcal{C}%
+\mathcal{C}^{\dag})+\xi_{flux}\right)  ^{2}\\
&  +\left\vert \lambda YY^{\prime}+M_{PQ}^{2}\kappa_{1}\cdot q\right\vert
^{2}+\left\vert \lambda XY^{\prime}\right\vert ^{2}+\left\vert \lambda
XY\right\vert ^{2}\text{.}%
\end{align}
By inspection, $V_{PQ}$ admits critical points where all fields other than $X$
and $\mathcal{C}$ vanish. \ Because a non-zero vev for such fields would break
at least part of the GUT group, we shall only consider vacua where $X$ and
$\mathcal{C}$ have non-trivial vevs.

Returning to equation (\ref{POLLIKE}), $\partial W/\partial X\neq0$ so that
supersymmetry is broken. \ Working to leading order in $q$, the vev
$\left\langle X\right\rangle =x+\theta^{2}F$ satisfies:%
\begin{align}
\left\vert x\right\vert ^{2}  &  =\xi_{PQ}\\
\frac{\overline{F}}{M_{PQ}^{2}}  &  =\kappa_{1}\cdot q\text{.} \label{FVAL}%
\end{align}
To estimate the value of $x$, recall that in terms of the background field
strength on $S_{PQ}$, we have:
\begin{equation}
\left\vert x\right\vert ^{2}=\xi_{PQ}=\xi_{flux}+\xi_{\ast}\text{.}%
\end{equation}
Note that in the absence of any matter curves, the Hermitian Yang-Mills
equations on $S_{PQ}$ would imply:
\begin{equation}
\omega_{PQ}\wedge F_{PQ}=0\text{.}%
\end{equation}
The presence of the matter curves allows
$\omega_{PQ}\wedge F_{PQ}$ to deviate away from zero. \ In vacua where $x$ is
non-zero, we therefore expect that the background $U(1)_{PQ}$ gauge field
configuration will effectively adjust itself so that $\xi_{PQ}$ is as close to
zero as possible. \ At generic points of $\Sigma_{X}$, the field strength
$F_{PQ}$ will scale as $M_{X}^{2}$. \ Assuming that the Planck length is the
minimal distance over which $F_{PQ}$ can vary by the amount $M_{X}^{2}$, it
follows that the minimal non-zero value of $\xi_{PQ}$ which can be attained
is:%
\begin{equation}
\xi_{\min}=\Delta_{\min}^{2}\cdot M_{X}^{2}%
\end{equation}
where:%
\begin{equation}
\Delta_{\min}=\frac{M_{X}}{M_{pl}}\sim10^{-3.5}%
\end{equation}
is the effective lattice spacing for Peccei-Quinn flux configurations.
\ Throughout this paper, we have assumed that $M_{X}\sim10^{15.5}$ GeV and
$M_{pl}\sim10^{19}$ which implies:
\begin{equation}
\left\vert x\right\vert \sim\Delta_{\min}\cdot M_{X}\sim10^{12}\text{ GeV.}%
\end{equation}
Remarkably, this simple estimate is in accord with the requirements of both
gauge mediation and axion physics! Recall that the solution of $\mu/B\mu$
problem using the Giudice-Masiero operator required that $\left\vert
x\right\vert /M_{X}\ll1$. We now see that this is related to
\begin{equation}
\frac{\left\vert x\right\vert }{M_{X}}\sim\Delta_{\min}\sim\frac{M_{GUT}%
}{M_{pl}}\ll1\text{.}%
\end{equation}
Indeed, this mild hierarchy between the GUT scale and the Planck scale played
a key role in some of the estimates of physical quantities (for example the
neutrino masses) in \cite{BHVII}.

Computing the scale of supersymmetry breaking $\sqrt{F}$ is somewhat more
delicate because the instanton action depends exponentially on the volume of
the K\"{a}hler surface $S_{PQ}$. \ For this reason, we shall instead determine
the mass scale $M_{PQ}$ required in order to achieve the value $F\sim10^{17}$
GeV$^{2}$. \ To estimate the size of the instanton factor, we introduce a
characteristic volume factor $V$ so that $q=\exp\left(  -M_{\ast}^{4}\cdot
V\right)  $. \ Similarly, we shall introduce a characteristic mass scale
$M\sim V^{-1/4}$. Returning to equation (\ref{FVAL}), the magnitude of $F$ is
then:%
\begin{equation}
\frac{\left\vert F\right\vert }{M_{PQ}^{2}}=\exp\left(  -M_{\ast}^{4}\cdot
V\right)  =\exp\left(  -\frac{2\pi}{\alpha_{GUT}}\frac{M_{GUT}^{4}}{M^{4}%
}\right)  \label{FSOLVE}%
\end{equation}
where in the first equality we have set $\kappa_{1}=1$, and in the second
equality we have used the relation between the volumes of surfaces and gauge
couplings for seven-branes discussed in \cite{BHVI,BHVII}. \ Assuming that
$M_{PQ}\sim M$, solving for $M$ thus yields:%
\begin{equation}
\frac{M}{M_{GUT}}=\left(  \frac{\alpha_{GUT}}{4\pi}\times W\left(  \frac{4\pi
}{\alpha_{GUT}}\cdot\frac{M_{GUT}^{4}}{F^{2}}\right)  \right)  ^{-1/4}\text{,}%
\end{equation}
where $W$ is the Lambert $W$-function. \ Setting $\alpha_{GUT}\sim1/25$,
$F\sim10^{17}$ GeV$^{2}$ and $M_{GUT}\sim3\times10^{16}$ GeV, we find:%
\begin{equation}
\frac{M}{M_{GUT}}\sim1.4\text{.}%
\end{equation}
In other words, with only a mild tuning of parameters in the geometry, the
anomalous $U(1)_{PQ}$ gauge theory achieves the scale of supersymmetry
breaking required to remain in accord with weak scale physics. \ Returning to
equation (\ref{FSOLVE}) and using the value $\left\vert F\right\vert
\sim10^{17}$ GeV$^{2}$ the instanton action is therefore given by:
\begin{equation}
\left\vert q\right\vert \sim\frac{\left\vert F\right\vert }{M_{PQ}^{2}}%
\sim5\times10^{-17}\text{.} \label{qest}%
\end{equation}

To conclude this subsection, we note that more generally, the value of
$\kappa_{1}$ could deviate from an order one number, so there is a certain
degree of tunability in such instanton contributions. \ Indeed, in general
F-theory compactifications, the value of $\tau_{IIB}$ can vary along the
three-fold base $B_{3}$.

\subsection{Stabilizing the Radial Mode}

In the previous sections we have seen that the parameter space of F-theory
GUTs is in principle compatible with a non-zero vev for $X$ and $q$ of the
form:%
\begin{align}
\left\vert X\right\vert  &  \sim10^{12}\text{ GeV},\\
\log\left\vert q\right\vert  &  \sim\log\frac{F}{M_{PQ}^{2}}\simeq
-38.\label{38now}%
\end{align}
On the other hand, the contribution from just the Polonyi terms and D-term
potential does not by itself stabilize the PQ invariant product:%
\begin{equation}
\widehat{X}\equiv qX.
\end{equation}
Indeed, at low energy scales, there is no D-term potential, as the effects of
the $U(1)_{PQ}$ gauge boson have been integrated out, leaving behind only an accidental global $U(1)$ at low energies. The
phase of $\widehat{X}$ corresponds to the QCD axion, which develops a
potential through QCD\ instanton effects, and possibly instanton contributions
from the\ PQ seven-brane. The norm corresponds to the other bosonic component of the chiral multiplet which is the saxion.

In this section we discuss how the norm of $\widehat{X}$ can develop a vev
compatible with the supersymmetry breaking conditions specified earlier.
Non-trivial contributions to the K\"{a}hler potential in theories with an anomalous $U(1)$ symmetry can often serve to stabilize the vevs of modes such as $\widetilde{X}$ which are not fixed by the Polonyi term alone \cite{ArkaniHamedANOM}. Our aim here will be to clarify the form of
fine-tuning necessary to achieve the required energy scales. To this end, we
first phrase in general terms the requisite conditions on the form of the
K\"{a}hler potential, and then show in a well-motivated example that these
conditions can be met.

To analyze the dynamics of $\widehat{X}$, it is convenient to consider the
unbroken PQ seven-brane gauge theory, which will contain the dynamical fields
$X$ and $q$. In principle, the FI\ parameter $\xi_{flux}$ should also
be included as a dynamical field, but this mode involves the dynamics of the
full ten-dimensional theory, and so for our purposes can effectively be
treated as a frozen parameter.

We begin by discussing the general features of the system defined by $q$, $X$,
and the PQ\ gauge boson. The holomorphic coupling $\tau$ of the PQ seven-brane
theory is related to $q$ through:%
\begin{equation}
q=\exp(2\pi i\tau)\equiv\exp(-S)\text{.}%
\end{equation}
For simplicity, we work in units where the PQ charge of the $X$ field is $+1$,
so that under a gauge transformation which shifts the PQ vector multiplet as:%
\begin{equation}
V_{PQ}\rightarrow V_{PQ}-\Lambda-\Lambda^{\dag},
\end{equation}
the $X$ field transforms as:%
\begin{equation}
X\rightarrow\exp\left(  \Lambda\right)  \cdot X.
\end{equation}
Gauge invariance of the product $qX$ then requires $q$ to transform as:%
\begin{equation}
q\rightarrow\exp\left(  -\Lambda\right)  \cdot q.
\end{equation}
or:%
\begin{equation}
S\rightarrow S+\Lambda.
\end{equation}
The theory with the PQ gauge boson can then be parameterized in terms of the
contribution from the K\"{a}hler potential and superpotential for $S$ and $X$:%
\begin{equation}
L\supset\int d^{4}\theta K(X^{\dag}e^{V_{PQ}}X;S+S^{\dag}+V_{PQ}%
)+\xi_{flux}V_{PQ}+\int d^{2}\theta M_{PQ}^{2}e^{-S}X+h.c.,
\end{equation}
where the general form we have taken is automatically invariant under gauge transformations.

To leading order, we approximate $K$ as the sum of two contributions, $K_{X}$
and $K_{S}$ such that:%
\begin{equation}
K(X^{\dag}e^{V_{PQ}}X;S+S^{\dag}+V_{PQ})=K_{X}(X^{\dag}e^{V_{PQ}}%
X)+K_{S}(S+S^{\dag}+V_{PQ}).
\end{equation}
Expanding to quadratic order in $V_{PQ}$, the mass of the PQ gauge boson is
given as:%
\begin{equation}
M_{U(1)_{PQ}}^{2}=X^{\dag}X\cdot K_{X}^{\prime}+\left(  X^{\dag}X\right)
^{2}\cdot K_{X}^{\prime\prime}+K_{S}^{\prime\prime},
\end{equation}
where the primes on $K_{X}$ and $K_{S}$ respectively denote derivatives of the K\"{a}hler potentials. To leading order, the mass of the PQ gauge boson
is controlled by $K_{S}^{\prime\prime}$, which can in principle be tuned from
a scale close to the GUT\ scale, to somewhat lower values.

Since the mass of the PQ gauge boson is far heavier than the scale of
supersymmetry breaking, we can view the D-term potential as imposing a
constraint on the norms $\left\vert X\right\vert $ and $\left\vert
q\right\vert $. Explicitly, the D-term constraint requires:%
\begin{equation}
X^{\dag}X \cdot K_{X}^{\prime}+K_{S}^{\prime}+\xi_{flux}%
=0.\label{Dconstraint}%
\end{equation}
Assuming $K_{X}^{\prime}>0$ in the regime of interest (which will be the case
when $K_{X}$ is to leading order given by the canonical K\"{a}hler potential),
note that $\xi_{flux}>0$ favors field configurations such that $X=0$
and $K_{S}^{\prime}\simeq-\xi_{flux}$. Indeed, this is just a
rephrasing of the Bousso-Polchinski flux scanning argument that $K_{S}%
^{\prime}$ and $\xi_{flux}$ should both be large and nearly cancel in
order for $\left\vert X\right\vert $ to remain below the GUT\ scale.

Next consider the F-term contribution to the potential:%
\begin{equation}
V_{F-term}=\left\vert M_{PQ}^{2}e^{-S}\right\vert ^{2}\left(  g^{X\overline
{X}}+g^{S\overline{S}}\left\vert X\right\vert ^{2}\right)  ,
\end{equation}
The overall multiplicative factor by $\exp(-S-S^{\dag})$ indicates the
tendency for this mode to approach $S\rightarrow\infty$, which is a possible
\textquotedblleft runaway direction\textquotedblright. This is counteracted,
however, by the D-term constraint of equation (\ref{Dconstraint}). We are
interested in the form of $V_{F-term}$ such that $\left\vert X\right\vert $ is
small in $M_{\ast}$ units, and $S$ is large. This can be achieved provided the
contributions from $g^{X\overline{X}}$ and $g^{S\overline{S}}\left\vert
X\right\vert ^{2}$ are roughly comparable in size. Assuming a roughly
canonical form for the K\"{a}hler potential of $X$, this means that the
K\"{a}hler metric $g_{S\overline{S}}$ must be small in $M_{\ast}$ units so
that $\left\vert X\right\vert ^{2}/g_{S\overline{S}}$ is comparable to
$g^{X\overline{X}}$. Note, however, that $g_{S\overline{S}}$ also enters into
the mass of the PQ gauge boson, and so cannot be too small.

The mass of the mode stabilized by $V_{F-term}$ is naturally close to the weak
scale. Indeed, the characteristic mass scale for fluctuations of $\widehat{X}$
are given as:%
\begin{equation}
m_{\widehat{X}}^{2}\sim\left\vert M_{PQ}^{2}e^{-S}\right\vert ^{2}%
g^{S\overline{S}}=\left\vert \frac{F}{M_{U(1)_{PQ}}}\right\vert ^{2}%
\propto\Delta_{PQ}^{2}.
\end{equation}
Thus, we can naturally expect the mass of the radial mode to be determined by
the scale of the PQ deformation. The precise numerical coefficient depends on
details of precisely how the vevs of $X$ and $q$ are fixed, but the basic
point remains that the mass of the radial component is near the weak scale,
while the phase, corresponding to the axion will have a much lower mass
induced by QCD\ instanton effects.

Having spelled out the general form of the required conditions, we now show
that well-motivated K\"{a}hler potentials $K_{X}$ and $K_{S}$ exhibit the
behavior required to achieve the desired minimum. This will also help to
clarify at least in this particular example the precise type of fine-tuning at
work in the Bousso-Polchinski flux scanning argument. For simplicity, we
consider the case where the K\"{a}hler potential of $S$ is given as:%
\begin{equation}
K_{S}=-M_{\ast}^{2}\log(S+S^{\dag}).
\end{equation}
This type of logarithmic behavior for $K_{S}$ is typically present for complex
surfaces which contract to a point at infinite distance in moduli space. A
different form based on a power law dependence is instead more natural if this
contraction occurs at finite distance in moduli space. In addition, we shall
consider a roughly canonical form for $K_{X}$ which allows for higher order
corrections:%
\begin{equation}
K_{X}=AX^{\dag}X+B\frac{\left(  X^{\dag}X\right)  ^{2}}{M_{X}^{2}}+O\left(
\frac{\left(  X^{\dag}X\right)  ^{3}}{M_{X}^{4}}\right)  ,\label{KXAB}%
\end{equation}
where the suppression scale $M_{X}$ is associated with integrating out massive
modes of size $M_{X}$ localized on the $X$ curve. The specific value of the
coefficient $B$ will in general receive corrections of both signs. In the
context of similar couplings between $X$ and MSSM fields, the dominant
contribution of this type is from heavy PQ\ gauge boson exchange. In the
context of purely $X$ contributions, however, the somewhat lower mass scale
$M_{X}$ allows for a more general possibility.

We now proceed to analyze the form of the effective potential in this case.
First consider the special case where we drop all higher order corrections to
$K_{X}$ in equation (\ref{KXAB}). The form of the D-term constraint is now:%
\begin{equation}
A\left\vert X\right\vert ^{2}-\frac{M_{\ast}^{2}}{S+S^{\dag}}+\xi
_{\text{flux}}=0
\end{equation}
while the F-term potential is given as:%
\begin{equation}
V_{F-term}=M_{PQ}^{4}\exp(-S-S^{\dag})\left(  \frac{1}{A}+(S+S^{\dag}%
)^{2}\frac{\left\vert X\right\vert ^{2}}{M_{\ast}^{2}}+O\left(  \left\vert
X\right\vert ^{4}\right)  \right)  ,
\end{equation}
where $V_{F-term}$ is to be viewed as a potential for the mode unfixed by the
D-term constraint. Parameterizing the form of $V_{F-term}$ in terms of
$\left\vert X\right\vert $, we obtain:%
\begin{align}
V_{F-term} &  =M_{PQ}^{4}\exp\left(  -\frac{M_{\ast}^{2}}{A\left\vert
X\right\vert ^{2}+\xi_{flux}}\right)  \cdot\left(  \frac{1}{A}%
-\frac{M_{\ast}^{2}}{A\left\vert X\right\vert ^{2}+\xi_{flux}}%
\frac{\left\vert X\right\vert ^{2}}{M_{\ast}^{2}}+...\right)  \\
&  \simeq M_{PQ}^{4}\exp\left(  -\frac{M_{\ast}^{2}}{\xi_{flux}%
}\right)  \cdot\left(  \frac{1}{A}+\frac{(M_{\ast}^{2}-\xi_{flux})}%
{\xi_{flux}^{2}}\left\vert X\right\vert ^{2}+...\right)  ,
\end{align}
where we have dropped all terms of order $\left\vert X\right\vert ^{4}$ to
obtain a consistent approximation which neglects the terms proportional to $B$
in $K_{X}$. The form of $V_{F-term}$ is quadratic, leading to a stable minimum
at $|X|=0$. To leading order, this is encouraging, because we are interested
in potentials with $X$ stabilized at values below the GUT\ scale.

%
\begin{figure}
[ptb]
\begin{center}
\includegraphics[
height=3.9228in,
width=6.6046in
]%
{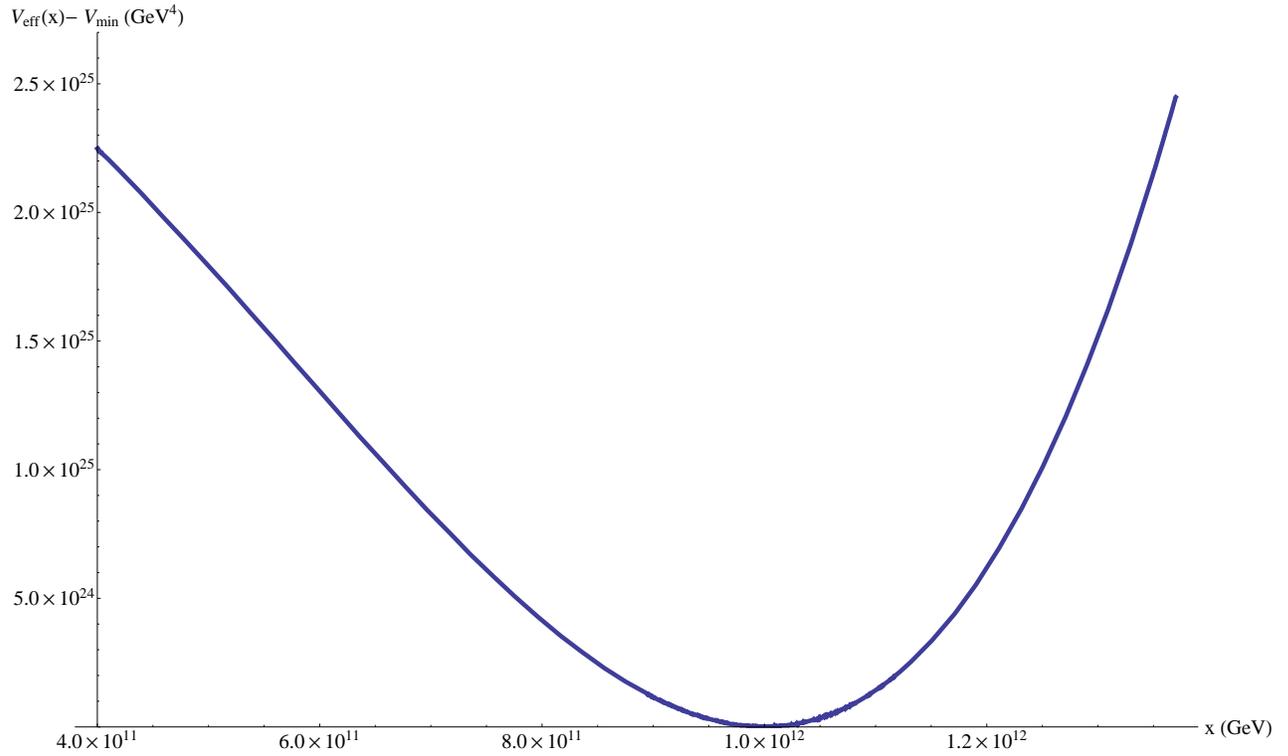}%
\caption{Plot of the effective potential for the saxion. With notation as in
section 8.4, the specific choice of parameters used in this plot are $M_{\ast
}=10^{17}$ GeV, $M_{X}=10^{15.5}$ GeV, $A=1$, $B=1.425$. By construction, the
value of $B$ has been chosen so that the minimum is located at $x_{\ast
}=10^{12}$ GeV. The shallow variation of the potential as a function of energy
scale illustrates that the mass of this radial mode is much smaller than $10^{12}$ GeV, and is instead closer to the weak scale.}%
\label{effsaxpot}%
\end{center}
\end{figure}

Including higher order corrections can shift the minimum of $V_{F-term}$ so
that a small non-zero vev for $X$ is indeed realized. Including the first
correction proportional to $B$ and expanding to order $\left\vert X\right\vert
^{4}$ now yields:%
\begin{equation}
V_{F-term}\simeq M_{PQ}^{4}\exp\left(  -\frac{M_{\ast}^{2}}{\xi_{flux}%
}\right)  \cdot\left(  \alpha+\beta\left\vert X\right\vert ^{2}+\gamma
\left\vert X\right\vert ^{4}+...\right)  ,
\end{equation}
where the coefficients $\alpha$, $\beta$ and~$\gamma$ are:%
\begin{align}
\alpha &  =\frac{1}{A}\\
\beta &  =\frac{(M_{\ast}^{2}-\xi_{flux})}{\xi_{flux}^{2}}%
-\frac{4B}{A^{2}}\frac{1}{M_{X}^{2}}\\
\gamma &  =\frac{A}{\xi_{flux}^{4}}\left(  (M_{\ast}^{2}-\xi
_{\text{flux}})^{2}-\frac{M_{\ast}^{4}}{2}\right)  -\frac{2B}{A^{2}}%
\frac{M_{\ast}^{2}}{M_{X}^{2}\cdot\xi_{flux}^{2}}+\frac{16B^{2}}{A^{3}%
}\frac{1}{M_{X}^{4}}.
\end{align}
An important feature of the above relations is that depending on the size of
$B$, the coefficient $\beta$ can be either positive or negative. In
particular, this illustrates that provided $B$ is large enough, we can expect
a shift in the minimum of $V_{F-term}$ to a small, nearby value $x_{\ast}$
given by:%
\begin{equation}
\left\vert x_{\ast}\right\vert ^{2}=\frac{A^{2}M_{X}^{2}\xi_{\text{flux}%
}+4B\xi_{flux}^{2}-A^{2}M_{X}^{2}M_{\ast}^{2}}{2A^{3}M_{X}^{2}%
+32\frac{B^{2}\xi_{flux}^{2}}{AM_{X}^{2}}-4ABM_{\ast}^{2}-\frac
{4A^{3}M_{X}^{2}M_{\ast}^{2}}{\xi_{flux}}+\frac{A^{3}M_{X}^{2}M_{\ast
}^{4}}{\xi_{flux}^{2}}}.\label{xcrit}%
\end{equation}

To obtain representative values for the size of the various coefficients, note
that the mass scale $M_{X}\sim10^{15.5}$ GeV is typically somewhat smaller
than $M_{\ast}\sim10^{17}$ GeV. As a representative example, we shall take:%
\begin{equation}
\frac{M_{X}}{M_{\ast}}\sim10^{-1.5}.
\end{equation}
Numerically, the value of $\xi_{flux}$ is fixed by the condition that
the value of $S$ generates an appropriate instanton action. Taking the rough
value specified by equation (\ref{38now}), this imposes the condition set by
the D-term constraint:%
\begin{equation}
\xi_{flux}\sim\frac{M_{\ast}^{2}}{76}.
\end{equation}
Achieving the required value of $\left\vert x_{\ast}\right\vert \sim10^{12}$
GeV then leads to a specific relation between $A$ and $B$. For example, with
$M_{X}\sim10^{15.5}$ GeV and $M_{\ast}\sim10^{17}$ GeV and $A=1$, $B\sim1.4$.
The coefficient $A$ is typically a number somewhat larger than
one, since the kinetic term for $X$ scales as $A\sim M_{\ast}^{2}%
Vol(\Sigma_{X})$, which is naturally on the order of $100-1000$. When
$A\sim100$, we instead find $B\sim1.4\times10^{4}$. Although this constitutes
a fine-tuning, note that the ratio $B/A^{2}$ remains an order one number.

\section{Region of MSSM Parameter Space\label{MSSMREGION}}

In previous sections, we have shown that low energy considerations constrain
the UV boundary conditions for the supersymmetry breaking sector. \ In
particular, we have found that for a broad class of F-theory
compactifications, the vev $\left\langle X\right\rangle =x+\theta^{2}F$
satisfies crude constraints from the weak scale when $x\sim10^{12}$ and
$F\sim10^{17}$ GeV$^{2}$. \ Based on such considerations, we can deduce that
if supersymmetry breaking is communicated to the MSSM\ via gauge mediation,
then the LSP\ will be the gravitino. \ In a purely top down approach, this
level of analysis is essentially all that can be obtained with any degree of
certainty. \ Indeed, deriving detailed features of the low energy spectrum
would require specifying all relevant interaction terms in both the F- and
D-terms, which at a purely practical level may not even be feasible.

Even if such a computation were in principle possible, there is no guarantee
that the resulting low energy physics would match to observation. \ Adhering
to the bottom up approach advocated in the Introduction, in this section we
show that simply achieving the correct low energy behavior in the Standard
Model strongly constrains both the sparticle spectrum, as well as properties
of the Higgs potential. \ Because we do not know their precise values, our
strategy will be to scan over a range of UV\ boundary conditions in gauge
mediation models close to the crude values specified in previous sections.
\ In addition, we also include the effects of the one parameter deformation
away from gauge mediation determined by heavy $U(1)_{PQ}$ gauge boson
exchange. \ We find that over this region, some parameters of the low energy
theory do not vary much, while others are more sensitive to high energy
inputs. \ To simplify the presentation, we shall confine our discussion to the
case of a single vector-like pair of messenger fields in the $5 \oplus
\overline{5}$ of $SU(5)$. \ The primary change in increasing the number of
messengers is that as the number of messenger fields increase, the NLSP\ can
transition from the bino-like neutralino to the stau.

To determine the low energy behavior of the theory, we have developed a small
modification of the program \texttt{SOFTSUSY} \cite{SOFTSUSYAllanach}. \ Once
an appropriate subset of UV boundary conditions has been specified,
\texttt{SOFTSUSY} performs a renormalization group flow of the parameters of
the theory down to the weak scale, adjusting the values of parameters such as
the $\mu$ and $B\mu$ terms as well as $\tan\beta$ near the weak scale to
remain consistent with electroweak symmetry breaking. \ More precisely, this
is accomplished by requiring that the masses squared for the Higgs fields
develop a suitable tachyonic value. \ The program then evolves these
parameters back up to the messenger scale, and iterates this procedure until
the results of these adjustments converge. \ In a certain sense, this
represents a microcosm for the entire bottom up approach to string
phenomenology. \ Indeed, the reason why the bottom up approach is in principle
quite predictive is that while crude considerations from UV or IR\ physics may
only serve to partially fix some details of the theory, iterating back and
forth between high and low scale physics can effectively constrain both
sectors to a high level of precision. \ Indeed, using this fact, we will be
able to extract remarkably detailed information which directly correlates the
high energy behavior of the theory with low energy physics.

The remainder of this section is organized as follows. \ In subsection
\ref{BOUND}, we specify in greater detail the UV\ boundary conditions which we
shall scan over, including contributions from heavy $U(1)_{PQ}$ gauge boson
exchange. \ Next, we turn to the low energy behavior of the theory in
subsection \ref{IRBOUND}. \ In particular, we show that achieving consistent
electroweak symmetry breaking effectively allows us to reliably extract the
value of $\mu$ and $\tan\beta$ near the scale of electroweak symmetry
breaking. \ Having determined the precise values of all UV boundary
conditions, we next present the low energy spectra for this class of models.
\ In nearly all gauge mediation models, the LSP is the gravitino, and our
situation is no different.\footnote{For an interesting recent example of a
gauge mediation scenario where the gravitino is not the LSP, see
\cite{CraigGreen}.} \ Similarly, with a single vector-like pair of messenger
fields in the $5 \oplus\overline{5}$ of $SU(5)$, for the most part, the NLSP
is given by a bino-like neutralino, although the one parameter deformation
from gauge mediation can also allow the stau to become the NLSP. \ Finally, we
conclude this section with a brief discussion of the mini-hierarchy problem.
\ As in most models, we find that a fine-tuning on the order of one part in a
hundred is required to remain in accord with present experimental bounds, and
including our deformation away from gauge mediation appears to only improve
this situation slightly.

\subsection{UV\ Boundary Conditions\label{BOUND}}

The low energy content of any mediation model is completely fixed once all
soft breaking parameters have been specified at the messenger scale. \ For
gauge mediation models, this amounts to specifying the number of messenger
fields, the messenger mass scale(s) $M_{mess}$, the gaugino mass unification
scale $\Lambda=F/x$, and the values of $\mu$, $B\mu$ and $\tan\beta$. \ The PQ
deformation can also contribute to the soft terms, and in certain cases will
lead to important deviations from the usual predictions of gauge mediation.
\ As explained in subsection \ref{PQDEF}, the size of this contribution is a
priori not fixed by purely local considerations, and it is therefore
appropriate to scan over the allowed range of soft mass terms to the Higgs
fields $H_{i}$ and all other chiral superfields $\Phi$ of the MSSM induced by
the PQ deformation:
\begin{align}
\delta_{PQ}m_{H_{i}}^{2}(M_{mess})  &  =2\Delta_{PQ}^{2}\\
\delta_{PQ}m_{\Phi}^{2}(M_{mess})  &  =-\Delta_{PQ}^{2}%
\end{align}
where $\Delta_{PQ}$ denotes the mass scale associated with the PQ deformation,
and the relative sign and magnitude is completely fixed by the relative
$U(1)_{PQ}$ charges of these fields. \ The UV boundary conditions for our
model are given by:
\begin{align}
M_{mess}  &  \sim10^{11.5}-10^{12.5}\text{ GeV}\\
\Lambda &  \sim10^{5}-10^{6}\text{ GeV}\\
B\mu(M_{mess})  &  =0\\
\mu &  \sim\pm10^{2} - 10^{3}\text{ GeV}\\
\Delta_{PQ}  &  \sim0 - 10^{3}\text{ GeV.}%
\end{align}
In addition, the A-terms vanish at the messenger scale. In the above, our
conventions for the sign of the $\mu$ term are the same as in
\cite{LESHOUCHESSkands}. \ Here, we have specified all requisite inputs for
the theory at the messenger scale, and have also included a potential range of
values over which we shall perform our scan of UV boundary conditions.
\ Because these boundary conditions completely fix the soft breaking
parameters of the MSSM, some points in this parameter space may not be
consistent with current experimental bounds on the mass of either the Higgs or
the sparticles of the MSSM.

\subsection{Constraining the MSSM\label{IRBOUND}}

To extract detailed properties of the low energy spectrum which are consistent
with electroweak symmetry breaking, we have used the \texttt{SOFTSUSY} package
\cite{SOFTSUSYAllanach}, and have also included a small modification which
incorporates the specific form of the PQ\ deformation in the present class of
models. \ The inputs at the messenger scale are given in terms of $M_{mess}$,
$\Lambda$, $\tan\beta(M_{mess})$, $\Delta_{PQ}$ and the sign of the $\mu$
term. \ Once these parameters are fixed, all other UV boundary conditions are
automatically adjusted by the algorithm to remain in accord with electroweak
symmetry breaking. \ A cursory inspection of the output reveals that the
spectrum is relatively insensitive to the value of $M_{mess}$ over the range
of values $10^{11.5}-10^{12.5}$. \ With little loss of precision, it is
therefore sufficient to fix $M_{mess}=10^{12}$ GeV. \ Similarly, for the most
part the sign of $\mu$ does not appear to significantly alter the results of
our analysis, and we shall therefore restrict to the case $\mu>0$.

Within the three-dimensional subspace parameterized by $\Lambda$, $\tan
\beta(M_{mess})$ and $\Delta_{PQ}$, only a two-dimensional subspace will in
general preserve the required boundary condition $B\mu(M_{mess})=0$. \ To
simplify our analysis, we fixed a particular value of either $\Lambda$ or
$\Delta_{PQ}$ and then scanned over the remaining two parameters to determine
the value of $\tan\beta(M_{mess})$. \ While the value of $\tan\beta(M_{mess})$
does depend on $\Lambda$, it is relatively insensitive to the value of
$\Delta_{PQ}$. \ To determine the effect of the PQ deformation, it therefore
suffices to fix all other UV boundary conditions, and simply vary the size of
$\Delta_{PQ}$.

The particular choice of values for $\Lambda\sim10^{5}-10^{6}$ GeV which we
scan over is in large part a consequence of current experimental bounds on the
mass of the lightest neutral Higgs $h^{0}$ ($114.5$ GeV). \ In addition, the
size of the deformation $\Delta_{PQ}\sim0-1000$ GeV is also bounded above by
the requirement that none of the squarks or sleptons should develop a
tachyonic mode near the weak scale. \ We note that as $\Lambda$ increases, the
soft scalar masses will also increase, so that larger PQ deformations become viable.

\begin{figure}
[ptb]
\begin{center}
\includegraphics[
height=4.1468in,
width=6.4524in
]%
{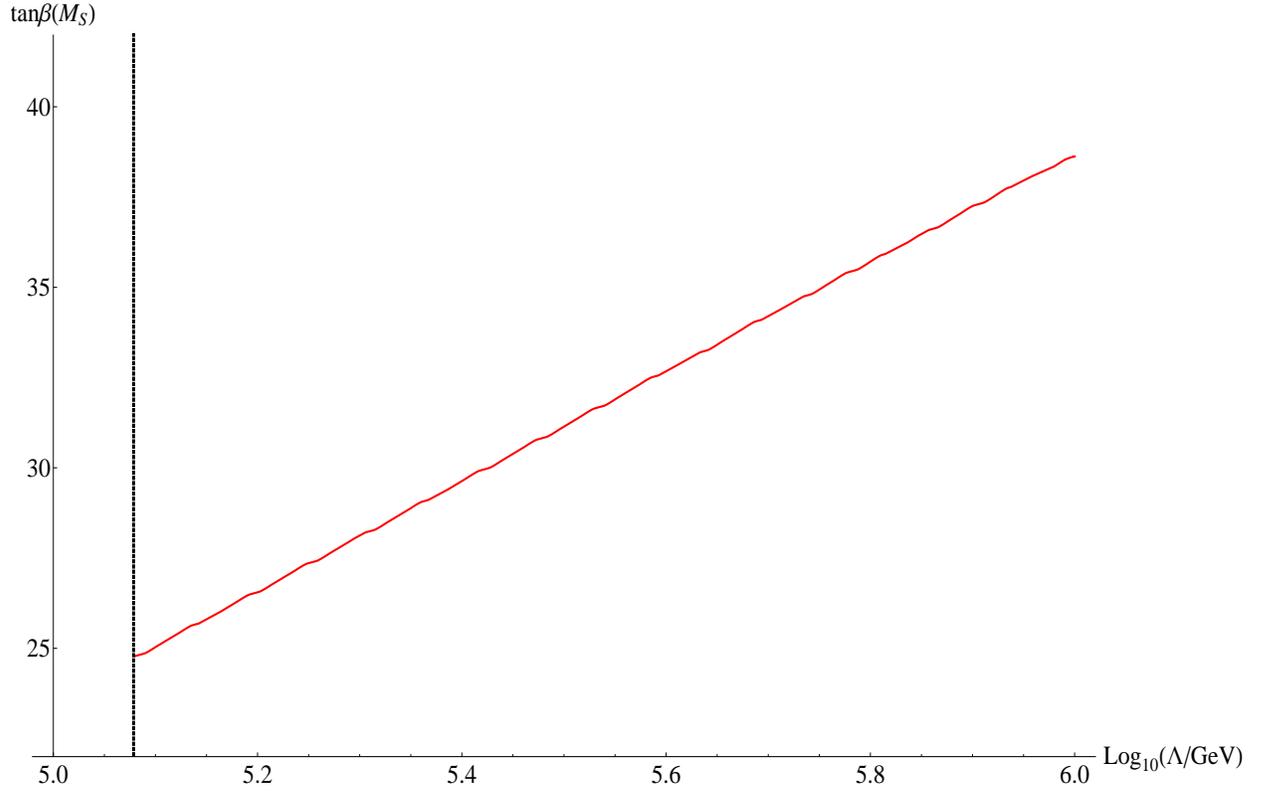}%
\caption{Plot of $\tan\beta$ at the scale $M_{S}$, the scale at which
electroweak symmetry breaking boundary conditions are imposed, as a function
of $\log_{10}(\Lambda/$GeV$)$. \ By inspection, $\tan\beta$ grows
logarithmically with the gaugino mass unification scale. \ The vertical line
at the left of the plot ($\Lambda = 10^{5.08}$ GeV) indicates
the experimentally excluded region based on current bounds on the mass of the Higgs.}%
\label{tanblambda}%
\end{center}
\end{figure}

\begin{figure}
[ptb]
\begin{center}
\includegraphics[
height=4.1468in,
width=6.4524in
]%
{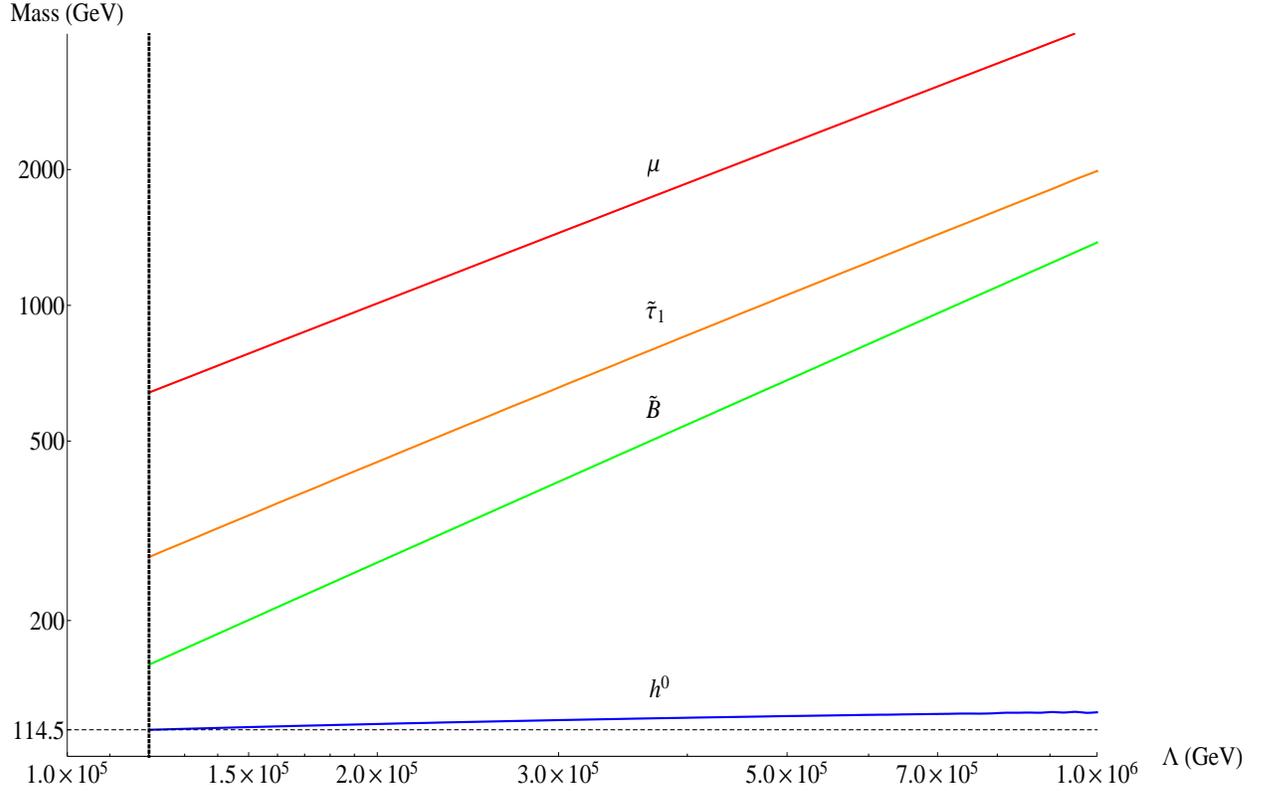}%
\caption{Plot of the $\mu$ term, stau mass, bino mass and Higgs mass as a
function of the gaugino mass unification scale $\Lambda$ in a single messenger
model with vanishing PQ\ deformation. \ The vertical line at the left ($\Lambda = 10^{5.08}$ GeV)
indicates the experimentally excluded region based on bounds on the mass of
the Higgs.}%
\label{mustauchilambda}%
\end{center}
\end{figure}

We present the results of various one parameter scans over $\Lambda$ and
$\Delta_{PQ}$. \ One scan is performed at vanishing PQ\ deformation so that
$\Delta_{PQ}=0$, with $\Lambda$ ranging from $10^{5}$ to $10^{6}$ GeV. \ At
the low end of this scan, the resulting Higgs mass in fact lies below the
experimental bound set by LEP\ that $m_{h^{0}}>114.4$ GeV at the 95 percent
confidence level \cite{LEPBOUND}. \ The mass of the Higgs monotonically
increases with $\Lambda$, and saturates the value $114.5$ GeV at
$\Lambda=10^{5.08}$ GeV. \ In figure \ref{tanblambda} we display $\tan
\beta(M_{S})$ as a function of $\Lambda$. \ Here, $M_{S}$ denotes the scale at
which electroweak symmetry breaking boundary conditions are imposed. \ By
inspection, $\tan\beta$ increases logarithmically with $\Lambda$ from $\sim25$
near $\Lambda=10^{5.08}$ GeV to $\sim38$ near $\Lambda=10^{6}$ GeV so that
$\tan\beta(M_{S})\sim31.5\pm6.5$. \ In the present single messenger model, the
NLSP is given by a bino-like neutralino, while the stau is the next lightest
MSSM sparticle. \ See figure \ref{mustauchilambda} for a plot of the bino
mass, stau mass and Higgs mass as a function of $\Lambda$. \ In this same
plot, we also present the value of the $\mu$ term as a function of $\Lambda$.
\ At its lowest value, $\mu=10^{2.8}\sim\allowbreak630$ GeV for $\Lambda
=10^{5.08}$ GeV, and the value of $\mu$ increases to $10^{3.6}\sim
\allowbreak4000$ GeV for $\Lambda=10^{6}$ GeV.%

We have also scanned over a range of values for the PQ deformation with
$\Lambda$ kept fixed. \ To leading order this deformation does not alter the
masses of any gauginos of the theory but does lower the masses of all sleptons
and squarks whilst increasing the soft masses squared of the Higgs fields.
\ Recall that in order for the Higgs potential to contain a tachyonic mode,
the mass squared of the Higgs fields undergoes a renormalization group flow
from positive values to negative values. \ As the PQ deformation increases,
the magnitude of this tachyonic mode consequently also decreases.

\begin{figure}
[ptb]
\begin{center}
\includegraphics[
height=4.1468in,
width=6.4524in
]%
{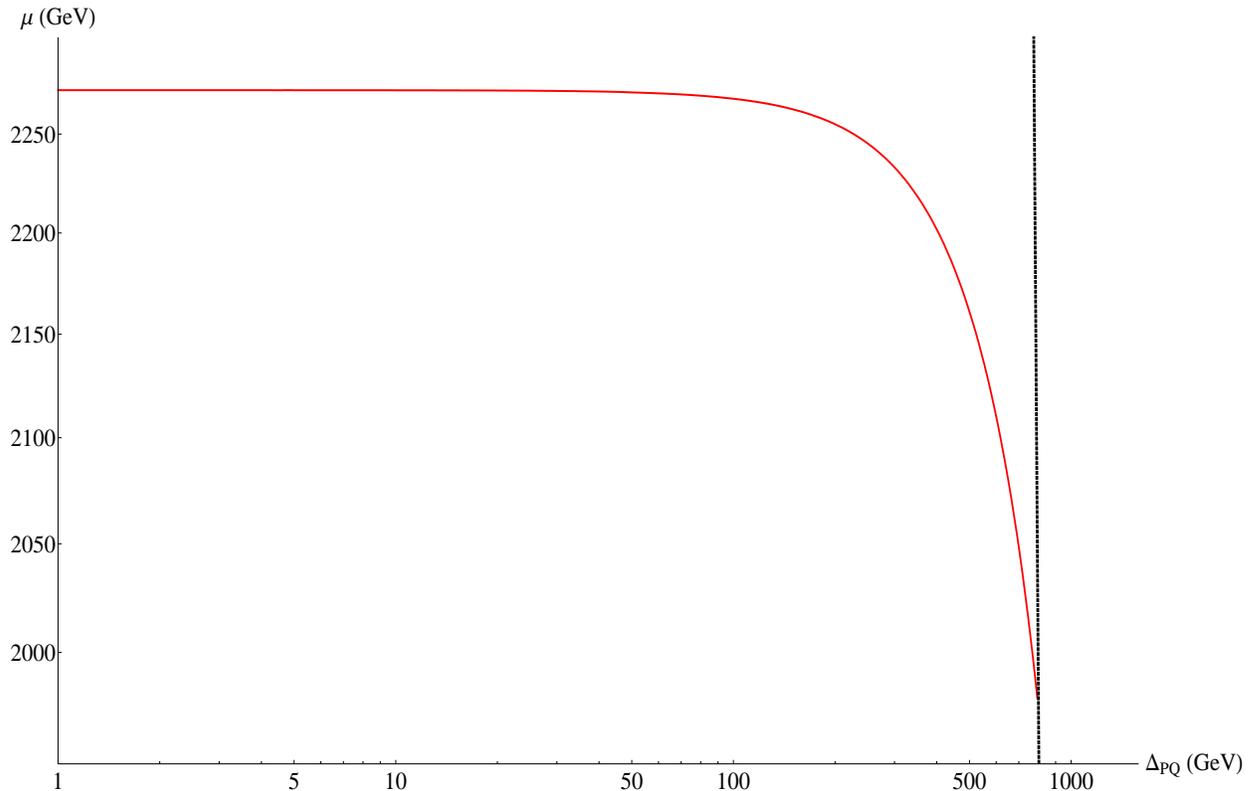}%
\caption{Plot of the parameter $\mu$ as a function of the PQ\ deformation
$\Delta_{PQ}$ in a single messenger model with $\Lambda=5\times10^{5}$ GeV.
Large values of $\Delta_{PQ}$ produce a tachyon in the effective potential,
which appears near the vertical line at $\Delta_{PQ} = 10^{2.9}$ GeV.}%
\label{pqmuplot}%
\end{center}
\end{figure}

\begin{figure}
[ptb]
\begin{center}
\includegraphics[
height=4.1468in,
width=6.4524in
]%
{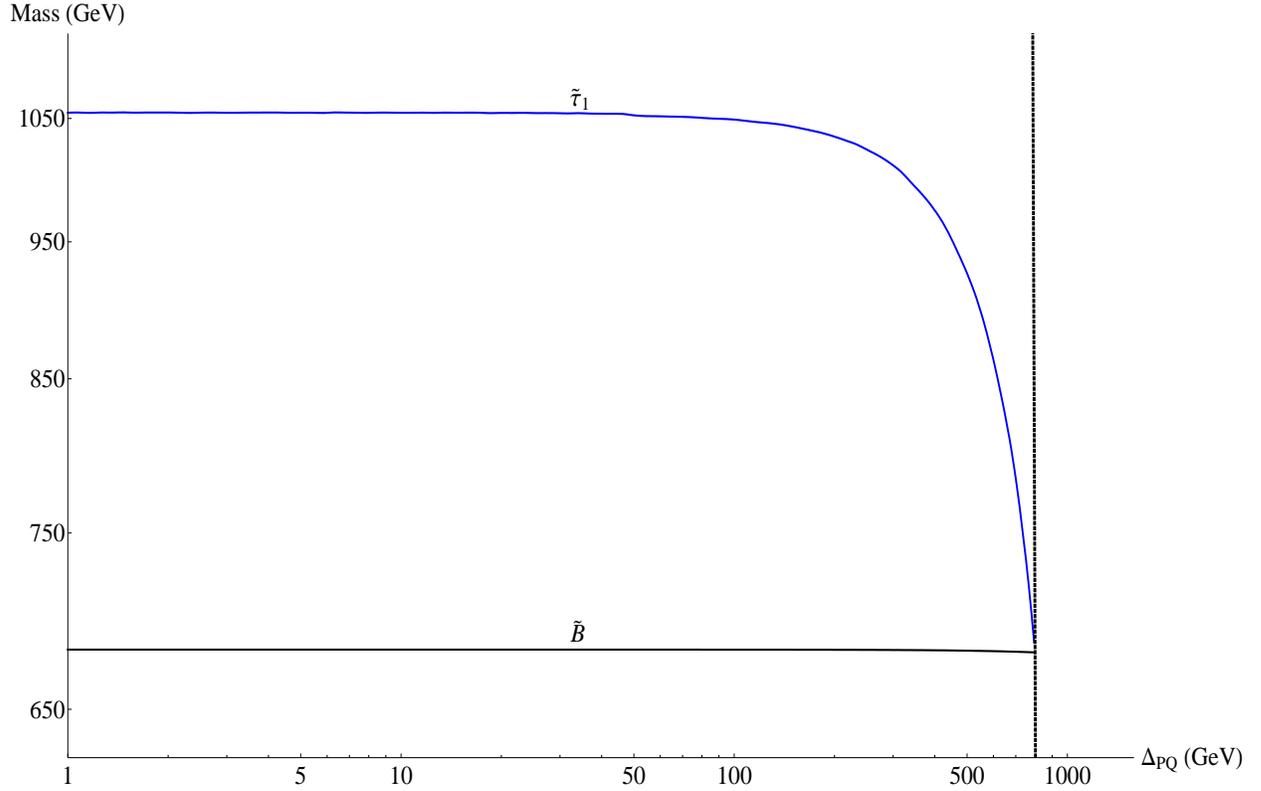}%
\caption{Plot of the stau mass and bino mass as a function of the
PQ\ deformation in a single messenger model with $\Lambda=5\times10^{5}$ GeV.
\ Whereas
the mass of the bino remains constant, for large values of $\Delta_{PQ}$, the
stau mass is comparable in mass to the bino. \ Near this region, the
effective potential develops a tachyonic mode at $\Delta_{PQ}=10^{2.9}$ GeV,
which is indicated by the vertical line at the right of the plot.}%
\label{pqstauchiplot}%
\end{center}
\end{figure}

As a representative example, we scanned over the PQ\ deformation for a fixed
value of $\Lambda=5\times10^{5}$ GeV. \ This scan begins at $\Delta_{PQ}=0$
and proceeds up to $\Delta_{PQ}=10^{2.9}$ GeV. \ For larger values of
$\Delta_{PQ}$, the corresponding effective potential for the sfermions
contains a tachyonic mode. \ Over this entire range, the resulting value of
$\tan\beta(M_{S})$ is $34\pm1$, and the mass of the lightest Higgs $h^{0}$ is
approximately $123\pm3$ GeV. \ See figure \ref{pqmuplot} for a plot of the
parameter $\mu$ as a function of $\Delta_{PQ}$. As the value of $\Delta_{PQ}$
increases, $\mu$ decreases slightly. \ A more dramatic consequence of the PQ
deformation is shown in figure \ref{pqstauchiplot} which shows that the mass
of the stau becomes nearly equal to the bino at large values of the PQ deformation.

\begin{figure}
[ptb]
\begin{center}
\includegraphics[
height=4.1468in,
width=6.4524in
]%
{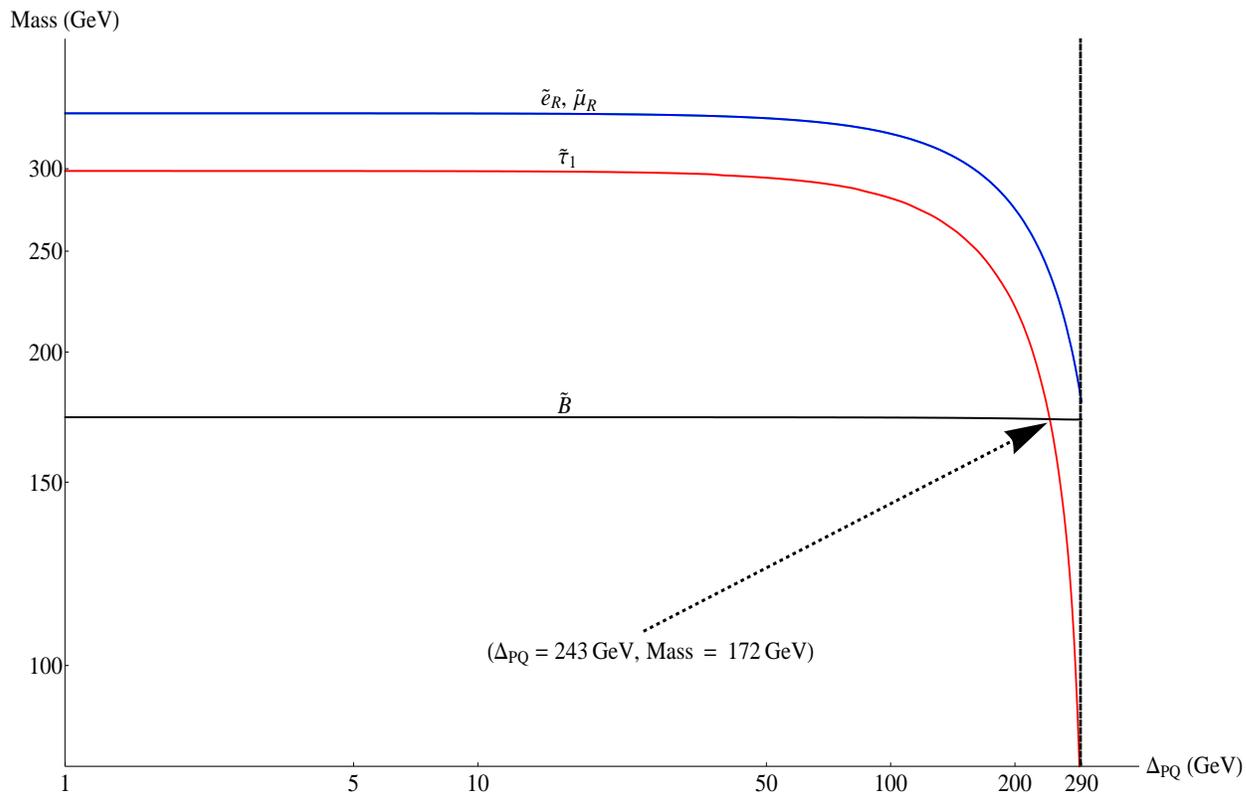}%
\caption{Plot of the stau, smuon, selectron and bino masses as a function of the
PQ deformation in a single messenger model with $\Lambda=1.3\times10^{5}$ GeV. \ Whereas
the mass of the bino remains constant, for large values of $\Delta_{PQ}$, the
other sparticles become lighter, and for a narrow range of values, the stau can
in fact become the NLSP. \ The mass of the stau equals the mass of the bino ($172$ GeV)
at $\Delta_{PQ} = 243$ GeV. \ Near this region, the effective potential develops
a tachyonic mode at $\Delta_{PQ}=290$ GeV, which is indicated by the vertical
line at the right of the plot.}%
\label{pqlowlambda}%
\end{center}
\end{figure}

For smaller values of $\Lambda$, the stau can become the NLSP at large values
of the PQ deformation. Performing a scan over the PQ deformation at the value
$\Lambda= 1.3 \times10^{5}$ GeV, we find that $\Delta_{PQ}$ can range from
zero up to $290$ GeV, beyond which point a tachyon develops in the scalar
potential. Over this entire range, the resulting value of $\tan\beta(M_{S})$
is $26 \pm1$, and the mass of the lightest Higgs $h^{0}$ is $115\pm1$ GeV. The
profile of the parameter $\mu$ is similar in shape to that given in figure
\ref{pqmuplot}, although the ranges of the scales are different. At large
values of the PQ deformation, the value of $\mu$ is $550$ GeV. In figure
\ref{pqlowlambda} we plot the four lightest sparticles and find that at a
value of $\Delta_{PQ} = 243$ GeV, the mass of the stau and bino-like
neutralino are identical. For a narrow range of values, the stau becomes
lighter. In addition, the mass of the right-handed selectron and smuon become
comparable in mass to the bino-like neutralino at large PQ deformation. \ This
example explicitly shows that that for a fixed value of $\Lambda$, increasing
the size of $\Delta_{PQ}$ can allow the stau to become the NLSP.

\begin{figure}
[ptb]
\begin{center}
\includegraphics[
height=4.1468in,
width=6.4524in
]%
{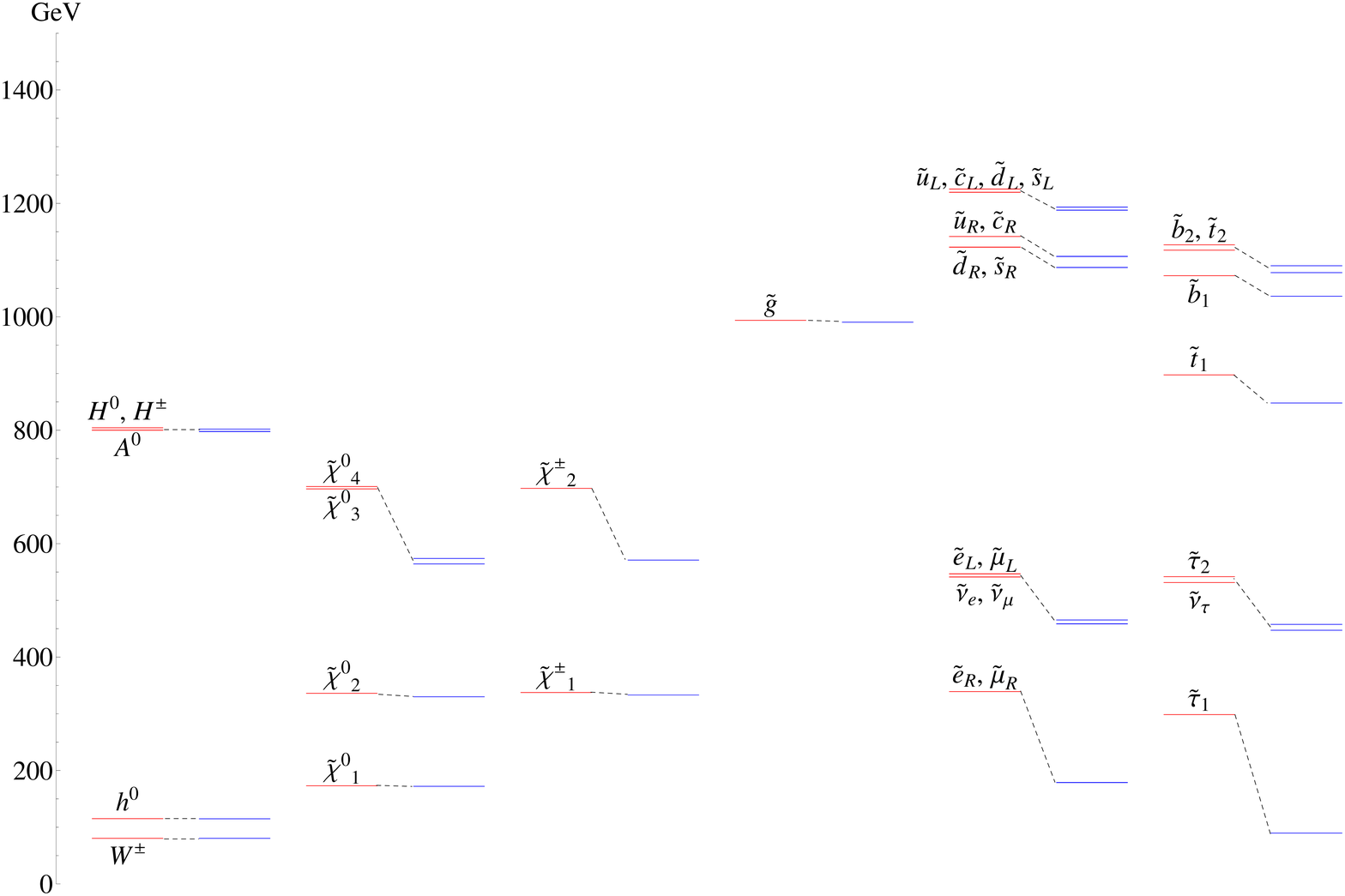}%
\caption{Plot of the sparticle masses separated by pairs of columns
in a single messenger model with gaugino mass unification scale $\Lambda
=1.3\times10^{5}$ GeV for vanishing PQ\ deformation (left red columns) and
for a deformation of $\Delta_{PQ}=290$ GeV (right blue columns). \ For this
particular value of $\Lambda$, larger values of $\Delta_{PQ}$ produce
a tachyonic mode in the effective potential. \ For large
deformations, the stau ($\tilde{\tau}_{1}$) is the NLSP. \ Further, the
right-handed selectron ($\tilde{e}_{R}$) and smuon ($\tilde{\mu}_{R}$) also
decrease in mass to the point where they are comparable to the mass of the
bino-like lightest neutralino ($\tilde{\chi}^{0}{}_{1}$). \ Note also that the
two Higgs-like neutralinos ($\tilde{\chi}^{0}{}_{3}$, $\tilde{\chi}^{0}{}_{4}%
$) and the Higgs-like charginos ($\tilde{\chi}^{\pm}{}_{2}$) both decrease in
mass in the PQ deformed theory.}%
\label{together}%
\end{center}
\end{figure}

It is also of interest to compare the entire sparticle spectrum in the limit
of vanishing PQ deformation, as well as in the presence of maximal
PQ\ deformation. \ Because the resulting value of $\tan\beta(M_{S})$ is
relatively insensitive to PQ\ deformations, in figure \ref{together} we
directly compare the sparticle spectrum at $\Lambda=1.3\times10^{5}$ GeV for
$\Delta_{PQ}=0$ with the spectrum obtained with a PQ\ deformation of
$\Delta_{PQ}$ $=290$ GeV. \ Beyond this value, a tachyonic mode is generically
present in the sfermion effective potential. \ At such large values of the PQ
deformation, we observe that the stau is now the NLSP. \ In fact, this plot
also shows that the masses of the right-handed selectron $(\tilde{e}_{R})$ and
smuon $(\tilde{\mu}_{R})$ also have significantly lower masses, which are
nearly comparable to the mass of the bino-like lightest neutralino $\left(
\tilde{\chi}^{0}{}_{1}\right)  $. \ While we certainly expect the scalar
masses of the MSSM\ to receive corrections from the PQ deformation, it is also
of interest to note that the Higgs-like charginos $\left(  \tilde{\chi}^{\pm
}{}_{2}\right)  $ also decrease in mass. \ This is primarily due to the fact
that the PQ\ deformation also alters the Higgs potential.

As mentioned above, large values of the PQ deformation decrease the magnitude
of the tachyonic mode in the Higgs potential, which alleviates some of the
fine-tuning present in the Higgs sector. \ To a certain extent, this
fine-tuning can be quantified as in \cite{FineTuneCarlos,FineTuneBarbieri}.
\ Using an existing routine in \texttt{SOFTSUSY}, we have computed the amount
of fine-tuning in the mass of the $Z$ boson with respect to $\mu$ using the
definition adopted in \cite{FineTuneCarlos}:%
\begin{equation}
\delta_{\mu}=\left\vert \frac{\partial\log M_{Z}^{2}}{\partial\log\mu
}\right\vert \text{.}%
\end{equation}
In general, while the PQ deformation appears to decrease the amount of
fine-tuning, it does not address in any substantial way the mini-hierarchy
problem. \ This is because at larger values of the PQ deformation, we
generically encounter a tachyonic mode in the slepton effective potential
which limits the amount of fine-tuning that this deformation can eliminate.
\ For example, with $\Lambda=1.3\times10^{5}$ GeV, we find $\delta_{\mu
}=2.3\times10^{2}$ when $\Delta_{PQ}=0$, while $\delta_{\mu}=1.5\times10^{2}$
for $\Delta_{PQ}=290$ GeV. \ In this way, the usual problem of fine-tuning at
the percent level is present here as well. \ Nevertheless, with the aim of
potentially further reducing the amount of fine-tuning present in this class
of models, it would be interesting to determine whether more elaborate
F-theory based models could contribute additional soft masses to the Higgs
fields, without introducing any effect on the other soft masses of the MSSM.
\ While we do not have an explicit realization of such a scenario, we have
also considered the phenomenology of models where the sign of the
PQ\ deformation is reversed so that the soft masses squared of the sleptons
and squarks increase while those of the Higgs fields decreases. \ In this
case, the masses of the gauginos effectively remain constant, and the
sfermions all increase in mass. \ This alternate PQ\ deformation also
increases the amount of fine-tuning in the Higgs potential. \ For example,
along this branch, we find that with $\Lambda=1.3\times10^{5}$ GeV and
$\Delta_{PQ}=10^{4}$ GeV, the fine-tuning measure $\delta_{\mu}=9.7\times
10^{4}$, and as is to be expected, the slepton and squarks achieve masses of
order $10^{4}$ GeV. \ For larger values of $\Delta_{PQ}$, a perturbative
analysis breaks down. \ Nevertheless, this at least suggests that for large
values of $\Delta_{PQ}$ along this branch, many of the scalar bosons of the
spectrum could effectively remain out of present observational reach, but not
too far away, perhaps realizing a less extreme version of the split
supersymmetry scenario advocated in
\cite{WellsSplit,ArkaniHamedSplit,GiudiceSplit}.

Having determined detailed features of the sparticle spectrum, it is important
to extract potential experimental signatures from this class of models. \ One
immediate avenue of interest would be to determine possible collider
signatures at the LHC \cite{ColliderStudy}. \ Indirect cosmological tests
could also serve to constrain the behavior of this class of models. \ As one
example, we note that the mass of the gravitino is roughly given by:%
\begin{equation}
m_{3/2}=\sqrt{\frac{4\pi}{3}}\frac{\left\vert F \right\vert }{M_{pl}}%
\sim10^{-2}-10^{-1}\text{ GeV}%
\end{equation}
where $M_{pl}\sim1.2\times10^{19}$ GeV. \ We find it encouraging that this
range of masses for the gravitino appears to be in accord with constraints
from big bang nucleosynthesis \cite{KawasakiBBN}. \ In this regard, it is
curious to observe that for slightly larger values of the gravitino mass near
$1$ GeV such as has been advocated in the sweet spot model of supersymmetry
breaking \cite{KitanoIbeSweetSpot}, there appears to be a slight tension with
such constraints. \ In any case, \ it would be of great interest to study
cosmological constraints on this class of models \cite{CosmoStudy}.

\section{Conclusions\label{CONCLUDE}}

Low scale supersymmetry breaking provides a window into the high energy
behavior of local F-theory GUT\ models. \ From a bottom up perspective,
correlating the scale of supersymmetry breaking with the weak scale imposes
additional restrictions on the ultraviolet behavior of the theory. \ In a
broad class of local F-theory models where the Higgs fields and Goldstino
chiral superfield $X$ localize on matter curves, correlating these two energy
scales translates into the simple geometric condition that these curves must
form a triple intersection. \ In this paper we have shown that $X$ either
couples to the Higgs fields through an F-term proportional to $XH_{u}H_{d}$ or
through a D-term proportional to $X^{\dag}H_{u}H_{d}$ which is generated by
integrating out the Kalua-Klein modes on the same curve as the zero mode $X$.
\ In the former case, the resulting vev $\left\langle X\right\rangle
=x+\theta^{2}F$ pushes the masses of the Higgs fields far above the weak
scale, exacerbating the $\mu/B\mu$ problem. \ In the latter case, the
resulting D-term realizes a variant of the Giudice-Masiero mechanism.
\ However, the suppression scale for the operator $X^{\dag}H_{u}H_{d}$ is
typically at its largest a few orders of magnitude below the Planck scale.
\ This has the important consequence that in these local models,
gravity/moduli mediated supersymmetry breaking generates a $\mu$ term which is
far too large. \ Instead, the scale of supersymmetry breaking must be
sufficiently low to solve the $\mu$ problem so that $F\sim10^{17}$ GeV$^{2}$.
\ Assuming that the dominant mediation mechanism is instead gauge mediated
supersymmetry breaking, bottom up considerations also determine the value of
$x\sim10^{12}$ GeV. \ We have also provided an explicit configuration of
intersecting seven-branes which realizes gauge mediated supersymmetry breaking
as well as a variant of the Giudice-Masiero mechanism. \ In this explicit
case, all of the fields of the MSSM in addition to the messenger sector are
charged under an ambient $U(1)$ Peccei-Quinn gauge symmetry which is typically
anomalous. \ In fact, we have also seen that the phase of $X$ can potentially
play the role of the QCD\ axion. \ Remarkably, purely bottom up considerations
connected to the weak scale automatically determine the axion decay constant
to be of order $f_{a}=\sqrt{2}\left\vert x\right\vert \sim10^{12}$ GeV, which
fits within the available window for invisible axion models. \ Motivated by
these considerations, we next explained why the existence of this $U(1)_{PQ}$
symmetry is particularly natural in many F-theory constructions which contain
$E_{6}$ singularities. \ In addition we have shown that instanton effects in
the anomalous $U(1)_{PQ}$ symmetry realized on the worldvolume of a
Peccei-Quinn seven-brane can break supersymmetry. \ The resulting value of $F$
is on the order of $10^{17}$ GeV$^{2}$ when the scale of the Peccei-Quinn
seven-brane is close to the GUT scale. \ We have also seen that the D-term
potential determines $x$ as a function of the minimal amount of flux through
the curve supporting $X$, and that this minimal value is $x\sim10^{12}$ GeV,
in beautiful agreement with purely bottom up considerations. \ Combining all
of these elements, we have also characterized the region of MSSM parameter
space determined by this class of compactifications. \ In the remainder of
this section we describe some further directions of potential interest.

In this paper we have seen that there is a preferred range of energy scales
available for deformed gauge mediated supersymmetry breaking in many local
F-theory GUT models. \ We have also given a precise description of the region
of MSSM\ parameter space defined by these models. \ It would be of great
interest to extract the collider signatures associated with this narrow region
of the MSSM parameter space.

We have also shown how higher rank enhancements in the singularity type of
F-theory GUT models can make the `diamond ring model' more natural. \ Even
though we have sketched many elements of this setup, some issues, including
how to avoid excess matter fields from additional adjoint representation
associated with the higher rank of enhancement remain to be settled in this
scenario. \ It would be important to address these issues in future work.

One of the most elegant features of gravity/moduli mediation models is that
the Giudice-Masiero mechanism automatically correlates the scale of
supersymmetry breaking with the value of the $\mu$ term. \ On the other hand,
we have also seen that in a broad class of models, integrating out the
Kaluza-Klein modes of the $X$ field will generate a Giudice-Masiero operator.
\ Insofar as the mass of these heavy modes is below the Planck scale, it is
therefore natural to ask whether gauge mediation is always preferred in such
cases. \ Depending on the origin of the $U(1)_{PQ}$ gauge symmetry, there
appear to be at least two ways that gravity mediation models could still
potentially yield an appropriate value for the $\mu$ term.

When the $U(1)_{PQ}$ gauge symmetry originates from the worldvolume theory of
seven-branes, $X$ and the Higgs fields most likely originate from matter
curves. \ In this paper we have assumed that the profile of the zero mode wave
function for the $X$ field near a point of triple intersection is an order one
number. \ On the other hand, depending on the choice of signs for gauge
fluxes, the local curvature of the del Pezzo surface can repel this gauge
singlet wave function away from the point of triple intersection \cite{BHVII}.
\ In this case, the coefficient of the Giudice-Masiero operator could
naturally be much smaller, effectively increasing the size of the suppression
scale. \ In this paper we have avoided exploiting this mechanism because it is
less predictive, but it is still a viable possibility.

It is also possible that the $U(1)_{PQ}$ gauge symmetry does not originate
from the worldvolume theory of a seven-brane. \ Indeed, the bulk gravity modes
of a generic compactification will typically contain several $U(1)$ factors
obtained by Kaluza-Klein reduction. \ In such a scenario, though, it is less
clear whether all of the matter content of the visible and messenger sectors
possess the correct $U(1)_{PQ}$ charge assignments to allow all required
couplings. \ For these models, the $X$ field may not originate from a matter
curve, but could simply be some generic modulus of the compactification. \ In
either case, it would be interesting to study additional properties of such
gravity/moduli mediated scenarios.

\section*{Acknowledgements}

We thank B.C. Allanach, M. Dine, A.L. Fitzpatrick, G. Giudice, T. Hartman, J. Marsano,
N. Saulina, S. Sch\"{a}fer-Nameki, P. Svr\v{c}ek, A.M. Uranga, H. Verlinde,
and E. Witten for helpful discussions. We would also like to thank the Sixth
Simons Workshop in Mathematics and Physics for hospitality while some of this
work was performed. \ JJH would also like to thank the 2008 Amsterdam Summer
Workshop on String Theory for hospitality while some of this work was
performed. \ The work of the authors is supported in part by NSF grant
PHY-0244821. \ The research of JJH was also supported by an NSF Graduate Fellowship.

\appendix

\section{Higher Order Instanton Corrections and the Axion
Potential\label{AXCONT}}

In section \ref{AXIONGMSB} we observed that the phase $a_{x}=\arg x$ directly
couples to the QCD\ instanton density. \ In order to solve the strong
CP\ problem, the minimum of the effective potential for this field must be
sufficiently close to zero. \ In this regard, the gauged $U(1)_{PQ}$ symmetry
shields the axion from many contributions which could a priori have shifted
the minimum of its potential. \ More generally, however, instanton effects
could violate this symmetry. \ In the context of the Fayet-Polonyi model, note
that the leading order behavior of the effective potential only depends on the
magnitude of $X$, which does not generate a potential for $a_{x}$. \ But
precisely because the experimental bounds on $\theta_{QCD}$ are so stringent,
subleading corrections from instanton effects could also potentially play an
important role. \ Such contributions can either correspond to terms in the
superpotential involving just the $X$ fields, or interaction terms between $X$
and fields such as the MSSM\ Higgs which develop an appropriate vev so that
they can contribute to the minimum of the axion potential. \ In this Appendix we appeal to the $U(1)_{PQ}$ symmetry of the seven-brane theory to
characterize the form of higher order instanton corrections, and then estimate
their effects on the axion potential. \ We show that if present, some of these
contributions can significantly alter the minimum of the axion potential.
\ But although considerations based on symmetry arguments can constrain the
form of possible contributions, they do not establish that such terms are
indeed present. To this end, we discuss potential means by which instanton
generated contributions to the axion potential can remain in accord with the
flatness required to solve the strong CP\ problem.

\subsection{Constraints From Symmetries}

Using symmetry arguments, we now provide a rough characterization of
additional instanton generated terms which involve the chiral superfields of
the MSSM. \ Although the linear term in $X$ in equation (\ref{POLLIKE})
appears to violate the anomalous $U(1)_{PQ}$ symmetry, the instanton factor
$q$ will also transform under this symmetry as well, and axion shifts provide
an important constraint on possible contributions \cite{HMSSNV}. \ Letting
$\alpha$ denote the PQ charge of $q$, because $X$ has PQ\ charge $-4$,
equation (\ref{POLLIKE}) implies:%
\begin{equation}
\alpha=+4
\end{equation}
Letting $H$ denote a generic Higgs field, and $\Phi$ any other MSSM\ chiral
superfield, the corresponding PQ\ charges are respectively $-2$ and $+1$. \ It
follows that a candidate contribution of the form:%
\begin{equation}
W_{inst}^{tot}\supset q^{k}H^{a}\Phi^{b}X^{c}%
\end{equation}
must satisfy the constraint:%
\begin{equation}
4k-2a+b-4c=0 \label{gencon}%
\end{equation}
where $a$, $b$, $c$ and $k$ are non-negative integers.

\subsection{Estimates on Higher Order Corrections}

We now estimate possible higher order instanton corrections to the minimum of
the axion potential. \ The leading order contribution to the axion potential
from additional superpotential terms can come from terms quadratic in $X$ such
as $X^{2}$ and $H^{2}X^{2}$ so that equation (\ref{gencon}) reduces to:%
\begin{align}
X^{2}  &  :k=2\\
H^{2}X^{2}  &  :k=1\text{.}%
\end{align}
As shown in detail in \cite{HMSSNV}, there is no contribution to the $k=1$
sector from quadratic terms in $X$. \ Expanding to leading order in $q$
therefore yields the X-dependant superpotential:\footnote{In a previous
version of this paper, a constant contribution from the instanton sector was
assumed to be present. A recent clarification of the analysis appearing in a
revised version of \cite{HMSSNV} illustrates that when an appropriate flux is
available such that the Polonyi term is generated, then this constant shift is
absent.}%
\begin{equation}
W(X)=M_{PQ}^{2}\kappa_{1}\cdot qX+\frac{M_{W}^{2}\kappa_{H}}{2M_{PQ}}\cdot
qX^{2}+\frac{M_{PQ}\kappa_{2}}{2}\cdot q^{2}X^{2}%
\end{equation}
where $M_{W}$ is shorthand for the mass scale associated with the Higgs vevs,
and the $\kappa_{i}$ are moduli dependent worldvolume determinant factors
which also include possible contributions from a possibly position dependent
axio-dilaton. \ Assuming that all effects from gravity decouple, the resulting
axion potential is:%
\begin{equation}
V_{ax}\left(  a_{x}\right)  =V_{QCD}\left(  a_{x}\right)  +\left\vert
M_{PQ}^{2}\kappa_{1}\cdot q+\frac{M_{W}^{2}\kappa_{H}}{M_{PQ}}\cdot
q\left\vert x\right\vert e^{ia_{x}}+M_{PQ}\kappa_{2}\cdot q^{2}\left\vert
x\right\vert e^{ia_{x}}\right\vert ^{2}\text{.}%
\end{equation}
Returning to the bound of line (\ref{VTHETABOUND}), the minimum of $V_{ax}$
remains sufficiently close to zero to solve the strong CP\ problem provided:%
\begin{equation}
\frac{V^{\prime\prime}(\theta_{0})}{V_{QCD}^{\prime\prime}(0)}<10^{-10}%
\end{equation}
so that:%
\begin{align}
\frac{\left(  M_{PQ}^{2}\kappa_{1}\cdot q\right)  \times\left(  \frac
{M_{W}^{2}\kappa_{H}}{M_{PQ}}\cdot q\left\vert x\right\vert \right)  }%
{10^{-4}\text{GeV}^{4}}  &  <10^{-10}\\
\frac{\left(  M_{PQ}^{2}\kappa_{1}\cdot q\right)  \times\left(  M_{PQ}%
\kappa_{2}\cdot q^{2}\left\vert x\right\vert \right)  }{10^{-4}\text{GeV}%
^{4}}  &  <10^{-10}\text{.}%
\end{align}
This amounts to the two conditions:%
\begin{align}
\left\vert \kappa_{1}\kappa_{H}\right\vert M_{PQ}M_{W}^{2}\left\vert
x\right\vert \cdot q^{3}  &  <10^{-14}\text{ GeV}^{4}\\
\left\vert \kappa_{1}\kappa_{2}\right\vert M_{PQ}^{3}\left\vert x\right\vert
\cdot q^{3}  &  <10^{-14}\text{ GeV}^{4}.
\end{align}
Plugging in the explicit values $q\sim5\times10^{-17}$, $M_{PQ}\sim
4.3\times10^{16}$ GeV, $\left\vert x\right\vert \sim10^{12}$ GeV, $M_{W}%
\sim10^{2}$ GeV, this requires:%
\begin{align}
\left\vert \kappa_{1}\kappa_{H}\right\vert  &
<2 \times 10^{2}\\
\left\vert \kappa_{1}\kappa_{2}\right\vert  &  <10^{-27}\text{.}%
\end{align}
Provided $\kappa_{1}$ and $\kappa_{H}$ are not very large, the contribution
from the Higgs dependent contribution will not shift the axion potential by a large amount.
The contribution from the purely $X$ sector is far more problematic. We will
return to possible means by which the coefficient $\kappa_{2}$ could be
arranged to be quite small, so as to satisfy this bound.

On similar grounds, note that there are also instanton contributions to the
axion potential from supergravity:
\begin{equation}
V_{ax}\supset-\frac{3}{M_{pl}^{2}}\left\vert W(X)\right\vert ^{2}=-\frac
{3}{M_{pl}^{2}}\left\vert W_{0}+M_{PQ}^{2}\kappa_{1}\cdot qX\right\vert
^{2}\text{,} \label{myvax}%
\end{equation}
where here, $W_{0}$ denotes a generic constant shift to the overall
superpotential of the theory. The precise value of this constant is tied up
with the value of the cosmological constant, and therefore a full discussion
of this type of contribution is somewhat beyond the scope of the present
paper. Nevertheless, the condition to avoid a significant shift in the axion
potential from this contribution is:%
\begin{equation}
\left\vert W_{0}\kappa_{1}\right\vert \frac{M_{PQ}^{2}\left\vert x\right\vert
}{M_{pl}^{2}}\cdot q<10^{-14}\text{ GeV}^{4}%
\end{equation}
or:%
\begin{equation}
\left\vert W_{0}\kappa_{1}\right\vert <\left(  10^{7}\text{ GeV}\right)
^{3}\text{.}%
\end{equation}
Thus, in order to avoid generating a large contribution to the axion
superpotential, we must assume that $W_{0}$ is set by an energy scale smaller
than that due to other scales appearing in the gauge mediation sector.

To summarize the discussion above, there is a potentially significant
contribution to the axion potential from superpotential terms of the form
$q^{2}X^{2}$. In addition, possible constant shifts in the form of the
superpotential can also potentially induce corrections to the form of the
axion potential. We now proceed to discuss possible mechanisms which can
rectify this issue.

\subsection{Achieving a Flat Axion Potential}

The above analysis establishes that in order for the phase of $X$ to play the
role of the QCD axion, several instanton contributions must be sufficiently
suppressed. In this subsection we discuss possible resolutions of this issue.

First consider the contribution related to $W_{0}$. As alluded to below line
(\ref{myvax}), the value of $W_{0}$ is also closely connected with the overall
value of the cosmological constant. In keeping with the approach espoused in this
paper we defer this and other issues related to gravity to a later stage of analysis.
Even so , we note that it is conceivable that a fine tuning in $W_{0}$ may be
available such that this term is sufficiently small.

Next consider instanton induced contributions of the form $q^{2}X^{2}$. The
primary question is whether the coefficient $\kappa_{2}$ can be arranged to be
sufficiently small to avoid possible issues with the disruption of the axion
potential. There are in principle a few ways in which this type of correction
could be arranged, which we now discuss in turn.

One natural possibility is that the characteristic scale multiplying the
various instanton contributions might involve a suppression scale closer to
$M_{pl}$ rather than $M_{PQ}\sim4\times10^{16}$ GeV. To determine the relevant
form of the contribution, note that the higher order instanton contributions
can be written as:%
\begin{equation}
W(X)=\kappa_{1}F_{X}\cdot X+\frac{\kappa_{2}}{2}\frac{\left(  F_{X}\cdot
X\right)  ^{2}}{M^{3}}+...\text{,}%
\end{equation}
with $F_{X}\sim10^{17}$ GeV$^{2}$, and $M$ some suppression scale. Although
local considerations naturally suggest the value $M\sim M_{PQ}$, it is in
principle possible to also consider larger values such as $M\sim M_{pl}%
\sim10^{19}$ GeV. Computing the overall constraint on the flatness of the
axion potential, we now obtain the constraint:%
\begin{equation}
\frac{\left\vert \kappa_{1}\kappa_{2}\right\vert \left\vert x\right\vert
F_{X}^{3}/M_{pl}^{3}}{10^{-4}\text{GeV}^{4}}<10^{-10}\text{,}%
\end{equation}
or:%
\begin{equation}
\left\vert \kappa_{1}\kappa_{2}\right\vert <10^{-20}\text{,}%
\end{equation}
which leads to a slightly less stringent constraint on the $\kappa$ coefficients.

As another possibility, it would be interesting to investigate whether a
suitable class of instanton configurations could be found such that only odd
powers of $q$ appear. In this case, form of the instanton expansion would be
of the form:%
\begin{equation}
W(X)=\kappa_{1}F_{X}\cdot X+\frac{\kappa_{3}}{3}\frac{\left(  F_{X}\cdot
X\right)  ^{3}}{M^{6}}...\text{.}%
\end{equation}
The overall constraint on the flatness of the axion potential would then be of
the form:%
\begin{equation}
\frac{\left\vert \kappa_{1}\kappa_{3}\right\vert \left\vert x\right\vert
^{2}F_{X}^{4}/M^{6}}{10^{-4}\text{GeV}^{4}}<10^{-10}\text{,}%
\end{equation}
so that even for $M\sim4\times10^{16}$ GeV, we obtain:%
\begin{equation}
\left\vert \kappa_{1}\kappa_{3}\right\vert <10^{-6}\text{,}%
\end{equation}
which again involves less fine tuning in the $\kappa$'s. For $M \sim M_{pl}$, the constraint is instead:
\begin{equation}
\left\vert \kappa_{1}\kappa_{3}\right\vert <10^{8}\text{,}%
\end{equation}
which is a far milder constraint.

In addition to the moduli dependent worldvolume determinant
factors, the profile of the axio-dilaton can also lead to an overall
suppression of the $\kappa$'s. Indeed, the zero mode contribution necessary
for a bound state of Euclidean D3-branes to contribute often requires the
presence of a flux threading the worldvolume of the Euclidean three-brane.
Integrating the profile of the axio-dilaton against the instanton density of
this flux, the overall position dependence of the axio-dilaton implies that in
generall, fluxes with different instanton numbers could produce quite
different suppression factors in the $\kappa$'s.

As a final possibility, it was found in \cite{HMSSNV} that in order for a Euclidean D3-instanton to
contribute to the superpotential at all, the flux threading the D3-brane must
lift to a trivial class in the threefold base $B_{3}$ of an F-theory
compactification. Although beyond the scope of this paper, it would be
interesting to determine whether it is possible to arrange for a
one-instanton sector to contribute, but to exclude contributions from higher
instantons due to a topological obstruction of some kind.

\newpage
\bibliographystyle{ssg}
\bibliography{fgutsgmsb}

\end{document}